\documentclass[10pt]{amsart}
\makeatletter
\tagsleft@false 
\makeatother
\usepackage[utf8]{inputenc}         
\usepackage{indentfirst}            
\usepackage{csquotes}              
\usepackage{graphicx}              
\usepackage{xcolor}                
\usepackage{epstopdf}              
\usepackage{textcomp}              
\usepackage{verbatim}              
\usepackage{amsmath, amssymb, amsthm, amsfonts}
\usepackage{booktabs}             
\usepackage{multirow}
\usepackage{array}
\usepackage{colortbl}             
\usepackage{tabularx}
\usepackage{threeparttablex}
\usepackage{longtable}
\usepackage{caption}
\usepackage{subfig}
\usepackage{float}                
\usepackage{stfloats} 
\usepackage{makecell}
\usepackage{algorithm}
\usepackage{algorithmic}
\usepackage{enumerate}
\usepackage{url}
\usepackage{breakurl}
\usepackage{flushend}             
\usepackage{soul}                 
\usepackage{pifont}               
\usepackage{cuted}                
\usepackage{picinpar}             
\usepackage{alltt}                
\usepackage{cite}
\usepackage[hidelinks]{hyperref}                 
\usepackage{etoolbox}

\hypersetup{ colorlinks=true, linkcolor=black, filecolor=black, urlcolor=black }

\usepackage{lipsum}
\DeclareGraphicsExtensions{.pdf}
\graphicspath{{path0/}{path1/}}

\begin{document}
\title{A Novel Two-Dimensional Smoothing Algorithm} 
\author[Initial Surname]{Xufeng Chen, Liang Yan, Xiaoshan Gao}
\email{chenxufeng@buaa.edu.cn, lyan1991@gmail.com, gaoshan0920@buaa.edu.cn}
\maketitle

\let\thefootnote\relax

\begin{abstract}
Smoothing and filtering two-dimensional sequences are fundamental tasks in fields such as computer vision. Conventional filtering algorithms often rely on the selection of the filtering window, limiting their applicability in certain scenarios. To this end, we propose a novel Two-Dimensional Smoothing (TDS) algorithm for the smoothing and filtering problem of two-dimensional sequences. Typically, the TDS algorithm does not require assumptions about the type of noise distribution. It is simple and easy to implement compared to conventional filtering methods, such as 2D adaptive Wiener filtering and Gaussian filtering. The TDS algorithm can effectively extract the trend contained in the two-dimensional sequence and reduce the influence of noise on the data by adjusting only a single parameter. In this work, unlike existing algorithms that depend on the filtering window, we introduce a loss function, where the trend sequence is identified as the solution when this loss function takes a minimum value. Therefore, within the framework of the TDS algorithm, a general two-dimensional sequence can be innovatively decomposed into a trend sequence and a fluctuation sequence, in which the trend sequence contains the main features of the sequence and the fluctuation sequence contains the detailed features or noise interference of the sequence. To ensure the reliability of the TDS algorithm, a crucial lemma is first established, indicating that the trend sequence and fluctuation sequence obtained by the TDS algorithm are existent and unique when the global smoothing parameter is determined. Three modified algorithms, namely TDS-\uppercase\expandafter{\romannumeral1}, TDS-\uppercase\expandafter{\romannumeral2}, TDS-\uppercase\expandafter{\romannumeral3}, are then proposed based on the TDS algorithm, with corresponding lemmas and corollaries demonstrating their reliability. Finally, the accuracy and effectiveness of the TDS algorithm are further verified through numerical simulations and image processing cases. The results show that the smoothing effect of the proposed TDS algorithm is excellent compared to traditional filtering algorithms in terms of MSE, PSNR and SSIM metrics in several challenging scenarios. This study provides a versatile and effective method for signal processing, pattern recognition, wireless communication, computer vision, image processing.
\end{abstract}

\section{Introduction}

Two-dimensional filtering is one of the most important tasks in computer vision, with numerous applications such as image analysis, action recognition, wireless communication, and autonomous driving \cite{761332, 10143284, 4409077,ma2021image,zhang2023underwater,5166446,7368945}, to name a few. In this work, we propose a novel TDS algorithm for the smoothing and filtering problem of two-dimensional sequences. However, to appreciate the contributions presented in this note, we first recall some of the key results that have been made pertaining to the field of two-dimensional filtering.

In the past few years, several elegant two-dimensional filtering methods, including nonlinear, linear, and partial differential equation (PDE) have attracted the attention of many researchers.

For nonlinear filtering methods, median filtering is a typical approach and can effectively remove noise from digital images or digital signals. It works well for low levels of salt-and-pepper noise (SPN). However, it is not satisfactory for high levels of SPN, and is also unable to preserve the edges of the processing image. Additionally, median filtering is relatively expensive with high time complexity. To address the challenges posed by high levels of SPN, several modified algorithms are proposed based on median filtering. These include the adaptive median filter (AMF) \cite{hwang1995adaptive}, the fast and efficient median filter (FEMF) \cite{hsieh2013fast}, the noise adaptive fuzzy switching median filter (NAFSM) \cite{toh2009noise}, adaptive center weighted median filter (ACWMF) \cite{lin2007new}, adaptive weighted mean filter (AWMF) \cite{zhang2014new}, and others. Among them, AMF is mainly used for eliminating impulse noise and SPN but may not be as effective for other types of noise, such as Gaussian noise. FEMF has a fast execution speed, but the denoising effect is greatly influenced by the window size, which requires adjustments based on specific applications. NAFSM employs fuzzy reasoning to handle uncertainty in the extracted local information caused by noise. ACWMF is an adaptive version of the median filter designed for noisy images with low contamination ratio. It performs varying levels of smoothing for noisy images owing to its adaptive nature. AWMF is used to detect and remove high levels of SPN, determining the adaptive window size by continuously expanding it. However, these methods still degrade image quality since they process each pixel in the noisy image without considering whether it is noise or not.

To address the aforementioned issue, several filtering algorithms integrated with noise detectors have been developed. Many of these algorithms are based on median or mean filters such as the directional weighted median filter (DWM) \cite{Dong2007193}, the adaptive switching median filter (ASWM) \cite{Akkoul2010587}, the optimal direction median filter \cite{Awad2011407}, the two-pass switching rank-ordered arithmetic mean (TSRAM) filter \cite{lin2010switching}, and the ROR non-local mean (ROR-NLM) \cite{Xiong20121663}. However, despite the above modified median filter algorithm has been widely used in some specific scenarios, they still face challenges when dealing with complex noise and heavily rely on the selection of the filtering window.
 
Other outstanding nonlinear filtering methods include the bilateral filter (BF) and non-local mean (NLM) filter, among others. While BF and NLM filters have good performance in edge preservation, they have poor denoising performance in Rayleigh distribution noise (speckle noise). Moreover, BF represents gradient distortion and high complexity \cite{2008Multiresolution}. The NLM filter based on BF are suitable for additive white Gaussian noise (AWGN) denoising and not for despeckling \cite{TORRES2014141}.
 
For linear filtering, the Gaussian filter based on Gaussian function is one of the most commonly used smoothing filters in signal and image processing applications. It is well-suited for tasks such as image denoising, image smoothing, and feature extraction. Singh \cite{9146980} conducted a detailed review for Gaussian filtering and its variants. Following the development of the unscented Kalman filter (UKF), many scholars have reexamined the Gaussian filter approach from different perspectives. Square-root filtering \cite{Bhaumik,4524036,Potnuru,6340302,TANG201284} and Gaussian-sum filtering \cite{6494405,6746072} have emerged as popular developments in this category. The objective of square-root filtering is to preclude the positive-definite requirement of error covariance matrix, which is essential in traditional Gaussian filters. On the other hand, Gaussian-sum filtering enhances estimation accuracy by approximating unknown probability density functions (pdf) with multiple Gaussian components instead of a single Gaussian. However, the objective of square-root filtering is limited to improving filter applicability. It does not influence the estimation accuracy significantly, and its computational efficiency is also similar to that of the ordinary filtering algorithm, but the multiple Gaussian approximates this arbitrary shape pdf with better precision compared to the ordinary filters. 

PDE methods are useful for noise elimination, where a noisy image is transformed into PDE forms to obtain a noise-free image \cite{Xu:17}. Another common filtering method is adaptive Wiener filtering (AWF) and its variants \cite{Abe}. AWF utilizes the center pixel value and local statistical values to determine weights, enabling adaptive filtering on the input image and resulting in a high-quality restored image. However, the computational complexity of AWF is high, demanding significant computing resources and time.

Motivated by the challenges mentioned earlier, we propose a novel TDS algorithm for the smoothing and filtering of two-dimensional sequences. Unlike existing methods that rely on the filtering window, the TDS model we propose considers the overall characteristics of the two-dimensional sequence and is independent of the filtering window. Within the TDS algorithm framework, a loss function is defined, and the trend sequence is identified as the solution when this loss function reaches its minimum value. This innovative approach allows a general two-dimensional sequence to be decomposed into a trend sequence and a fluctuation sequence, providing significant insights for the feature decomposition of the research object. Moreover, the TDS algorithm eliminates the need to assume the type of noise distribution and is simple and easy to implement. Consequently, the TDS algorithm is well-suited for various types of noise reduction, including additive noise and multiplicative noise, and it outperforms traditional filtering algorithms in several scenarios. Additionally, TDS proves effective for tasks such as image smoothing, sharpening, and enhancement. Our main contributions can be summarized as follows:

\begin{itemize}
	\item The proposed TDS algorithm offers a versatile and effective technology for smoothing and filtering two-dimensional sequences.
	\item Addressing the reliability of the TDS algorithm, a crucial lemma is obtained, affirming that the trend sequence and fluctuation sequence derived by the TDS algorithm are both existent and unique when the global smoothing parameter is determined.
	\item The TDS algorithm is extended to propose three modified algorithms: TDS-\uppercase\expandafter{\romannumeral1}, TDS-\uppercase\expandafter{\romannumeral2}, TDS-\uppercase\expandafter{\romannumeral3}. Meanwhile,  corresponding lemmas and corollaries are acquired and demonstrated, establishing the reliability of these three modifications.
	\item As the global smoothing parameter reaches a certain threshold, the impact of its variation on the smoothing result diminishes significantly, which shows that the TDS algorithm has better robustness.
	\item The TDS algorithm integrates multiple functions, including image smoothing, sharpening, enhancement, and edge extraction. Additionally, it provides a novel and effective approach for signal processing, pattern recognition, wireless communication, computer vision, and image processing.
\end{itemize}

The rest of this paper is organized as follows: Section \ref{sec 2} provides a detailed description of the proposed TDS algorithm method. In section \ref{sec 3}, we present the lemma and corollary, along with the corresponding proof for the TDS algorithm. Section \ref{sec 4} introduces the modified TDS algorithm, along with the associated lemmas and corollaries (TDS-\uppercase\expandafter{\romannumeral1}, TDS-\uppercase\expandafter{\romannumeral2}, and TDS-\uppercase\expandafter{\romannumeral3}). The numerical simulation and image processing cases of the TDS algorithm are discussed in section \ref{sec 5}. Section \ref{sec 6} contains the conclusion. Additionally, the specific proof of the TDS algorithm and other supporting proofs can be found in the Appendix.

\section{Algorithm Modeling}\label{sec 2}

In this section, we present the TDS algorithm model and some properties, which we will need in the sequel.

Consider a two-dimensional sequence, which can be regarded as a set of data $ z_{i,j} $ collected at different sampling points $ \left(x_i,y_j\right) $ ($ i=1,2,\cdots,m.\ j=1,2,\cdots,n $). However, due to the presence of noise interference, these data points often include noise in addition to the underlying real trend. Therefore, it becomes necessary to analyze the genuine trend of the two-dimensional sequence by eliminating the noise interference. In this case, the real trend of the sampling data $ z_{i,j} $ can be seen as the trend of the data after noise removal. Thus, the following general two-dimensional sequence is analyzed.

Assuming that the two-dimensional sequence is a $ m\times n $ matrix and can be expressed as $ \textit{\textbf{Z}}=\left(z_{i,j}\right)\in  \mathbb{R}^{m\times n} $. Besides, the sequence \textit{\textbf{Z}} can be observed but is generally interfered with by random noise. Thus, the matrix \textit{\textbf{Z}} is decomposed into a trend sequence \textbf{\textit{G}} and a noise sequence \textit{\textbf{C}}, i.e.,

\begin{align}
	\textit{\textbf{Z}}=\textit{\textbf{G}}+\textit{\textbf{C}}
\end{align}

\noindent where $ \textit{\textbf{G}}=\left(g_{i,j}\right)\in \mathbb{R}^{m\times n} $ and  $ \textit{\textbf{C}}=\left(c_{i,j}\right)\in  \mathbb{R}^{m\times n} $ are unobservable sequences. In fact, there are two cases for the trend sequence \textit{\textbf{G}} and the noise sequence \textit{\textbf{C}}, respectively.

\begin{figure}[!ht]
	\centering
	\subfloat[Original sequence $ \textit{\textbf{Z}} $. ]{\includegraphics[width=.31\columnwidth]{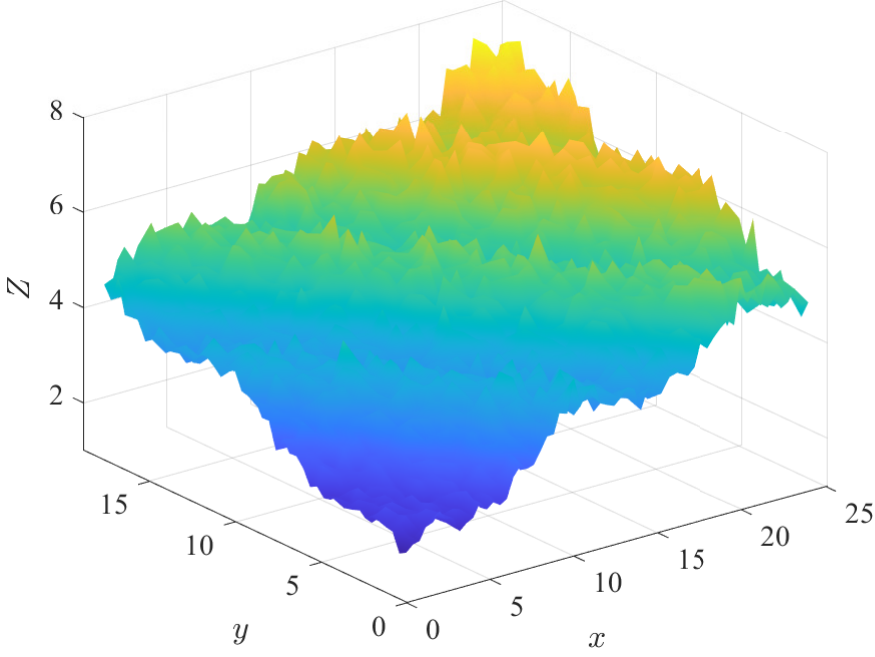}}\hspace{5pt}
	\subfloat[Trend $ \textit{\textbf{G}} $.  ]{\includegraphics[width=.31\columnwidth]{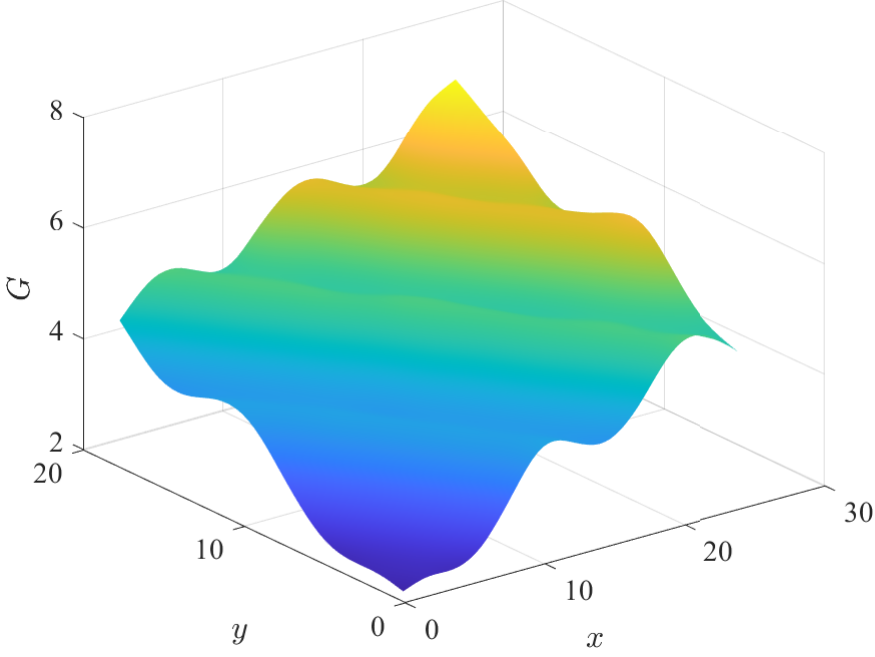}}\hspace{5pt}
	\subfloat[Noise $ \textit{\textbf{C}} $. ]{\includegraphics[width=.31\columnwidth]{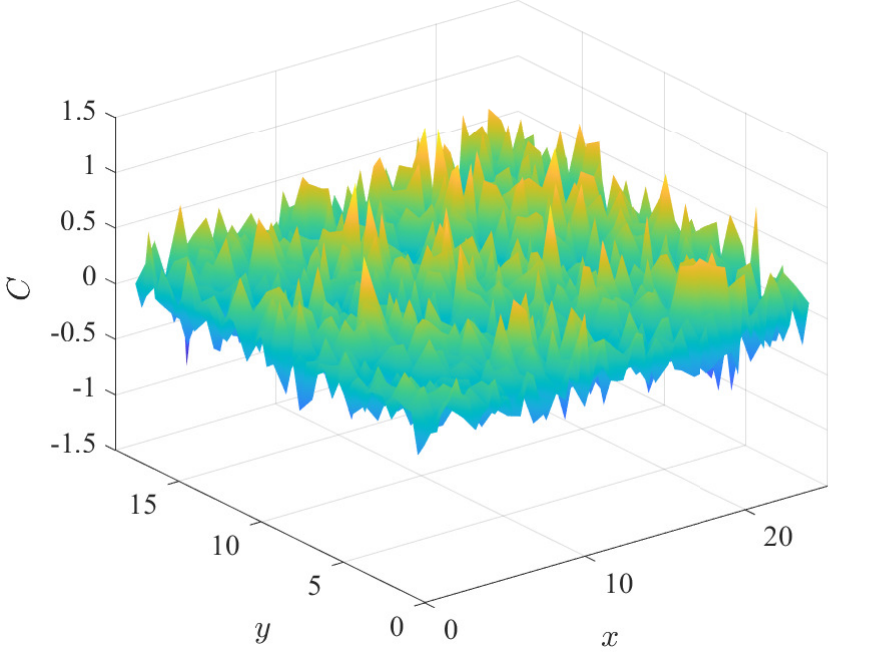}}
	\caption{There is noise interference.}
	\label{Fig1}
\end{figure}

\begin{figure}[!ht]
	\centering
	\subfloat[Original sequence $ \textit{\textbf{Z}} $. ]{\includegraphics[width=.31\columnwidth]{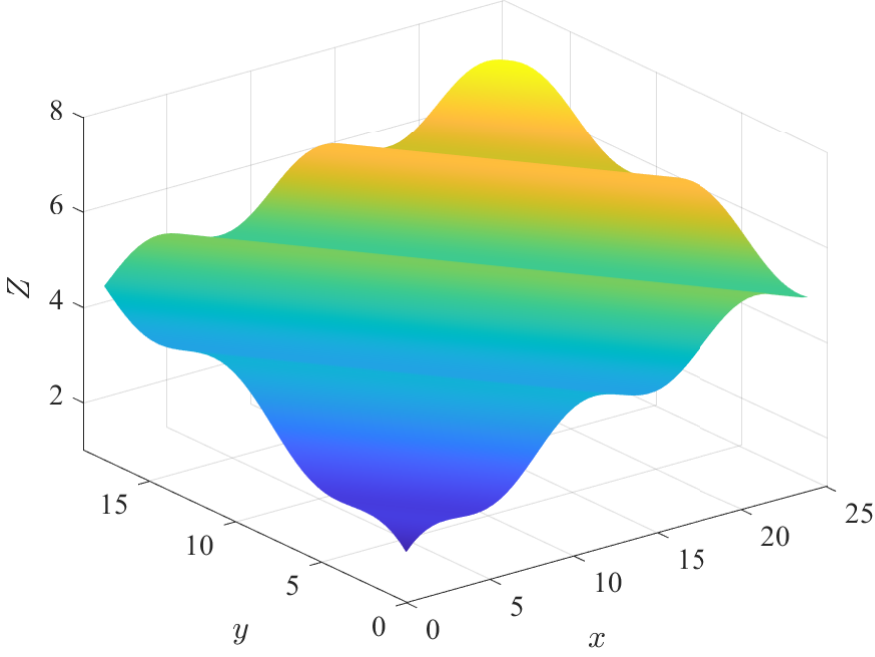}}\hspace{5pt}
	\subfloat[Trend $ \textit{\textbf{G}} $. ]{\includegraphics[width=.31\columnwidth]{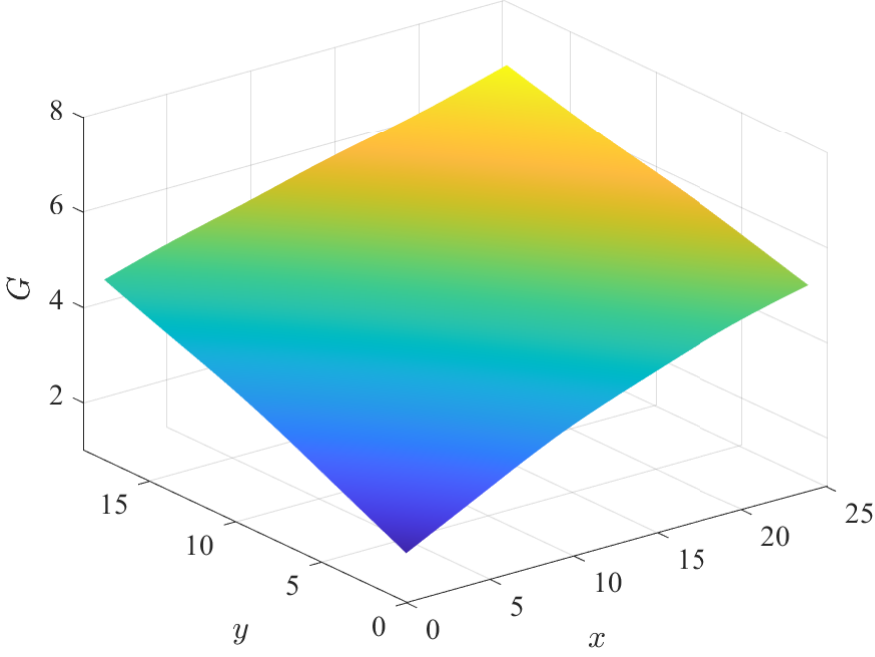}}\hspace{5pt}
	\subfloat[Fluctuation $ \textit{\textbf{C}} $. ]{\includegraphics[width=.31\columnwidth]{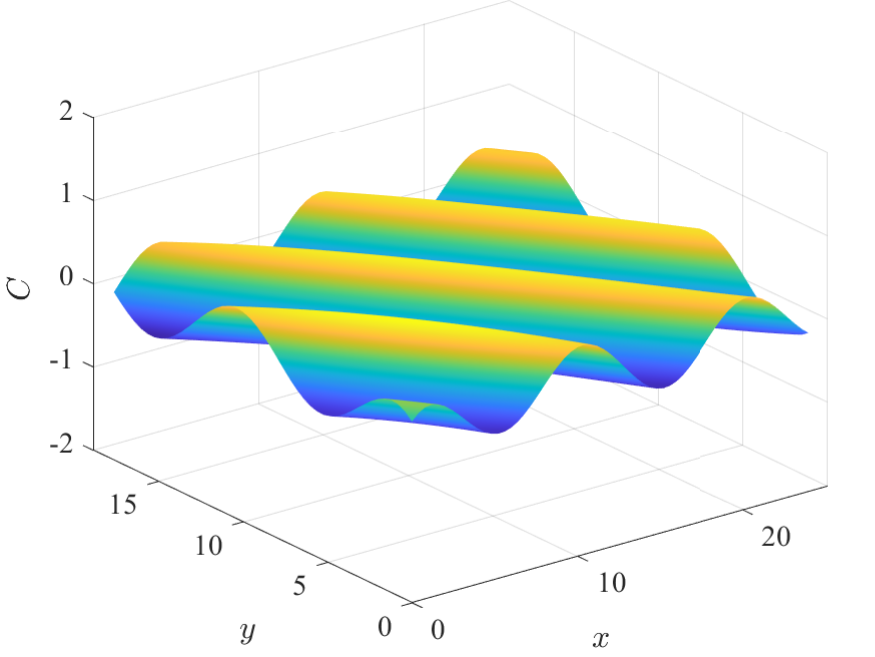}}
	\caption{There is no noise interference.}
	\label{Fig2}
\end{figure}

\begin{description}
	\item[\textit{\textbf{Case1}}] The sequence \textit{\textbf{G}} can be interpreted as all useful information contained in the two-dimensional sequence \textit{\textbf{Z}} when the two-dimensional sequence \textit{\textbf{Z}} is interfered with by noise. Namely, the sequence \textit{\textbf{G}} contains all useful information parts of the sequence \textit{\textbf{Z}}, whereas the sequence \textit{\textbf{C}} contains noise parts of the sequence \textit{\textbf{Z}}, as shown in Fig.\ref{Fig1}.
	\item[\textit{\textbf{Case2}}] When there is no noise interference, the sequence \textit{\textbf{G}} can be interpreted as the trend contained in the two-dimensional sequence \textit{\textbf{Z}}, and the sequence \textit{\textbf{C}} can be understood as the fluctuation of sequence \textit{\textbf{Z}}. In other words, the sequence \textit{\textbf{G}} contains main feature of the sequence \textit{\textbf{Z}}, whereas the sequence \textit{\textbf{C}} contains detailed feature of the sequence \textit{\textbf{Z}}, as shown in Fig.\ref{Fig2}.
\end{description} 

For convenience, in the following the sequence \textit{\textbf{C}} is uniformly referred to as fluctuation sequence.

Within the framework of this smoothing algorithm, the loss function $ M $  is firstly defined as

\begin{align}
	\begin{split}
		M=&\sum_{i=1}^{m}\sum_{j=1}^{n}\left(z_{i,j}-g_{i,j}\right)^2+\lambda\\\times&\Big\{\sum_{i=1}^{m}\sum_{j=3}^{n}\left[\left(g_{i,j}-g_{i,j-1}\right)-\left(g_{i,j-1}-g_{i,j-2}\right)\right]^2\\+&
		\sum_{j=1}^{n}\sum_{i=3}^{m}\left[\left(g_{i,j}-g_{i-1,j}\right)-\left(g_{i-1,j}-g_{i-2,j}\right)\right]^2 \Big\}
	\end{split}
\end{align}

The trend sequence \textit{\textbf{G}} is identified as the solution when the loss function $ M $ takes the minimum value. In other words, when the values of the trend sequence \textit{\textbf{G}} minimize the loss function $ M $, \textit{\textbf{G}} is the sought solution. Additionally, the number of rows $ m $ and columns $ n $ of the two-dimensional sequence \textit{\textbf{Z}} must satisfy $ m\geq3 $ and $ n\geq3 $.

For the purpose of analysis, we introduce the difference operators $ \Delta $ and $ \nabla $, where

\begin{align}
	\begin{split}
			\Delta g_{i,j}=g_{i,j}-g_{i,j-1}
			\\\nabla g_{i,j}=g_{i,j}-g_{i-1,j}
	\end{split}
\end{align}

Thereupon

\begin{align}
	\begin{split}
			\Delta^2g_{i,j}=\Delta\left(\Delta g_{i,j}\right)=g_{i,j}-2g_{i,j-1}+g_{i,j-2}
			\\\nabla^2g_{i,j}=\nabla\left(\nabla g_{i,j}\right)=g_{i,j}-2g_{i-1,j}+g_{i-2,j}
	\end{split}
\end{align}

\begin{figure}[htbp]
	\centering
	\subfloat[The difference operator and computing nodes. ]{\label{Fig3:a} \includegraphics[width=.47\columnwidth]{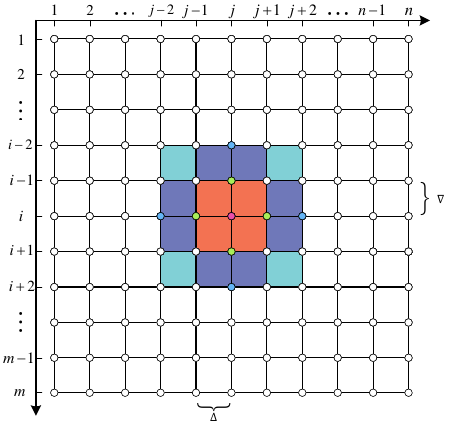}}\hspace{5pt}
	\subfloat[Algorithm structure.]{\label{Fig3:b}
	\includegraphics[width=.47\columnwidth]{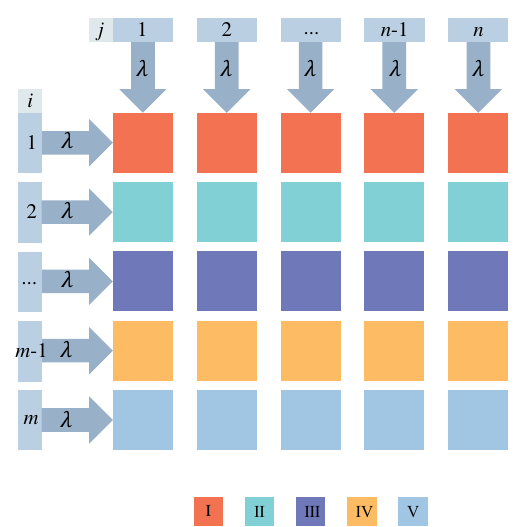}}
	\caption{The structure and computing nodes of TDS algorithm.}\label{Fig.3}
\end{figure}

The operating relationship of the difference operator $ \Delta $ and $ \nabla $ is shown in Fig.\ref{Fig.3}\subref{Fig3:a}.

Then

\begin{align}
	\begin{split}
			M=&\sum_{i=1}^{m}\sum_{j=1}^{n}\left(z_{i,j}-g_{i,j}\right)^2+\lambda \big\{\sum_{i=1}^{m}\sum_{j=3}^{n}\left[\Delta^2g_{i,j}\right]^2\\+&
		\sum_{j=1}^{n}\sum_{i=3}^{m}\left[\nabla^2g_{i,j}\right]^2\Big\}
	\end{split}
\end{align}

As shown in Fig.\ref{Fig.3}\subref{Fig3:b}, $ \lambda $ ($ \lambda>0 $) can be referred to as \textit{\textbf{the global smoothing parameter}} of this smoothing algorithm. Here are some characteristics:

\begin{enumerate}
	\item 	When $ \lambda=0 $, if the loss function $ M $ reaches the minimum value, then we have $ g_{i,j}=z_{i,j} $ ($ i=1,2,\cdots,m $. $ j=1,2,\cdots,n $), i.e., the trend sequence \textit{\textbf{G}} is the two-dimensional sequence \textit{\textbf{Z}} itself when $ \lambda\ =\ 0 $.
	\item 	When $ \lambda=+\infty $, if the loss function $ M $ reaches the minimum value, then the trend sequence \textit{\textbf{G}} converges to a definitive sequence independent of $ \lambda $.
	\item 	When $ \lambda $ is increases, the distribution of the trend sequence \textit{\textbf{G}} becomes smoother.
\end{enumerate}

In fact, $ \lambda $ also means the balance between fitting and smoothing in the loss function $ M $. For convenience, let

\begin{align}
	\begin{split}
	r\left(i,j\right)=&\left(z_{i,j}-g_{i,j}\right)^2,\left(i=1,\cdots, m.\ j=1,\cdots, n\right)
	\\p\left(i,j\right)=&\left[\Delta^2g_{i,j}\right]^2,\left(i=1,\cdots, m.\ j=3,\cdots, n\right)
	\\q\left(i,j\right)=&\left[\nabla^2g_{i,j}\right]^2,\left(i=3,\cdots,m.\ j=1,\cdots, n\right)
	\end{split}
\end{align}

Let $ R=\sum_{i=1}^{m}\sum_{j=1}^{n}r\left(i,j\right)$,  $P=\sum_{i=1}^{m}\sum_{j=3}^{n}p\left(i,j\right)$,  $Q=\sum_{j=1}^{n}\sum_{i=3}^{m}q\left(i,j\right)$. 
 
Then, $ M $ can be rewritten as

\begin{align} \label{13}
	M=R+\lambda\left(P+Q\right)
\end{align}

In the \textit{Eqn}.\eqref{13}
\begin{enumerate}
	\item \textit{R} portrays the extent to which the trend sequence \textit{\textbf{G}} tracks the original sequence \textit{\textbf{Z}}.
	\item When the number of rows $ i $ is determined, \textit{P} portrays the smoothing of the trend sequence \textit{\textbf{G}} on different columns.
	\item When the number of columns $ j $ is determined, \textit{Q} portrays the smoothing of the trend sequence \textit{\textbf{G}} on different rows.
\end{enumerate}

At the same time, to obtain the extreme value of the loss function $ M $, we take partial derivatives of  $ M $ with respect to $ g_{i,j} $, set $ \partial M/\partial g_{i,j}=0 $, and consequently, the trend sequence \textit{\textbf{G}} can be solved. The sequence \textit{\textbf{G}} is the solution of \textit{Eqn}.\eqref{14} (see Appendix \ref{Appendix A} for details).

\begin{align} \label{14}
	\textit{\textbf{G}}+\lambda\left(\textit{\textbf{GT}}+\textit{\textbf{HG}}\right)=\textit{\textbf{Z}}
\end{align}

\noindent where $ \textit{\textbf{T}}\in \mathbb{R}^{n\times n} $ and $ \textit{\textbf{H}}\in \mathbb{R}^{m\times m} $, and both are positive semidefinite and real symmetric matrices.

For matrices \textit{\textbf{T}} and \textit{\textbf{H}}, the following conclusion can be obtained.

Let $ k $-order square matrix \textit{\textbf{N}} be
\begin{enumerate}
	\item 	When $ k=3 $ 	
	\begin{align}
			\textit{\textbf{N}}=\left[\begin{matrix}1&-2&1\\-2&4&-2\\1&-2&1\\	\end{matrix}\right]
	\end{align}
	\item	When $ k=4 $
			\begin{align}
			\textit{\textbf{N}}=\left[\begin{matrix}1&-2&1&0\\-2&5&-4&1\\1&-4&5&-2\\0&1&-2&1\\\end{matrix}\right]
	\end{align}
	\item 	When $ k>4 $
	\begin{align}
	\textit{\textbf{N}} = \begin{bmatrix}
	1 & {- 2} & 1 & & & & \\
	{- 2} & 5 & {- 4} & 1 & & & \\
	1 & {- 4} & 6 & {- 4} & 1 & & \\
	& \ddots & \ddots & \ddots & \ddots & \ddots & \\
	& & 1 & {- 4} & 6 & {- 4} & 1 \\
	& & & 1 & {- 4} & 5 & {- 2} \\
	& & & & 1 & {- 2} & 1 \\
\end{bmatrix}
\end{align}
\end{enumerate}

The matrix \textit{\textbf{N}} is a $ k $-order square matrix, whose sum of each row and column is zero, respectively. In addition, the matrix \textit{\textbf{N}} is a positive semidefinite and real symmetric matrix, so all eigenvalues of the matrix \textit{\textbf{N}} are real numbers, and the eigenvalues $ \zeta $ of the matrix \textit{\textbf{N}} satisfy $ 0\le\zeta\le16 $ (see Appendix \ref{Appendix B} for details). Also, it is easy to know $ \mathrm{rank}\left(\textit{\textbf{N}}\right)=k-2 $ after elementary operation, and therefore the matrix \textit{\textbf{N}} has two eigenvalues of $ 0 $. Furthermore, the square matrix \textit{\textbf{N}} is denoted as \textit{\textbf{H}} when the order $ k $ of the matrix \textit{\textbf{N}} is equal to $ m $, and the square matrix \textit{\textbf{N}} is denoted \textit{\textbf{T}} when the order $ k $ of the matrix \textit{\textbf{N}} is equal to $ n $. In addition, \textit{\textbf{N}} is also a sparse matrix. For example, the proportion of nonzero elements is only $ 4.94 \% $ when $ k = 100 $.

In \textit{Eqn}.\eqref{14}, the elements of the matrix \textit{\textbf{G}} are all unobservable values, so it contains $ m\times n $ unknowns. Meanwhile, the \textit{Eqn}.\eqref{14} contains $ m\times n $ equations. Therefore, the trend sequence \textit{\textbf{G}} can be solved by \textit{Eqn}.\eqref{14}. Furthermore, the fluctuation sequence \textit{\textbf{C}} can be expressed as

\begin{align}
	\textit{\textbf{C}}=\textit{\textbf{Z}}-\textit{\textbf{G}}
\end{align}

According to \textit{Eqn}.\eqref{14}, then

\begin{align}
	\lambda\left(\textit{\textbf{GT}}+\textit{\textbf{HG}}\right)=\textit{\textbf{C}}
\end{align}

For ease of description, we refer to this algorithm as the TDS algorithm. It can be seen as an algorithm that is precisely set to extract a smooth surface from the data.

\section{Proof of existence and uniqueness of trend sequence}\label{sec 3}

In this section, we present a pivotal lemma and corollary of the TDS algorithm, both of which hold great significance for the application and reliability of the algorithm.

\textbf{Lemma 1}: A general two-dimensional sequence $ \textit{\textbf{Z}}\in \mathbb{R}^{m\times n} $ ($ m\geq3 $, $ n\geq3 $) can be decomposed into a trend sequence $ \textit{\textbf{G}}\in \mathbb{R}^{m\times n} $ and a fluctuation sequence $ \textit{\textbf{C}}\in \mathbb{R}^{m\times n} $. Moreover, the trend sequence \textit{\textbf{G}} and the fluctuation sequence \textit{\textbf{C}} satisfy the following matrix equations, respectively

\begin{gather} 
	\textit{\textbf{G}}+\lambda\left(\textit{\textbf{GT}}+\textit{\textbf{HG}}\right)=\textit{\textbf{Z}}\label{20}\\
	\lambda\left(\textit{\textbf{GT}}+\textit{\textbf{HG}}\right)=\textit{\textbf{C}}\label{21}
\end{gather}

\noindent where $ \lambda $ is \textit{\textbf{the global smoothing parameter}} of the TDS algorithm. Both $ \textit{\textbf{T}}\in \mathbb{R}^{n\times n} $ and $ \textit{\textbf{H}}\in \mathbb{R}^{m\times m} $ are positive semidefinite and real symmetric matrix. Besides, the eigenvalues of the matrix \textit{\textbf{T}} and \textit{\textbf{H}} satisfy $ 0\le\zeta\le16 $, and $ \mathrm{rank}\left(\textit{\textbf{T}}\right)=n-2 $, $ \mathrm{rank}\left(\textit{\textbf{H}}\right)=m-2 $.

Furthermore, the trend sequence \textit{\textbf{G}} and the fluctuation sequence \textit{\textbf{C}} are, respectively

\begin{gather}
	\mathrm{vec}\left(\textit{\textbf{G}}\right)=\left(\textit{\textbf{I}}_n\otimes\textit{\textbf{I}}_m+\lambda\textit{\textbf{T}}\otimes\textit{\textbf{I}}_m +\lambda\textit{\textbf{I}}_n\otimes\textit{\textbf{H}}\right)^{-1}\mathrm{vec}\left(\textit{\textbf{Z}}\right)
	\\\mathrm{vec}\left(\textit{\textbf{C}}\right)=\lambda\left(\textit{\textbf{T}}\otimes\textit{\textbf{I}}_m+\textit{\textbf{I}}_n\otimes\textit{\textbf{H}}\right)\mathrm{vec}\left(\textit{\textbf{G}}\right)
\end{gather}

\noindent where both $ \textit{\textbf{I}}_n $ and $ \textit{\textbf{I}}_m $ are identity matrices, and both \textit{\textbf{G}} and \textit{\textbf{C}} are unique and exist when $ \lambda $ is determined.

In addition, it is worth noting that \textit{\textbf{T}} and \textit{\textbf{H}} are sparse matrices, which greatly reduces the complexity of the TDS algorithm.

According to Lemma 1, we have the following corollary.

\textbf{Corollary 1}: Assume that $ \textit{\textbf{Z}}\in  \mathbb{R}^{m\times n} $ is a two-dimensional sequence, and $ \textit{\textbf{G}}\in \mathbb{R}^{m\times n} $ is the trend sequence of the sequence \textit{\textbf{Z}}, then the trend sequence \textit{\textbf{G}} satisfy the following Sylvester equation

\begin{align}\label{24}
	\textit{\textbf{GA}}+\textit{\textbf{BG}}=\textit{\textbf{Z}}
\end{align}

\noindent where $ \textit{\textbf{A}}=\textit{\textbf{I}}_n/2+\lambda\textit{\textbf{T}} $, $ \textit{\textbf{B}}=\textit{\textbf{I}}_m/2+\lambda\textit{\textbf{H}} $.

Furthermore, the trend sequence \textit{\textbf{G}} can be solved as

\begin{align}\label{19}
	\textit{\textbf{G}}=\int_{0}^{+\infty}{e^{-\textit{\textbf{B}}t}\textit{\textbf{Z}}e^{-\textit{\textbf{A}}t}\mathrm{d}t}
\end{align}

The above corollary shows that the trend sequence \textit{\textbf{G}} of the sequence \textit{\textbf{Z}} in the TDS algorithm is the solution of a certain kind of Sylvester equation.

\textit{Proof}:

\textit{Eqn}.\eqref{14} can be further expressed as

\begin{align}\label{26}
	\textit{\textbf{G}}\left(\textit{\textbf{I}}+\lambda\textit{\textbf{T}}\right)+\lambda\textit{\textbf{HG}}=\textit{\textbf{Z}}
\end{align}

\noindent where \textit{\textbf{I}} is $ n\times n $ identity matrix.

Let

\begin{gather}
	\textit{\textbf{A}}=\left(\textit{\textbf{I}}+\lambda\textit{\textbf{T}}\right)\\
	\textit{\textbf{B}}=\lambda\textit{\textbf{H}}
\end{gather}

\noindent where $ \textit{\textbf{A}}\in \mathbb{R}^{n\times n} $ and $ \textit{\textbf{B}}\in \mathbb{R}^{m\times m} $.

Then, the formula \eqref{26} can be rewritten as

\begin{align} \label{28}
	\textit{\textbf{GA}}+\textit{\textbf{BG}}=\textbf{\textit{Z}}
\end{align}

Next, make the matrix $ \mathrm{vec} $ operation on both sides of \textit{Eqn}.\eqref{28}, then

\begin{align} \label{29}
	\left(\textit{\textbf{A}}^T\otimes \textit{\textbf{I}}_m+\textit{\textbf{I}}_n\otimes\textit{\textbf{B}}\right)\mathrm{vec}\left(\textit{\textbf{G}}\right)=\mathrm{vec}\left(\textit{\textbf{Z}}\right)
\end{align}

\noindent where $ \textit{\textbf{A}}^T\otimes\textit{\textbf{I}}_m $ is Kronecker product of matrix $ \textit{\textbf{A}}^T $ and $ \textit{\textbf{I}}_m $.

Owing to \textit{\textbf{T}} is a real symmetric matrix, so

\begin{align}
	\textit{\textbf{A}}=\textit{\textbf{A}}^T
\end{align}

Further, the \textit{Eqn}.\eqref{29} can expressed as

\begin{align} \label{31}
	\left(\textit{\textbf{A}}\otimes \textit{\textbf{I}}_m+\textit{\textbf{I}}_n\otimes \textit{\textbf{B}}\right)\mathrm{vec}\left(\textit{\textbf{G}}\right)=\mathrm{vec}\left(\textit{\textbf{Z}}\right)
\end{align}

The necessary and sufficient condition for compatibility of matrix equation \eqref{31} is

\begin{align}
	\mathrm{rank}\left(\textit{\textbf{A}}\otimes \textit{\textbf{I}}_m+\textit{\textbf{I}}_n\otimes \textit{\textbf{B}},\mathrm{vec}\left(\textit{\textbf{Z}}\right)\right)=\mathrm{rank}\left(\textit{\textbf{A}}\otimes \textit{\textbf{I}}_m+\textit{\textbf{I}}_n\otimes\textit{\textbf{B}}\right)
\end{align}

According to appendix, \textit{\textbf{N }}is a positive semidefinite and real symmetric matrix, so that all eigenvalues of the matrix \textit{\textbf{N}} are nonnegative real numbers. Therefore, all eigenvalues of the matrix \textit{\textbf{T}} and \textit{\textbf{H}} are nonnegative real numbers.

It is assumed that the eigenvalues of matrix \textit{\textbf{T}} and \textit{\textbf{H}} are $ \mu_1,\mu_2,\cdots,\mu_n $ and $ \rho_1,\rho_2,\cdots,\rho_m $, respectively, and $ \mu_i\geq0 $ ($ i=1,2,\cdots,n $), $ \rho_j\geq0 $ ($ j=1,2,\cdots,m $). Therefore, the eigenvalues of \textit{\textbf{A}} and \textit{\textbf{B}} are $ 1+\lambda\mu_i $ and $ \lambda\rho_j $, respectively. Then, according to correlation theory, the eigenvalues of $ \textit{\textbf{A}}\otimes\textit{\textbf{I}}_m+\textit{\textbf{I}}_n\otimes\textit{\textbf{B}} $ are $ 1+\lambda\mu_i+\lambda\rho_j $ ($i=1,2,\cdots,n.\ j=1,2,\cdots,m $), and $ 1+\lambda\mu_i+\lambda\rho_j>0 $. Therefore, $ \textit{\textbf{A}}\otimes\textit{\textbf{I}}_m+\textit{\textbf{I}}_n\otimes\textit{\textbf{B}} $ is a full rank matrix, i.e., $ \mathrm{rank}\left(\textit{\textbf{A}}\otimes\textit{\textbf{I}}_m+\textit{\textbf{I}}_n\otimes\textit{\textbf{B}},\mathrm{vec}\left(\textit{\textbf{Z}}\right)\right)=\mathrm{rank}\left(\textit{\textbf{A}}\otimes\textit{\textbf{I}}_m+\textit{\textbf{I}}_n\otimes\textit{\textbf{B}}\right) $. In other words, $ \textit{\textbf{A}}\otimes\textit{\textbf{I}}_m+\textit{\textbf{I}}_n\otimes\textit{\textbf{B}} $ is also an invertible matrix. Consequently, the matrix \textit{Eqn}.\eqref{20} or \textit{Eqn}.\eqref{31} must have a unique solution.

Thereupon, the solution of the formula \eqref{28} is

\begin{align}
	\mathrm{vec}\left(\textit{\textbf{G}}\right)=\left(\textit{\textbf{A}}\otimes \textit{\textbf{I}}_m+\textit{\textbf{I}}_n\otimes \textbf{\textit{B}}\right)^{-1}\mathrm{vec}\left(\textit{\textbf{Z}}\right)
\end{align}

Then

\begin{align}
	\begin{split}
		\mathrm{vec}\left(\textit{\textbf{G}}\right)&=\left(\left(\textit{\textbf{I}}_n+\lambda\textit{\textbf{T}}\right)\otimes \textit{\textbf{I}}_m+\lambda \textit{\textbf{I}}_n\otimes\textit{\textbf{H}}\right)^{-1}\mathrm{vec}\left(\textit{\textbf{Z}}\right) \\&=\left(\textit{\textbf{I}}_n\otimes\textit{\textbf{I}}_m+\lambda\textit{\textbf{T}}\otimes\textit{\textbf{I}}_m+\lambda \textit{\textbf{I}}_n\otimes\textit{\textbf{H}}\right)^{-1}\mathrm{vec}\left( \textit{\textbf{Z}}\right)
	\end{split}
\end{align}

Further

\begin{align}
	\begin{split}
			\mathrm{vec}\left(\textit{\textbf{C}}\right)&=\lambda \mathrm{vec}\left(\textit{\textbf{GT}}+\textit{\textbf{HG}}\right) \\&=\lambda\left(\textit{\textbf{T}}\otimes\textit{\textbf{I}}_m+\textit{\textbf{I}}_n\otimes\textit{\textbf{H}}\right)\mathrm{vec}\left(\textit{\textbf{G}}\right)
	\end{split}
\end{align}

For the TDS algorithm, it is essential to emphasize that once the dimension of the matrix $ \textit{\textbf{Z}} $ is determined, the matrices $ \textit{\textbf{T}} $ and $ \textit{\textbf{H}} $ are completely determined and do not rely on the values within matrix $ \textit{\textbf{Z}} $. In other words, the results of the TDS algorithm are solely determined by the global smoothing parameter $ \lambda $. Therefore, it is necessary for us to analyze the determination of the global smoothing parameter $ \lambda $. Here, the following two cases are discussed.

\begin{description}
	\item[\textit{\textbf{Case1}}] When the distribution of noise is known, iterate $ \lambda $ until the fluctuation sequence satisfies the known distribution of noise. As the numerical simulation results provided in section \ref{sec 5.1.1}, $ \lambda=735 $ is determined in this way.
	\item[\textit{\textbf{Case2}}] When the distribution of noise is unknown, we need to determine the quality assessment metric $ \alpha $ in advance (see Section \ref{sec 5.1.5} for details), the determination of $ \lambda $ are summarized in algorithm \ref{alg1}.
\end{description} 

\begin{algorithm}[H]
	\caption{Global smoothing parameter $ \lambda $ for TDS.}\label{alg1}
	\renewcommand{\algorithmicrequire}{\textbf{Input:}}
	\renewcommand{\algorithmicensure}{\textbf{Output:}}
	\begin{algorithmic}[1]
		\REQUIRE The two-dimensional sequence $ \textit{\textbf{Z}} $ to be processed, and reference sequence $ \textit{\textbf{z}} $.\\
		\ENSURE Trend sequence $ \textit{\textbf{G}} $ and fluctuation sequence $ \textit{\textbf{C}} $.\\	
		\STATE Initialize parameters including determining the quality assessment metric and its value $ \alpha_{0} $, predefined threshold $ \epsilon $, iterative step $ k $, and upper limit $ \bar{\lambda} $ of the global smoothing parameter $ \lambda $.
		\STATE $ \lambda\gets0 $.
		\STATE Calculate $ \textit{\textbf{G}} $ according to \textit{Eqn}.\eqref{20}.
		\STATE Calculate $ \textit{\textbf{C}} $ according to \textit{Eqn}.\eqref{21}.
		\STATE Calculate quality assessment metric value $ \alpha $ according to $ \textit{\textbf{G}} $ and $ \textit{\textbf{z}} $.
		\WHILE {$ \left| \alpha-\alpha_{0} \right|>\epsilon $}
		\IF {$ \lambda\leq\bar{\lambda} $}	
		\STATE $\lambda\gets \lambda+k$.
		\STATE Update $ \textit{\textbf{G}} $ and $ \textit{\textbf{C}} $.
		\STATE Update $ \alpha $.
		\ELSE 
		\STATE Cannot find the appropriate trend sequence $ \textit{\textbf{G}} $ and fluctuation sequence $ \textit{\textbf{C}} $.
		\STATE \textbf{break}
		\ENDIF
		\ENDWHILE
		\RETURN {$ \textit{\textbf{G}} $, $ \textit{\textbf{C}} $, $ \lambda $.}
	\end{algorithmic}
\end{algorithm}

\section{Modified TDS Algorithm}\label{sec 4}

In this section, we introduce three modified algorithms derived from the TDS algorithm and provide the associated lemmas and corollaries.

\subsection{TDS-\uppercase\expandafter{\romannumeral1} Algorithm}

\begin{figure}[!ht]
	\centering
	\subfloat[TDS-\uppercase\expandafter{\romannumeral1}.]{\label{Fig4:a} \includegraphics[width=.47\columnwidth]{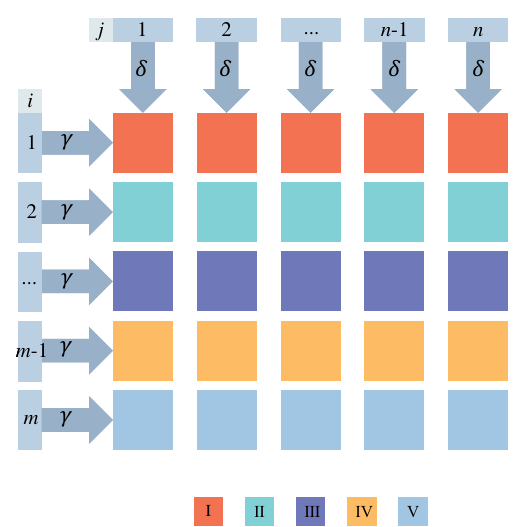}}\hspace{5pt}
	\subfloat[TDS-\uppercase\expandafter{\romannumeral2}.]{\label{Fig4:b}
		\includegraphics[width=.47\columnwidth]{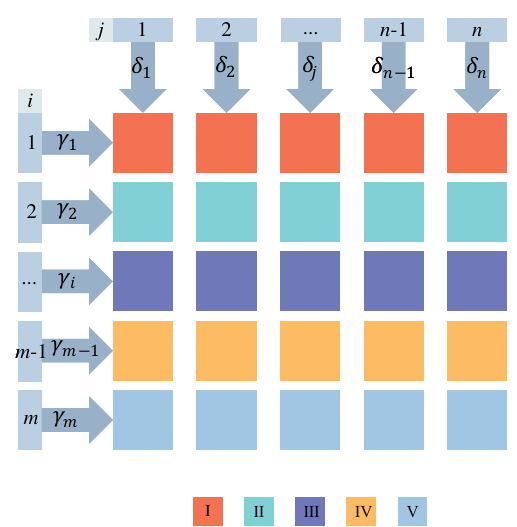}}
	\caption{Algorithm structure of TDS-\uppercase\expandafter{\romannumeral1} and TDS-\uppercase\expandafter{\romannumeral2}.}\label{Fig.4}
\end{figure}

In the TDS algorithm, the global smoothing parameter for the two-dimensional sequence is denoted as $ \lambda $. However, if the two-dimensional sequence \textit{\textbf{Z}} exhibits significant fluctuations across different columns, while having smaller fluctuations across different rows, or vice versa, it might be challenging to ensure optimal results when decomposing the two-dimensional sequence \textit{\textbf{Z}} into the trend sequence \textit{\textbf{G}} and the fluctuation sequence \textit{\textbf{C}} using the global smoothing parameter $ \lambda $. The structure of this smoothing algorithm is shown in Fig.\ref{Fig.4}\subref{Fig4:a}. Based on this, the loss function $ M $ can be readjusted as

\begin{align}
\begin{split}
		M=&\sum_{i=1}^{m}\sum_{j=1}^{n}\left(z_{i,j}-g_{i,j}\right)^2\\+&
	\gamma\sum_{i=1}^{m}\sum_{j=3}^{n}\left[\Delta^2g_{i,j}\right]^2+\delta\sum_{j=1}^{n}\sum_{i=3}^{m}\left[\nabla^2g_{i,j}\right]^2
\end{split}
\end{align}

\noindent where $ \gamma>0 $ and $ \delta>0 $.

At this point, we take partial derivative of $ M $ with respect to $ g_{i,j} $, set $ \partial M/\partial g_{i,j}=0 $, then

\begin{align} \label{38}
	\textit{\textbf{G}}+\gamma\textit{\textbf{GT}}+\delta\textit{\textbf{HG}}=\textit{\textbf{Z}}
\end{align}

Therefore, the trend sequence \textit{\textbf{G}} can be solved by \textit{Eqn}.\eqref{38}, and the fluctuation sequence \textit{\textbf{C}} is

\begin{align}
	\textit{\textbf{C}}=\gamma\textit{\textbf{GT}}+\delta\textit{\textbf{HG}}
\end{align}

For convenience, this modified TDS algorithm can be referred to as TDS-\uppercase\expandafter{\romannumeral1} algorithm. Therefore, in the TDS-\uppercase\expandafter{\romannumeral1} algorithm, it is considered that the smoothing parameters $ \gamma $ are the same for any row, and $ \gamma $ can be called \textit{\textbf{the global row smoothing parameter}}. Similarly, the smoothing parameters $ \delta $ are the same for any column, and $ \delta $ can be called \textit{\textbf{the global column smoothing parameter}}.

Similarly, we give the following lemma.

\textbf{Lemma 2}: A general two-dimensional sequence $ \textit{\textbf{Z}}\in \mathbb{R}^{m\times n} $ ($ m\geq3 $, $ n\geq3 $) can be decomposed into a trend sequence $ \textit{\textbf{G}}\in \mathbb{R}^{m\times n} $ and a fluctuation sequence $ \textit{\textbf{C}}\in \mathbb{R}^{m\times n} $ by TDS-\uppercase\expandafter{\romannumeral1} algorithm. The trend sequence \textit{\textbf{G}} and the fluctuation sequence \textit{\textbf{C}} satisfy the following matrix equations, respectively

\begin{gather}
	\textit{\textbf{G}}+\gamma\textit{\textbf{GT}}+\delta\textit{\textbf{HG}}=\textit{\textbf{Z}}\\
	\gamma\textit{\textbf{GT}}+\delta\textit{\textbf{HG}}=\textit{\textbf{C}}
\end{gather}

\noindent	where $ \gamma $ and $ \delta $ are \textit{\textbf{the global row smoothing parameter}} and \textit{\textbf{the global column smoothing parameter}}, respectively. Both $ \textit{\textbf{T}}\in \mathbb{R}^{n\times n} $ and $ \textit{\textbf{H}}\in \mathbb{R}^{m\times m} $ are positive semidefinite and real symmetric matrix. Besides, the eigenvalues of matrix $ \textit{\textbf{T}} $ and $ \textit{\textbf{H}} $ satisfy $ 0\le\zeta\le16 $, and $ \mathrm{rank}\left(\textit{\textbf{T}}\right)=n-2 $, $ \mathrm{rank}\left(\textit{\textbf{H}}\right)=m-2 $.

Further, the trend sequence \textit{\textbf{G}} and the fluctuation sequence \textit{\textbf{C}} are, respectively

\begin{gather}
	\mathrm{vec}\left(\textit{\textbf{G}}\right)=\left(\textit{\textbf{I}}_n\otimes\textit{\textbf{I}}_m+\gamma\textit{\textbf{T}}\otimes\textit{\textbf{I}}_m+\delta\textit{\textbf{I}}_n\otimes\textit{\textbf{H}}\right)^{-1}\mathrm{vec}\left(\textit{\textbf{Z}}\right)\\
	\mathrm{vec}\left(\textit{\textbf{C}}\right)=\left(\gamma\textit{\textbf{T}}\otimes\textit{\textbf{I}}_m+\delta\textit{\textbf{I}}_n\otimes\textit{\textbf{H}}\right)\mathrm{vec}\left(\textit{\textbf{G}}\right)
\end{gather}

\noindent	where both $ \textit{\textbf{I}}_n $ and $ \textbf{}\textit{\textbf{I}}_m $ are identity matrices, and both \textit{\textbf{G}} and \textit{\textbf{C}} are unique and exist when $ \gamma $ and $ \delta $ are determined.

The proof of the Lemma 2 is not repeated.

According to the Lemma 2, we still have the following simple corollary.

\textbf{Corollary 2}: Assume that $ \textit{\textbf{Z}}\in \mathbb{R}^{m\times n} $ is a two-dimensional sequence, and $ \textit{\textbf{G}}\in \mathbb{R}^{m\times n} $ is the trend sequence of the sequence \textit{\textbf{Z}}, then the trend sequence \textit{\textbf{G}} satisfies the following Sylvester equation

\begin{align}
	\textit{\textbf{GA}}+\textit{\textbf{BG}}=\textit{\textbf{Z}}
\end{align}

\noindent	where $ \textit{\textbf{A}}=\textit{\textbf{I}}_n/2+\gamma\textit{\textbf{T}} $, $ \textit{\textbf{B}}=\textit{\textbf{I}}_m/2+\delta\textit{\textbf{H}} $.

Furthermore, the trend sequence \textit{\textbf{G}} can be solved as

\begin{align}
	\textit{\textbf{G}}=\int_{0}^{+\infty}{e^{-\textit{\textbf{B}}t}\textit{\textbf{Z}}e^{-\textit{\textbf{A}}t}\mathrm{d}t}
\end{align}

The above corollary shows that the trend sequence \textit{\textbf{G}} of the sequence \textit{\textbf{Z}} in the TDS-\uppercase\expandafter{\romannumeral1} algorithm is also the solution of a certain kind of Sylvester equation.

\subsection{TDS-\uppercase\expandafter{\romannumeral2} Algorithm}

In the TDS-\uppercase\expandafter{\romannumeral1} algorithm, it is assumed that the smoothing parameters are uniform for any row (denoted as $ \gamma $) and uniform for any column (denoted as $ \delta $). However, in certain two-dimensional sequences \textit{\textbf{Z}}, fluctuations may exhibit significant disparities in different rows as well as in different columns. In such cases, ensuring favorable results for the decomposition of the two-dimensional sequence \textit{\textbf{Z}} into the trend sequence \textit{\textbf{G}} and the fluctuation sequence \textit{\textbf{C}} using the global row smoothing parameter $ \gamma $ and the global column smoothing parameter $ \delta $ could be challenging. The structural representation of this modified smoothing algorithm is illustrated in Fig.\ref{Fig.4}\subref{Fig4:b}. Based on this consideration, the loss function $ M $ can be redefined as follows

\begin{align}
	\begin{split}
	M=&\sum_{i=1}^{m}\sum_{j=1}^{n}\left(z_{i,j}-g_{i,j}\right)^2
	\\+&\sum_{i=1}^{m}\sum_{j=3}^{n}{\gamma_i\left[\Delta^2g_{i,j}\right]^2}+\sum_{j=1}^{n}\sum_{i=3}^{m}{\delta_j\left[\nabla^2g_{i,j}\right]^2}
	\end{split}
\end{align}

\noindent where $ \gamma_i>0 $ ($ i=1,\cdots, m $) and $ \delta_j>0 $ ($ j=1,\cdots, n $).

Let

\begin{align}
	\boldsymbol{\mathit{\Gamma}}=\left[\begin{matrix}\gamma_1&0&\cdots&0&0\\0&\gamma_2&\cdots&0&0\\\vdots&\vdots&\ddots&\vdots&\vdots\\0&0&\cdots&\gamma_{m-1}&0\\0&0&\cdots&0&\gamma_m\\\end{matrix}\right]
\end{align}
\begin{align}
	\boldsymbol{\mathit{\Delta}}=\left[\begin{matrix}\delta_1&0&\cdots&0&0\\0&\delta_2&\cdots&0&0\\\vdots&\vdots&\ddots&\vdots&\vdots\\0&0&\cdots&\delta_{n-1}&0\\0&0&\cdots&0&\delta_n\\\end{matrix}\right]
\end{align}

At this point, we take partial derivative of $ M $ with respect to $ g_{i,j} $, set $ \partial M/\partial g_{i,j}=0 $, then

\begin{align} \label{55}
	\textit{\textbf{G}}+\boldsymbol{\mathit{\Gamma}}\textit{\textbf{GT}}+\textit{\textbf{HG}}\boldsymbol{\mathit{\Delta}}=\textit{\textbf{Z}}
\end{align}

Hence, the trend sequence \textit{\textbf{G}} can be solved by \textit{Eqn}.\eqref{55}. Subsequently, the fluctuation sequence \textit{\textbf{C}} can be solved as

\begin{align}
	\boldsymbol{\mathit{\Gamma}}\textit{\textbf{GT}}+\textit{\textbf{HG}}\boldsymbol{\mathit{\Delta}}=\textit{\textbf{C}}
\end{align}

For descriptive purposes, this modified TDS algorithm can be referred to as the TDS-\uppercase\expandafter{\romannumeral2} algorithm. In the TDS-\uppercase\expandafter{\romannumeral2} algorithm, the smoothing parameters $ \gamma_i $ vary for different rows and are referred to as \textit{\textbf{the smoothing parameter of the $ i $-th row}}. Similarly, the smoothing parameters $ \delta_j $ vary for different columns and are called \textit{\textbf{the smoothing parameter of the $ j $-th column}}.

Similarly, the following lemma is provided.

\textbf{Lemma 3}: A general two-dimensional sequence $ \textit{\textbf{Z}}\in \mathbb{R}^{m\times n} $ ($ m\geq3 $, $ n\geq3 $) can be decomposed into a trend sequence $ \textit{\textbf{G}}\in \mathbb{R}^{m\times n} $ and a fluctuation sequence $ \textit{\textbf{C}}\in \mathbb{R}^{m\times n} $ by TDS-\uppercase\expandafter{\romannumeral2} algorithm. The trend sequence \textit{\textbf{G}} and the fluctuation sequence \textit{\textbf{C}} satisfy the following matrix equations, respectively

\begin{gather}
	\textit{\textbf{G}}+\boldsymbol{\mathit{\Gamma}} \textit{\textbf{GT}}+\textit{\textbf{HG}}\boldsymbol{\mathit{\Delta}}=\textit{\textbf{Z}}\\
	\boldsymbol{\mathit{\Gamma}}\textit{\textbf{GT}}+\textit{\textbf{HG}}\boldsymbol{\mathit{\Delta}}=\textit{\textbf{C}}
\end{gather}

\noindent where $ \boldsymbol{\mathit{\Gamma}}=\mathrm{diag}\left(\gamma_1,\cdots,\gamma_m\right) $ and $ \boldsymbol{\mathit{\Delta}}=\mathrm{diag}\left(\delta_1,\cdots,\delta_n\right) $ are diagonal matrices composed of a series of the smoothing parameter respectively. Both $ \textit{\textbf{T}}\in \mathbb{R}^{n\times n} $ and $ \textit{\textbf{H}}\in \mathbb{R}^{m\times m} $ are positive semidefinite and real symmetric matrix. Besides, the eigenvalues of matrix $ \textit{\textbf{T}} $ and $ \textit{\textbf{H}} $ satisfy $ 0\le\zeta\le16 $, and $ \mathrm{rank}\left(\textit{\textbf{T}}\right)=n-2 $, $ \mathrm{rank}\left(\textit{\textbf{H}}\right)=m-2 $.

Further, the trend sequence \textit{\textbf{G}} and the fluctuation sequence \textit{\textbf{C}} are, respectively

\begin{gather}
	\mathrm{vec}\left(\textit{\textbf{G}}\right)=\left(\textit{\textbf{I}}_n\otimes\textit{\textbf{I}}_m+\textit{\textbf{T}}\otimes\boldsymbol{\mathit{\Gamma}}+\boldsymbol{\mathit{\Delta}}\otimes\textit{\textbf{H}}\right)^{-1}\mathrm{vec}\left(\textit{\textbf{Z}}\right)\\
    \mathrm{vec}\left(\textit{\textbf{C}}\right)=\left(\textit{\textbf{T}}\otimes\boldsymbol{\mathit{\Gamma}}+\boldsymbol{\mathit{\Delta}}\otimes\textit{\textbf{H}}\right)\mathrm{vec}\left(\textit{\textbf{G}}\right)
\end{gather}

\noindent	where both $ \textit{\textbf{I}}_n $ and $ \textit{\textbf{I}}_m $ are identity matrices, and both $ \textit{\textbf{G}} $ and $ \textit{\textbf{C}} $ are unique and exist when $ \boldsymbol{\mathit{\Gamma}} $ and $ \boldsymbol{\mathit{\Delta}} $ are determined.

The proof of the Lemma 3 is not repeated.

For the TDS algorithm, TDS-\uppercase\expandafter{\romannumeral1} algorithm, and TDS-\uppercase\expandafter{\romannumeral2} algorithm, we can obtain the following conclusions:

\begin{enumerate}
	\item 	When $ \gamma=\gamma_1=\gamma_2=\cdots=\gamma_m $ and  $ \delta=\delta_1=\delta_2=\cdots=\delta_n $, the TDS-\uppercase\expandafter{\romannumeral2} algorithm degenerates into the TDS-\uppercase\expandafter{\romannumeral1} algorithm.
	\item	When $ \gamma=\delta $, the TDS-\uppercase\expandafter{\romannumeral1} algorithm degenerates into the TDS algorithm.
\end{enumerate}

\subsection{TDS-\uppercase\expandafter{\romannumeral3} Algorithm}

On the basis of the TDS-\uppercase\expandafter{\romannumeral1} algorithm and the TDS-\uppercase\expandafter{\romannumeral2} algorithm,  if a two-dimensional sequence \textit{\textbf{Z}} exhibits significant variation in fluctuation across different columns, while the fluctuation remains relatively consistent across different rows, or vice versa, there arises a challenge in selecting appropriate smoothing parameters. Utilizing the TDS-\uppercase\expandafter{\romannumeral2} algorithm would require determining $ m $ row smoothing parameters and $ n $ column smoothing parameters in advance, adding to the complexity. On the other hand, employing the TDS-\uppercase\expandafter{\romannumeral1} algorithm necessitates determining a global row smoothing parameter and a global column smoothing parameter in advance, which may pose difficulties in ensuring optimal results when decomposing the two-dimensional sequence \textit{\textbf{Z}} into the trend sequence \textit{\textbf{G}} and the fluctuation sequence \textit{\textbf{C}}. In light of this, the loss function $ M $ can be redefined as follows

\begin{align}
		\begin{split}
		M=&\sum_{i=1}^{m}\sum_{j=1}^{n}\left(z_{i,j}-g_{i,j}\right)^2\\+&
		\gamma\sum_{i=1}^{m}\sum_{j=3}^{n}\left[\Delta^2g_{i,j}\right]^2+\sum_{j=1}^{n}\sum_{i=3}^{m}{\delta_j\left[\nabla^2g_{i,j}\right]^2}
	\end{split}
\end{align}

Or

\begin{align}
	\begin{split}
			M=&\sum_{i=1}^{m}\sum_{j=1}^{n}\left(z_{i,j}-g_{i,j}\right)^2\\+&
		\sum_{i=1}^{m}\sum_{j=3}^{n}{\gamma_i\left[\Delta^2g_{i,j}\right]^2}+\delta\sum_{j=1}^{n}\sum_{i=3}^{m}\left[\nabla^2g_{i,j}\right]^2
	\end{split}
\end{align}

At this point, we take partial derivative of $ M $ with respect to $ g_{i,j} $, set $ \partial M/\partial g_{i,j}=0 $, then

\begin{align} \label{64}	\textit{\textbf{G}}+\gamma\textit{\textbf{GT}}+\textit{\textbf{HG}}\boldsymbol{\mathit{\Delta}}=\textit{\textbf{Z}}
\end{align}

Or

\begin{align} \label{65}
    \textit{\textbf{G}}+\boldsymbol{\mathit{\Gamma}} \textit{\textbf{GT}}+\delta\textit{\textbf{HG}}=\textit{\textbf{Z}}
\end{align}

Therefore, the trend sequence \textit{\textbf{G}} can be solved by \textit{Eqn}.\eqref{64} and \textit{Eqn}.\eqref{65}, and the fluctuation sequence \textit{\textbf{C}} as

\begin{align}
\textit{\textbf{C}}=\gamma\textit{\textbf{GT}}+\textit{\textbf{HG}}\boldsymbol{\mathit{\Delta}}
\end{align}

Or

\begin{align}
	\textit{\textbf{C}}=\boldsymbol{\mathit{\Gamma}} \textit{\textbf{GT}}+\delta\textit{\textbf{HG}}
\end{align}

Consequently, according to the above analysis, TDS, TDS-\uppercase\expandafter{\romannumeral1} and TDS-\uppercase\expandafter{\romannumeral3} are only special case of TDS-\uppercase\expandafter{\romannumeral2}.

Finally, extensive numerical simulation analysis and image filtering cases show that the TDS algorithm can achieve satisfactory results in various challenging scenarios (see Section \ref{sec 5} for details). Besides, the modified TDS algorithm can fine-tune the results of the TDS algorithm.

\section{Example Analysis} \label{sec 5}
Here, we provide implementation and numerical details of the TDS and modified TDS algorithm described in section \ref{sec 2} and \ref{sec 4} respectively. This includes comprehensive numerical simulation analyses subjected to various types of noise, applications in image filtering, and examples of image processing cases.
\subsection{Numerical Simulation Analysis}\label{sec 5.1}

\subsubsection{Additive white Gaussian noise (AWGN)}\label{sec 5.1.1}
In an experiment, we assume that the ideal experimental data follows the function $ z\left(x,y\right)=x+y+2\sin{\left(x+y\right)}+10 $. However, the presence of errors makes the measured experimental data $ \ Z(x,y)=z(x,y)+n(x,y) $ during the experimental process, where $ n(x,y) $ represents the random experimental error at the sampling point $ (x,y) $. Meanwhile, the sampling point are denoted as $ (x_i,y_j) $, where $ x_1=1.0 $, $ x_2=1.1 $, $ \cdots $, $ x_{30}=3.9 $, $ x_{31}=4.0 $, and $ y_1=1.0 $, $ y_2=1.1 $, $ \cdots $, $ y_{20}=2.9 $, $ y_{21}=3.0 $, so that the total number of sampling points is $ 31\ \times\ 21 $. Then, the image of the ideal experimental data $ z\left(x,y\right) $ is plotted, as shown in Fig.\ref{Fig8}\subref{Fig8:a}.

Randomly generate numbers $ n(x_i,y_j) $ that follows standard normal distribution at sample $ (x_i,y_j) $, i.e., $ n(x_i,y_j) $ is white Gaussian noise (WGN). Consequently, the measured experimental data $ Z(x_i,y_j)=z(x_i,y_j)+n(x_i,y_j) $ is obtained. A plot of the noise $ n(x,y) $ is plotted, as shown in Fig.\ref{Fig8}\subref{Fig8:b}. Meanwhile, a plot of $ Z(x,y) $ is plotted, as shown in Fig.\ref{Fig8}\subref{Fig8:c}.
\addtocounter{figure}{1}
\begin{figure}[htbp]
	\centering
	\subfloat[$ z(x,y) $.]{\label{Fig8:a}
	\includegraphics[width=.47\columnwidth]{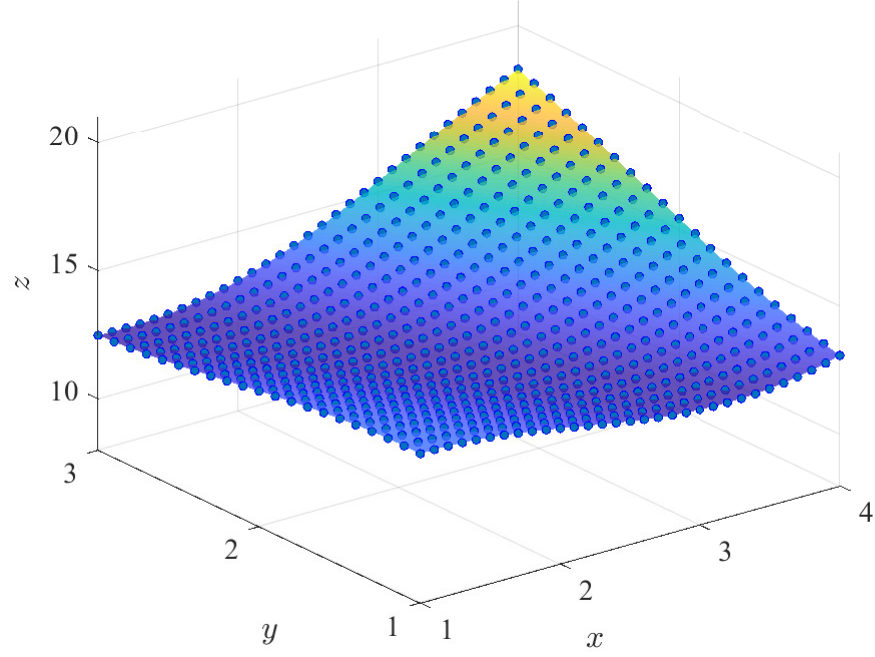}}\hspace{4pt}
	\subfloat[$ n(x,y) $.]{\label{Fig8:b}
	\includegraphics[width=.47\columnwidth]{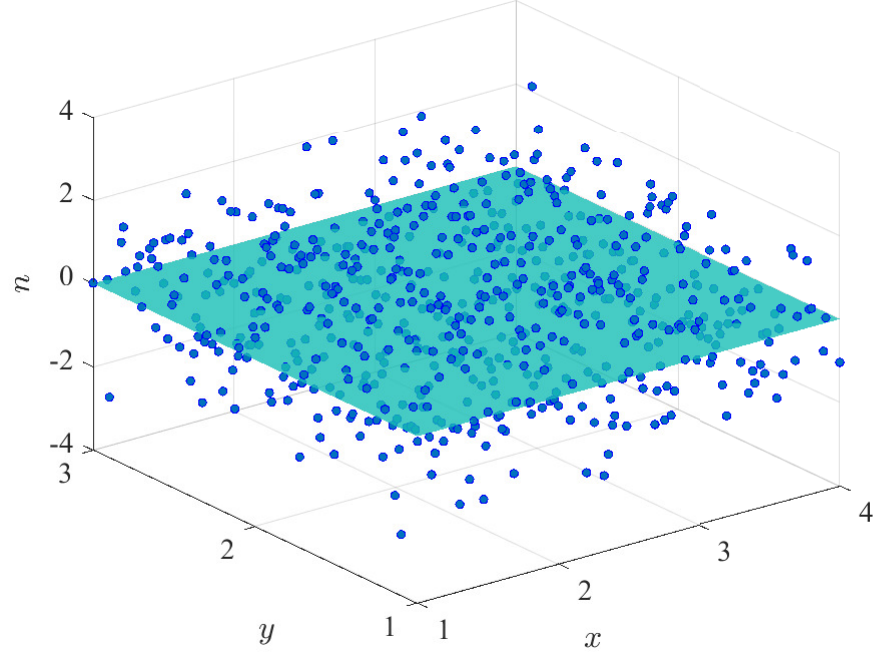}}\\
\end{figure}
\begin{figure}[H]\ContinuedFloat*   
	\subfloat[$ Z(x,y) $.]{\label{Fig8:c}
	\includegraphics[width=.47\columnwidth]{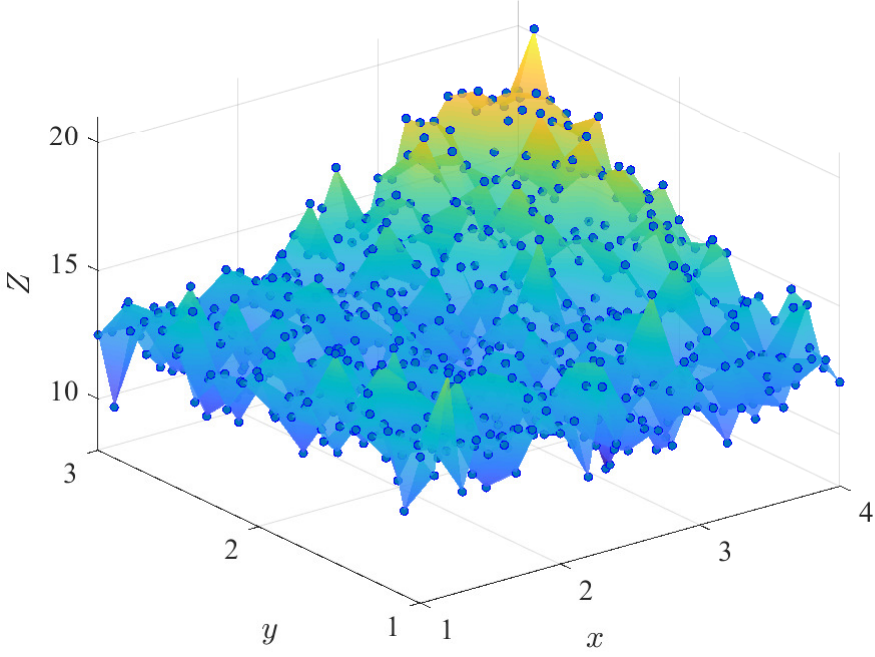}}
	\subfloat[$ z(x,y) $ and $ Z(x,y) $.]{\label{Fig8:d}
	\includegraphics[width=.47\columnwidth]{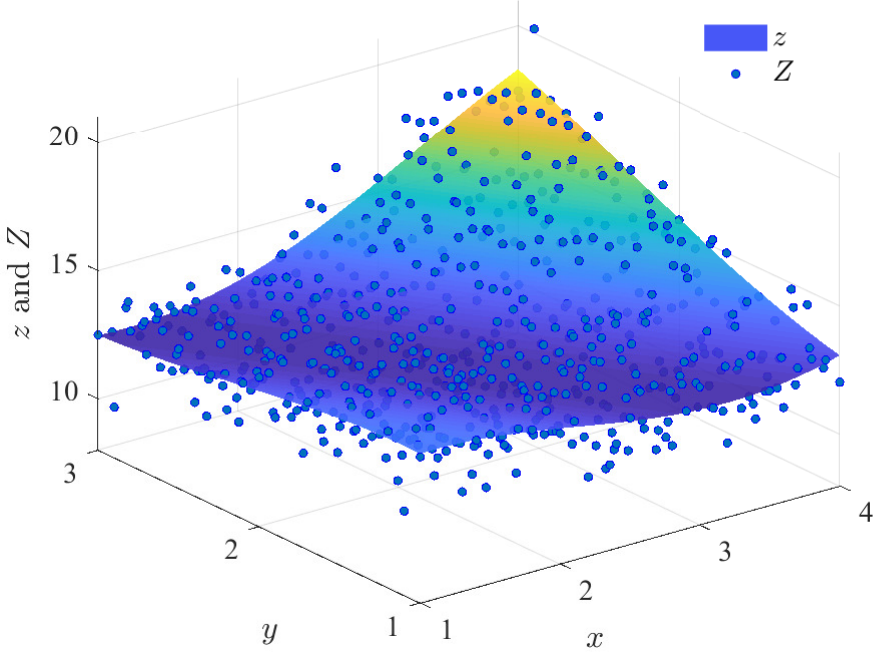}}\\
	\caption{Experimental and real data under AWGN.}\label{Fig8}
\end{figure}

Based on the TDS algorithm, the measured experimental data \textit{\textbf{Z}} is smoothed, the global smoothing parameters $ \lambda $ is taken as $ 0.1 $, $ 1 $, $ 10 $, $ 100 $, $ 735 $, $ 10000 $, respectively. The trend sequence $ \textbf{\textit{G}} $ and distribution of the fluctuation sequence $ \textit{\textbf{C}} $ is plotted, as shown in Fig.\ref{Fig9}.

\addtocounter{figure}{2}
\renewcommand{\thesubfigure}{\alph{subfigure}}  
\setcounter{subfigure}{0}
\begin{figure}[H]
	\centering
	\subfloat[$ G $ ($\lambda=0.1$).]{\includegraphics[width=.47\columnwidth]{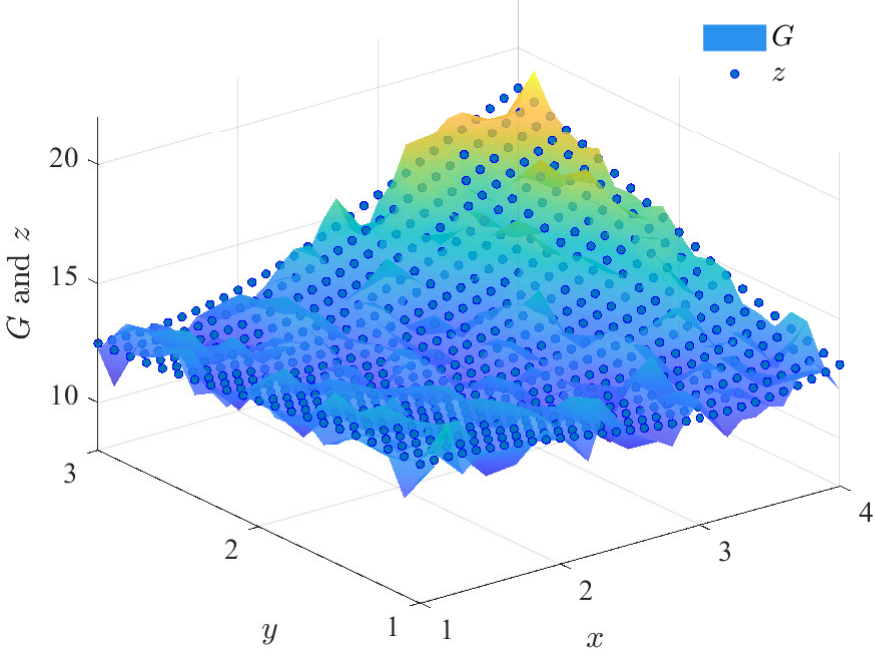}}\hspace{5pt}
	\subfloat[$ C $ ($\lambda=0.1$), $ \mu=0, \sigma=0.50 $.]{\includegraphics[width=.47\columnwidth]{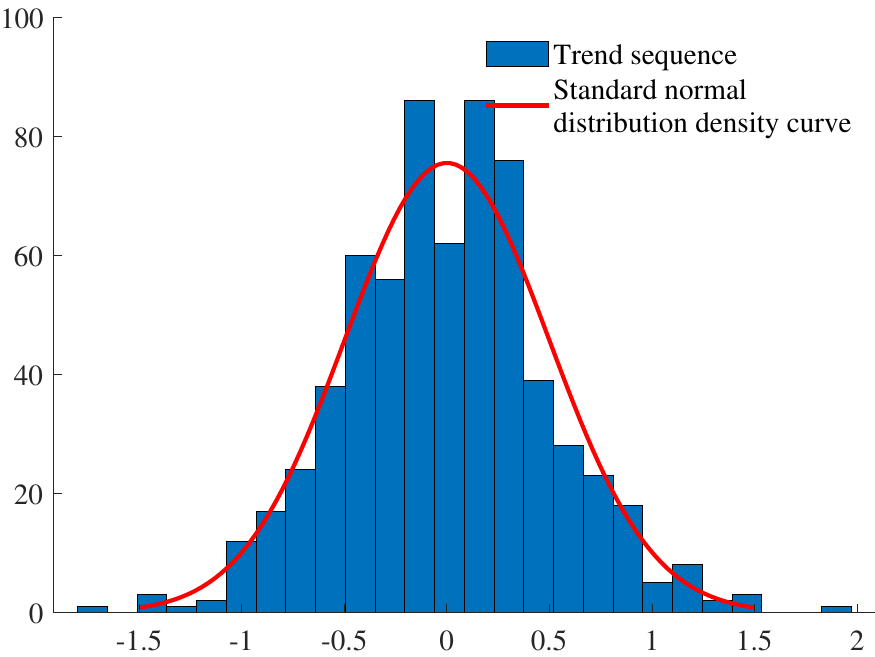}}\\
\end{figure}
\begin{figure}[H]\ContinuedFloat*
	\subfloat[$ G $ ($\lambda=1$). ]{\includegraphics[width=.47\columnwidth]{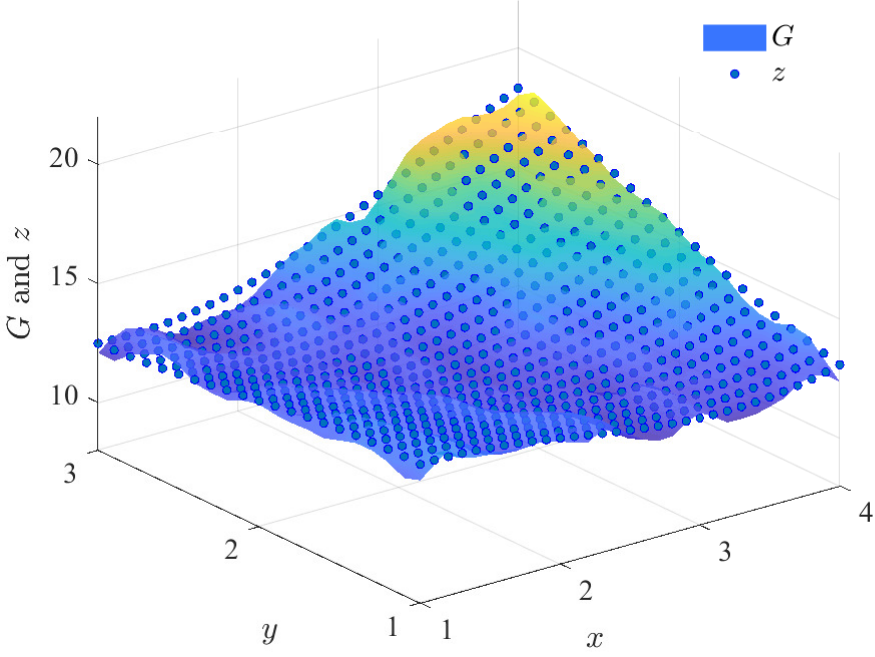}}\hspace{5pt}
	\subfloat[$ C $ ($\lambda=1$), $ \mu=0, \sigma=0.84 $.]{\includegraphics[width=.47\columnwidth]{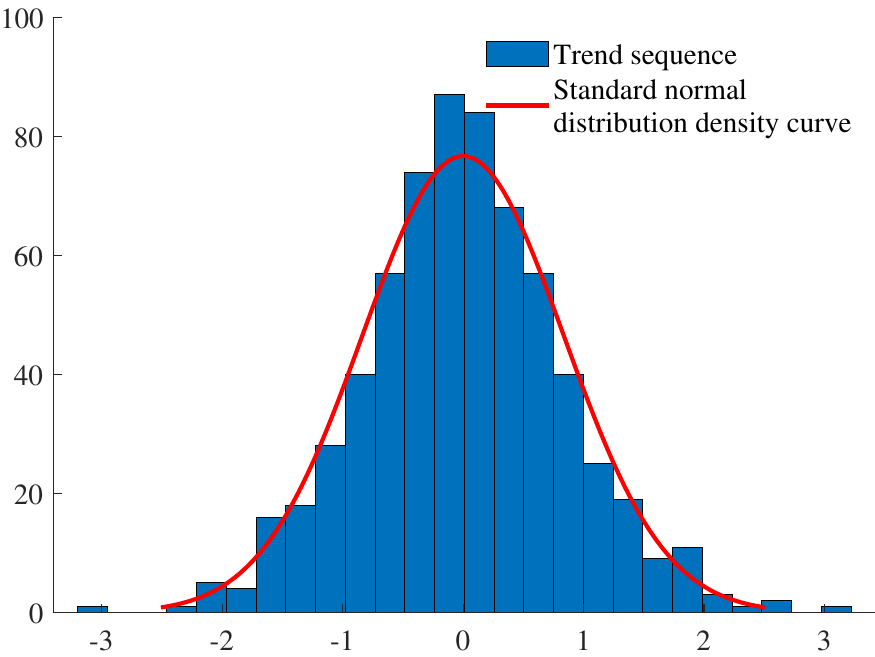}}\\
	\subfloat[$ G $ ($\lambda=10$).]{\includegraphics[width=.47\columnwidth]{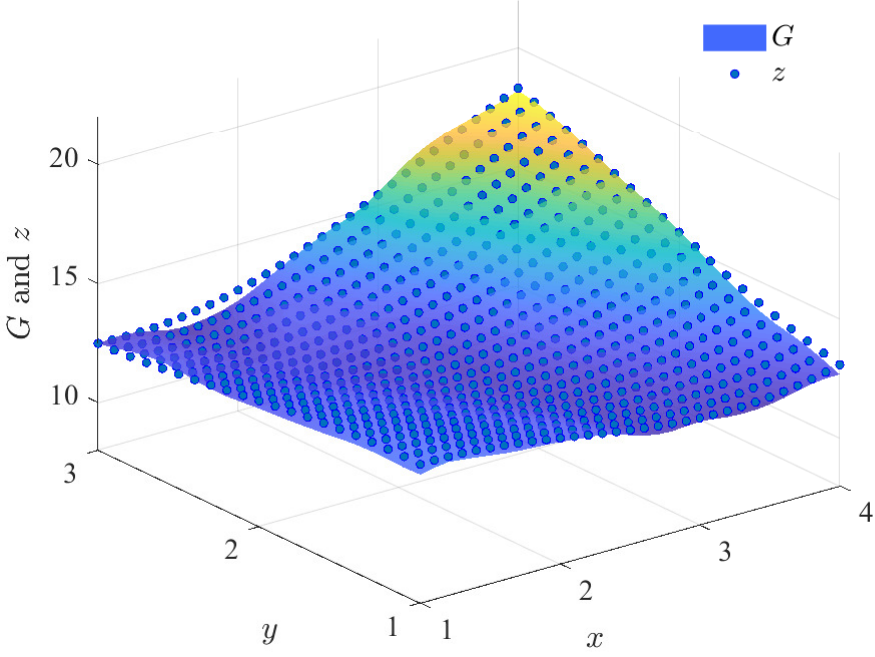}}\hspace{5pt}
	\subfloat[$ C $ ($\lambda=10$), $ \mu=0, \sigma=0.95 $.]{\includegraphics[width=.47\columnwidth]{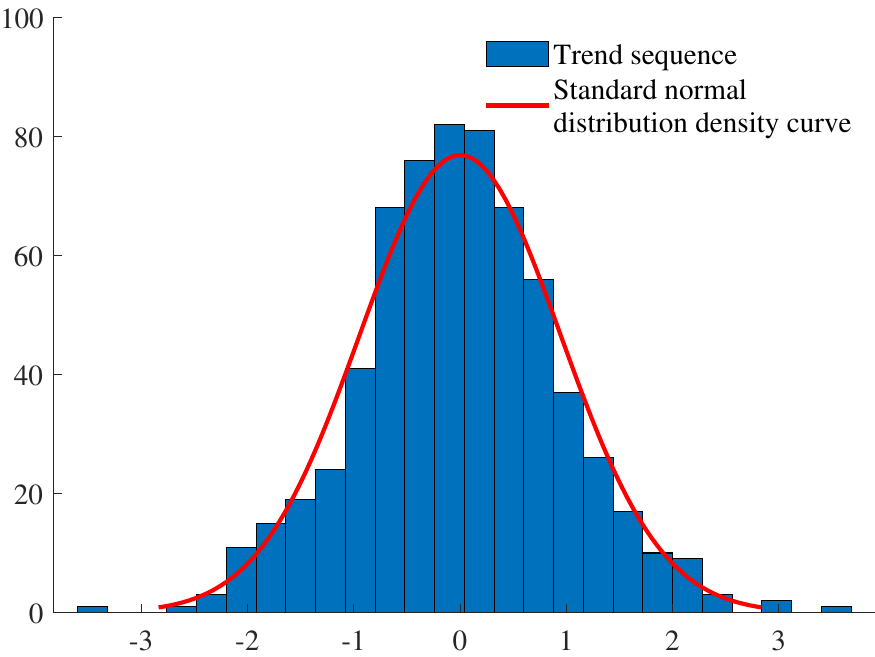}}\\
	\subfloat[$ G $
	($\lambda=100$).]{\includegraphics[width=.47\columnwidth]{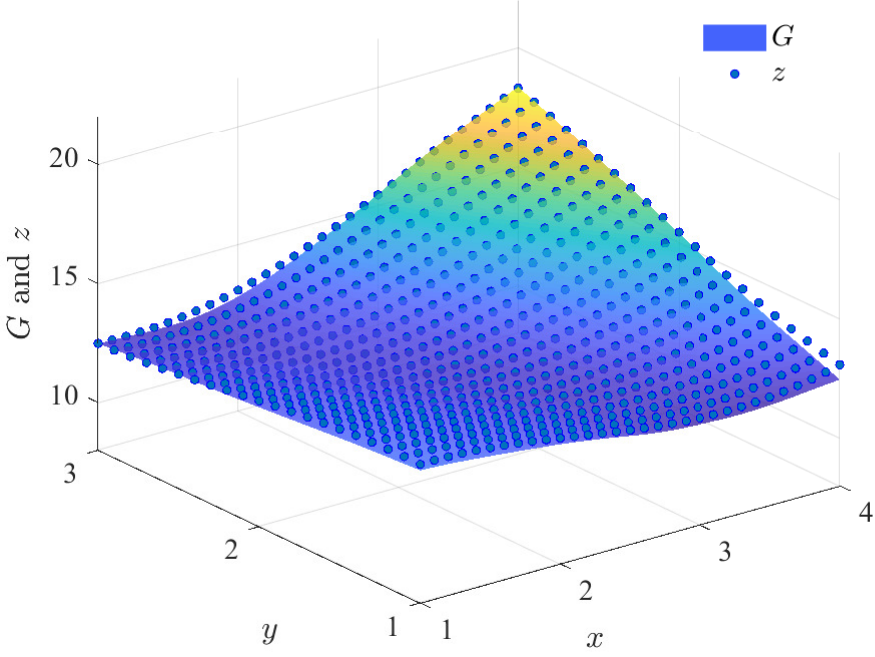}}\hspace{5pt}
	\subfloat[$ C $ ($\lambda=100$), $ \mu=0, \sigma=0.97 $.]{\includegraphics[width=.47\columnwidth]{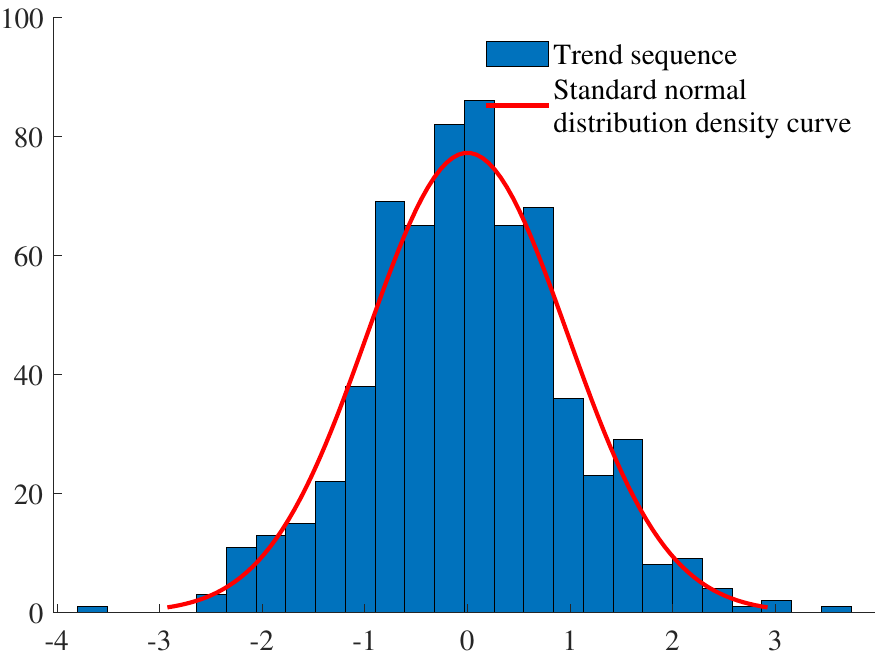}}\\
\end{figure}
\begin{figure}[H]\ContinuedFloat
	\subfloat[$ G $
	($\lambda=735$).]{\includegraphics[width=.47\columnwidth]{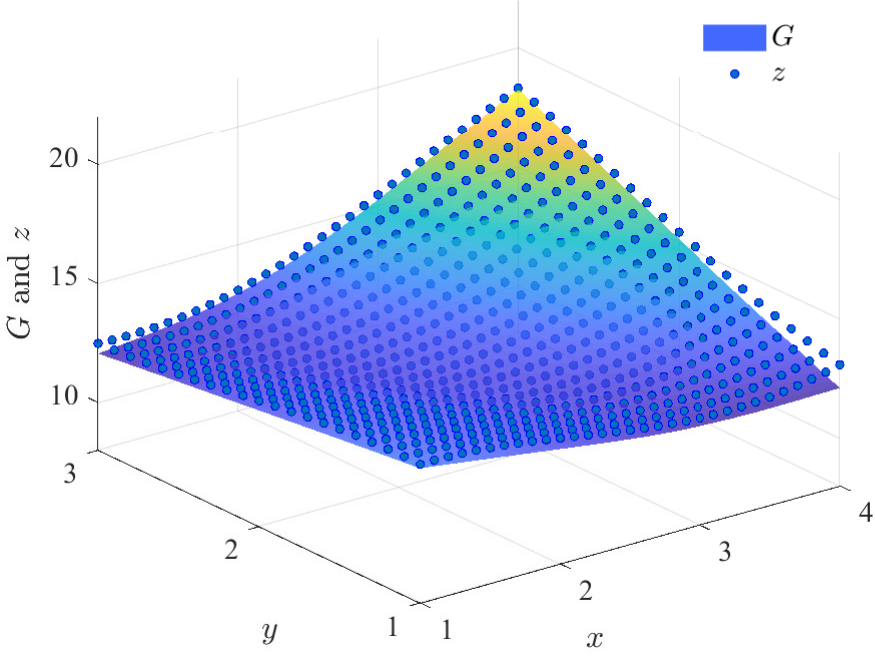}}\hspace{5pt}
	\subfloat[$ C $ ($\lambda=735$), $ \mu=0, \sigma=1.00 $.]{\includegraphics[width=.47\columnwidth]{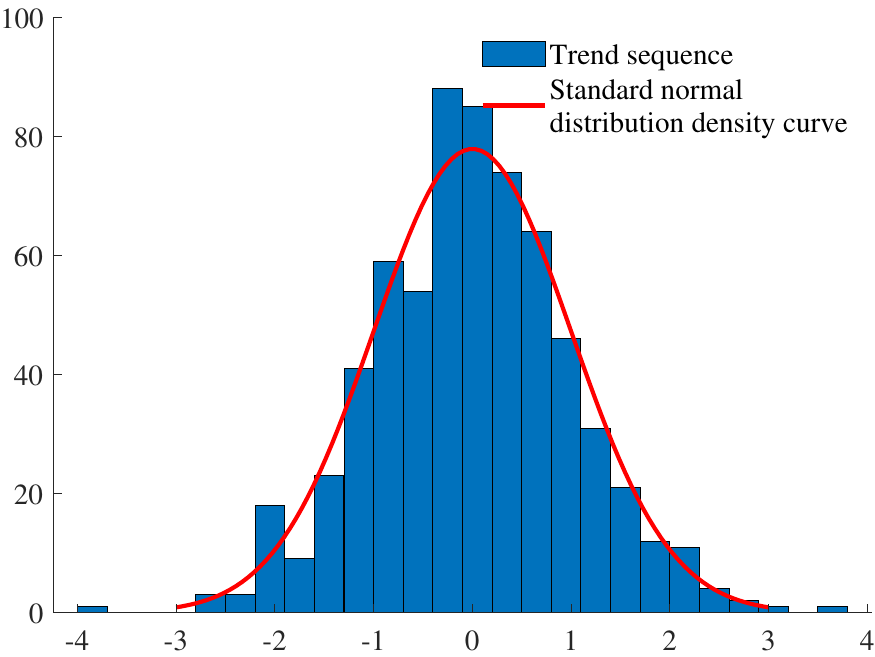}}\\	
	\subfloat[$ G $ ($\lambda=10000$).]{\includegraphics[width=.47\columnwidth]{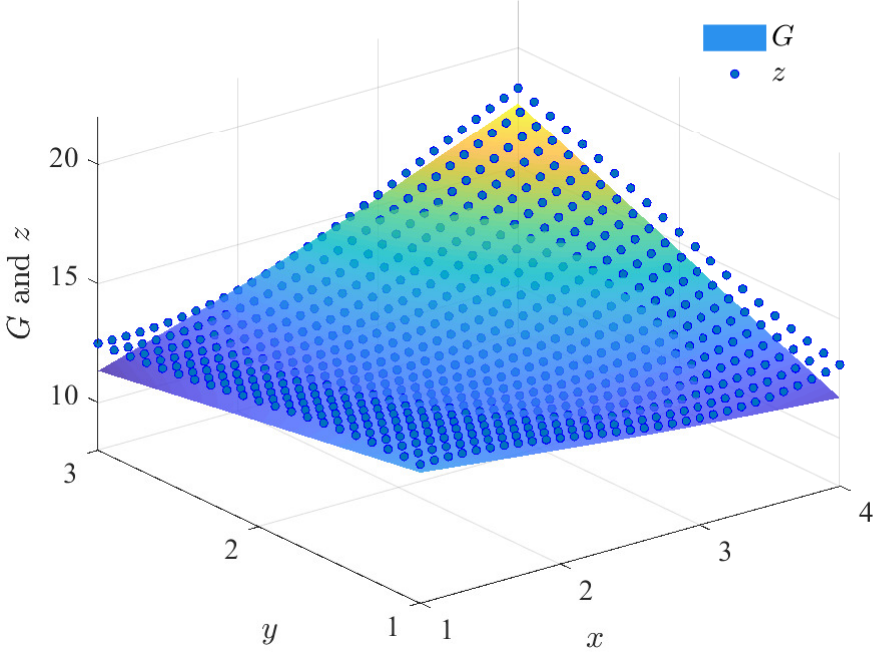}}\hspace{5pt}
	\subfloat[$ C $ ($\lambda=10000$), $ \mu=0, \sigma=1.08 $.]{\includegraphics[width=.47\columnwidth]{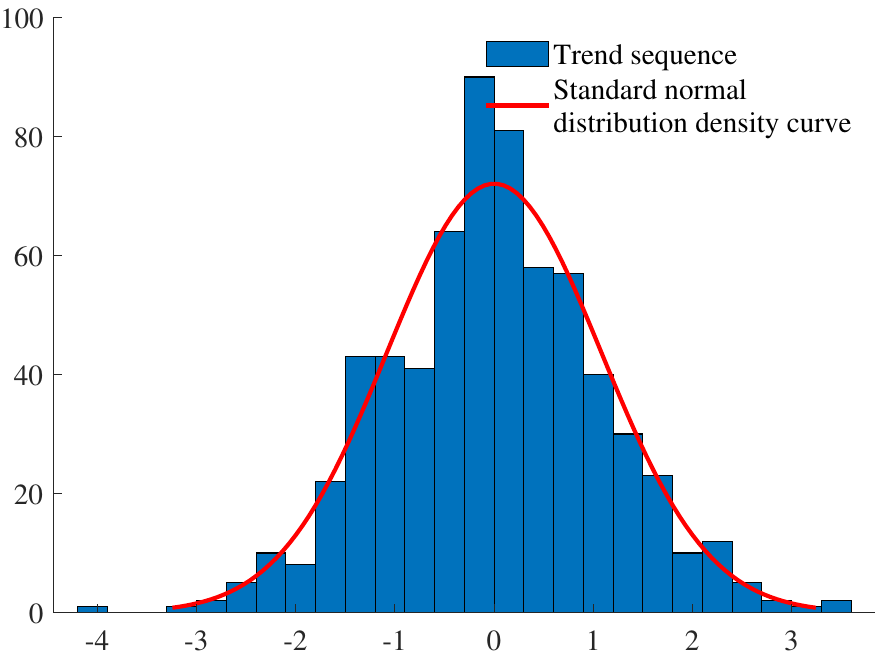}}
	\caption{The trend sequence $ \textit{\textbf{G}} $ and distribution of fluctuation sequence $ \textit{\textbf{C}} $ under AWGN.}\label{Fig9}
\end{figure}

It can be seen from Fig.\ref{Fig9} that the distribution of the trend sequence $ \textit{\textbf{G}} $ is smoother with the gradual increase of $ \lambda $. In this numerical case, the prior information of the noise is given in advance, that is, the noise obeys the standard normal distribution. Consequently, the TDS algorithm can completely filter out the given noise when $ \lambda=735 $.

\begin{figure}[htbp]
	\centering
	\subfloat[Median filtering.]{\includegraphics[width=.47\columnwidth]{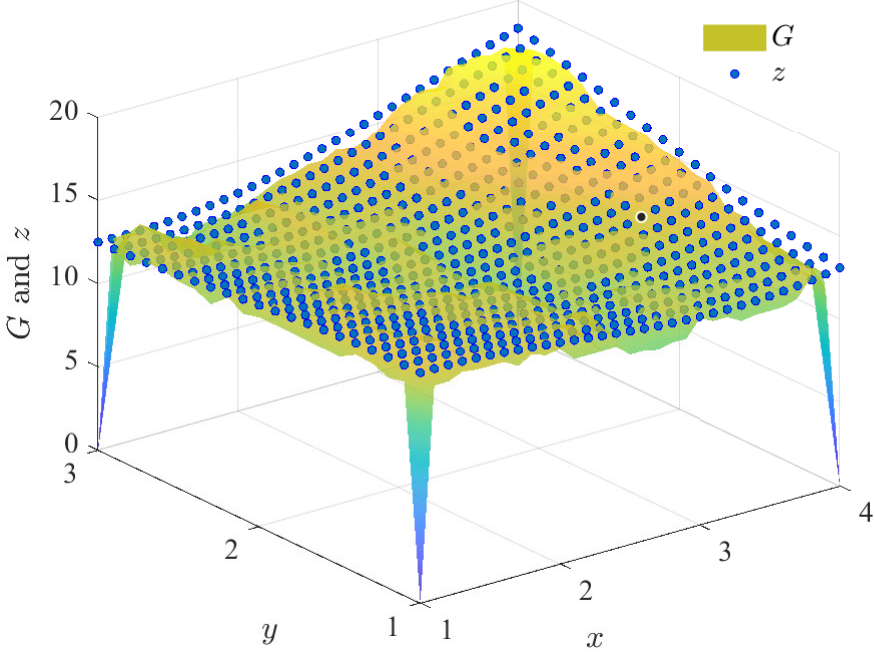}}\hspace{5pt}
	\subfloat[Mean filtering.]{\includegraphics[width=.47\columnwidth]{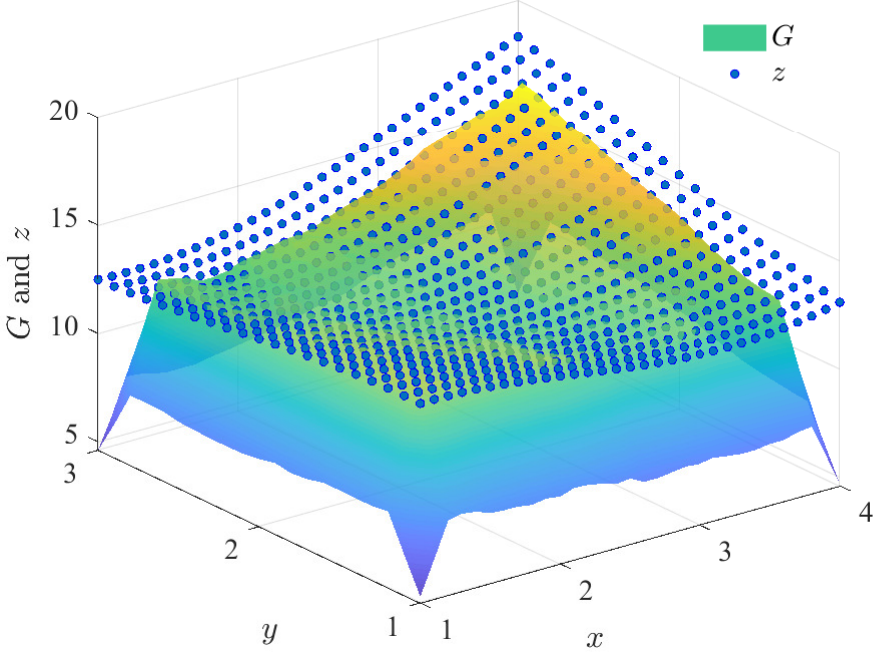}}\\
	\subfloat[Gaussian filtering.]{\includegraphics[width=.47\columnwidth]{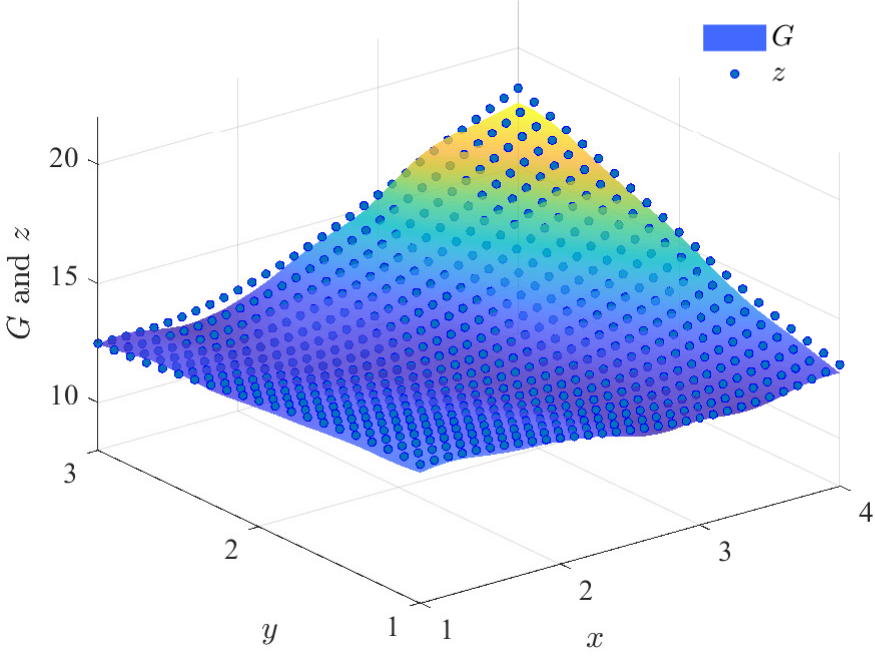}}\hspace{5pt}
	\subfloat[2D adaptive Wiener filtering.]{\includegraphics[width=.47\columnwidth]{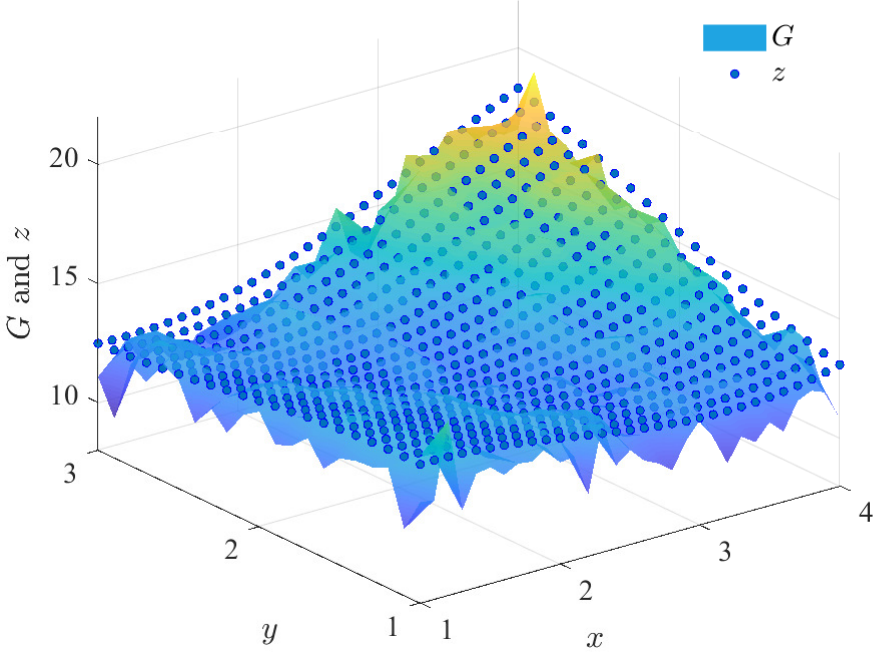}}\\
	\caption{Denoising results of filtering algorithms under AWGN.}\label{Fig9_1}
\end{figure}

Aiming at this problem, we apply various filtering techniques, including median filtering, mean filtering, Gaussian filtering, and 2D adaptive Wiener filtering, to smooth the sequence $ \textit{\textbf{Z}} $. In the median filtering algorithm, the default $ 3 \times 3 $ template size is selected. For the mean filtering algorithm, we create a $ 5 \times 5 $ mean filter template. The standard deviation is set to $ 2 $ in the Gaussian filtering algorithm. In the 2D adaptive Wiener filtering, the filtering size is $ 3 \times 3 $. The smoothing results of these different algorithms are shown in Fig.\ref{Fig9_1}.
\renewcommand{\thesubfigure}{\alph{subfigure}}  
\setcounter{subfigure}{0}

As shown in Fig.\ref{Fig9} and Fig.\ref{Fig9_1}, the effect of TDS algorithm is much better than the median, mean and 2D adaptive Wiener filtering algorithm intuitively when the noise is AWGN (see Table.\ref{tab1} for details). Although the TDS algorithm has no a very significant advantage over the Gaussian filtering algorithm (see Table.\ref{tab1} for details, the TDS algorithm is still superior to the Gaussian filtering algorithm in terms of MSE, PSNR, and SSIM.), the implementation of the Gaussian filtering algorithm is complex and the processing effect for non-Gaussian noise is not ideal (see Table.\ref{tab1} for details, the TDS algorithm has faster execution efficiency over the Gaussian filtering algorithm.).

\subsubsection{Complex noise (CN)}
\begin{figure}[htbp]
	\centering
	\subfloat[The measured experimental data.]{\includegraphics[width=.47\columnwidth]{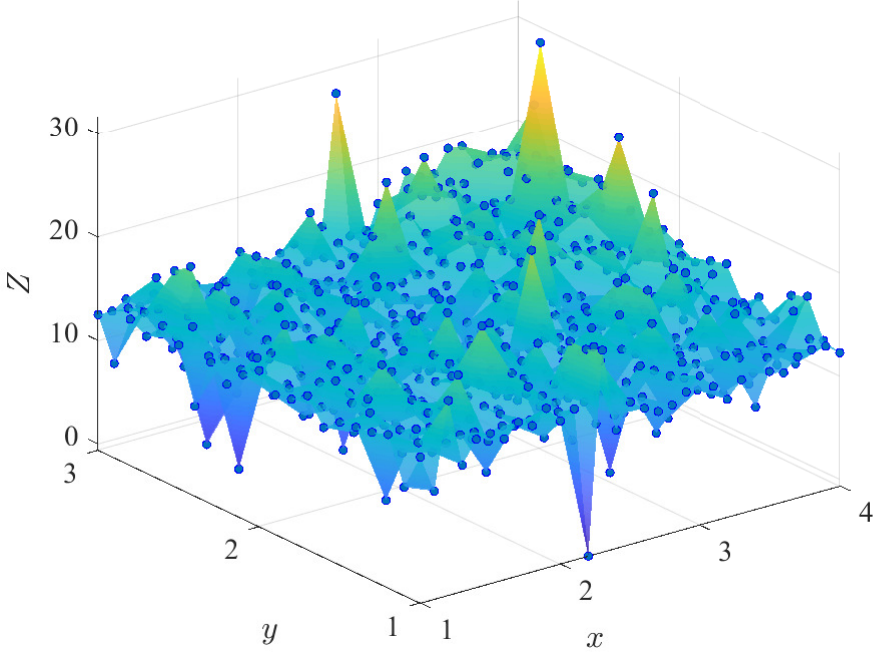}}\hspace{5pt}
	\subfloat[TDS.]{\includegraphics[width=.47\columnwidth]{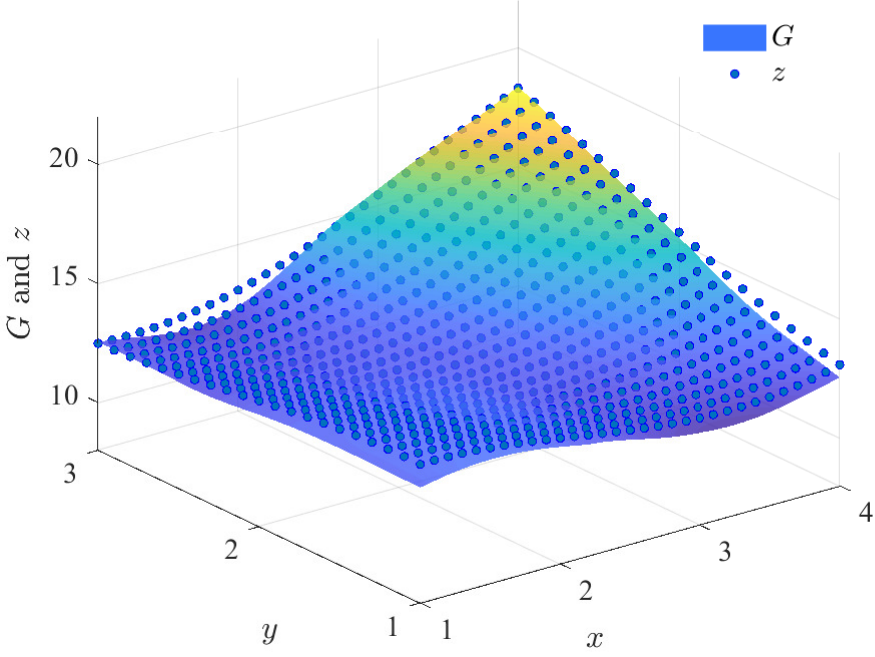}}\\
	\subfloat[Median filtering.]{\includegraphics[width=.47\columnwidth]{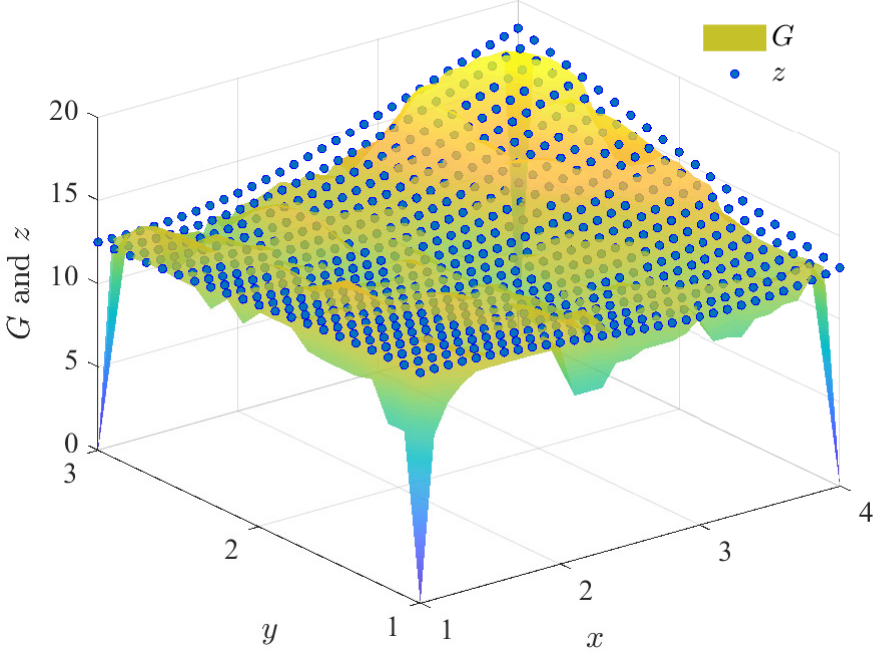}}\hspace{5pt}
	\subfloat[Mean filtering.]{\includegraphics[width=.47\columnwidth]{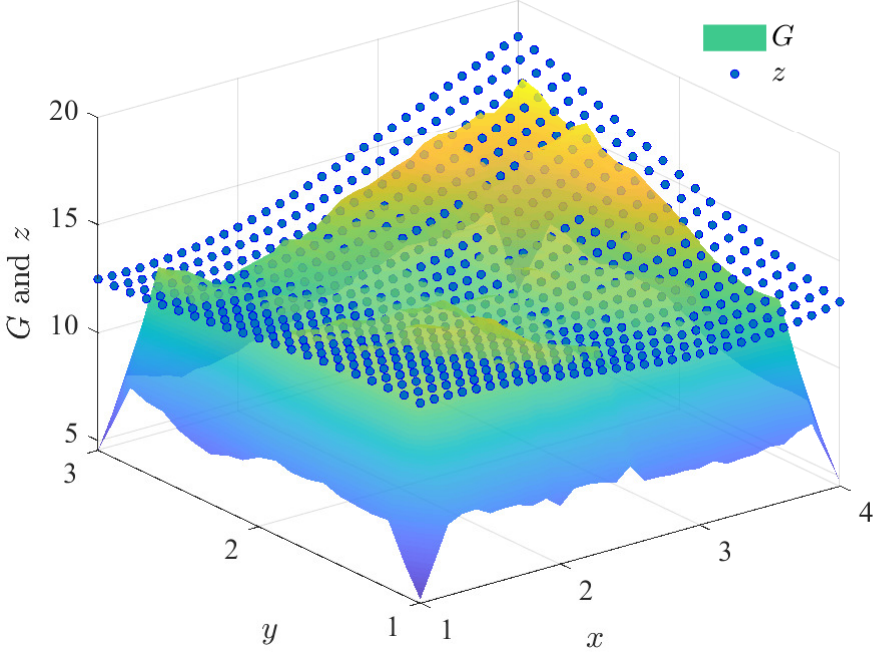}}\\
	\subfloat[Gaussian filtering.]{\includegraphics[width=.47\columnwidth]{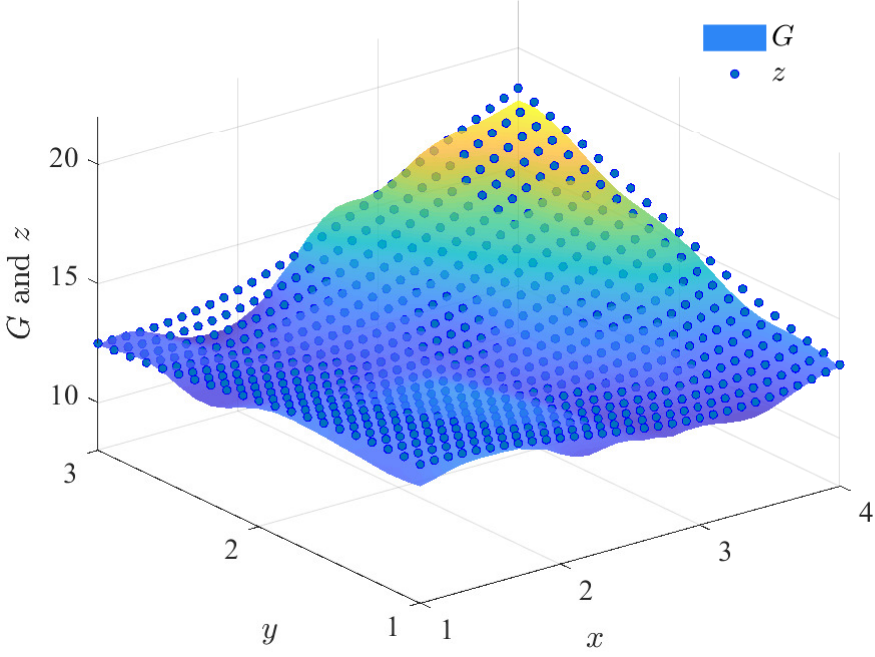}}\hspace{5pt}
	\subfloat[2D adaptive Wiener filtering.]{\includegraphics[width=.47\columnwidth]{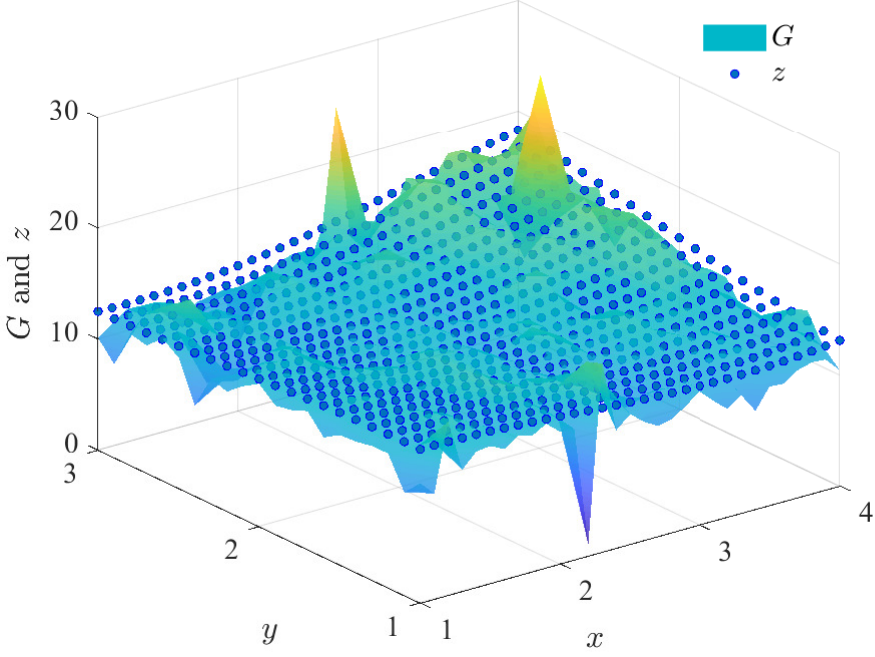}}\\
	\caption{Denoising results of filtering algorithms under CN.}\label{Fig9_2}
\end{figure}

We assume that the noise $ N(x,y)=n_{1}(x,y)\times n_{2}(x,y) $, where $ n_{1}(x,y) $ follows a standard normal distribution, and $ n_{2}(x,y) $ follows a gamma distribution with shape parameter $ 2 $ and scale parameter $ 1 $. The other details remain unchanged, and $ \ Z(x,y)=z(x,y)+N(x,y) $. The effects of different algorithms are shown in Fig.\ref{Fig9_2}.

As shown in Fig.\ref{Fig9_2}, compared with other filtering algorithms, the TDS algorithm can effectively smooth the original sequence well under CN interference and is easy to implement.

\subsubsection{Multiplicative white Gaussian noise (MWGN)}

\begin{figure}[!ht]
	\centering
	\subfloat[The measured experimental data.]{\includegraphics[width=.47\columnwidth]{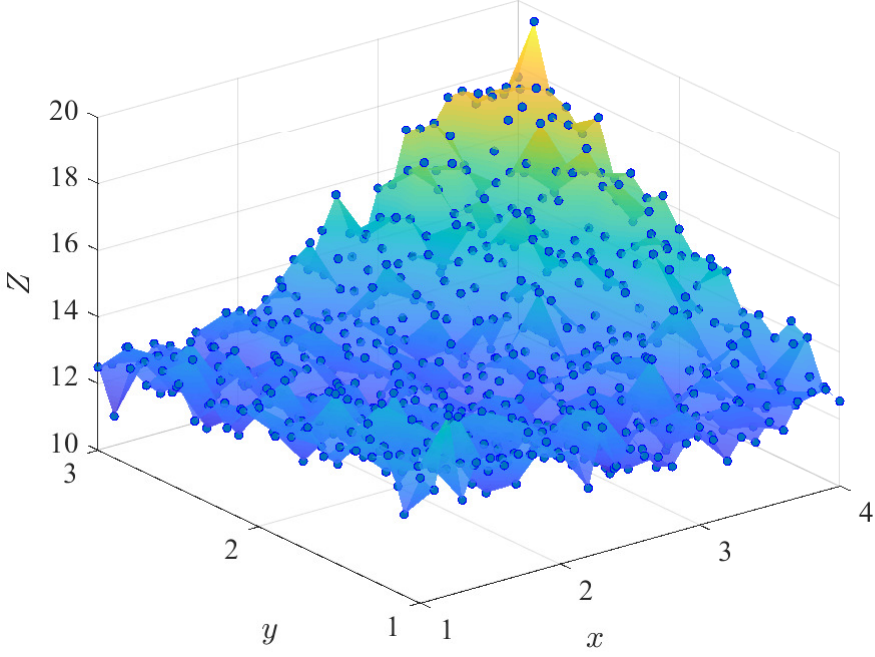}}\hspace{5pt}
	\subfloat[TDS.]{\includegraphics[width=.47\columnwidth]{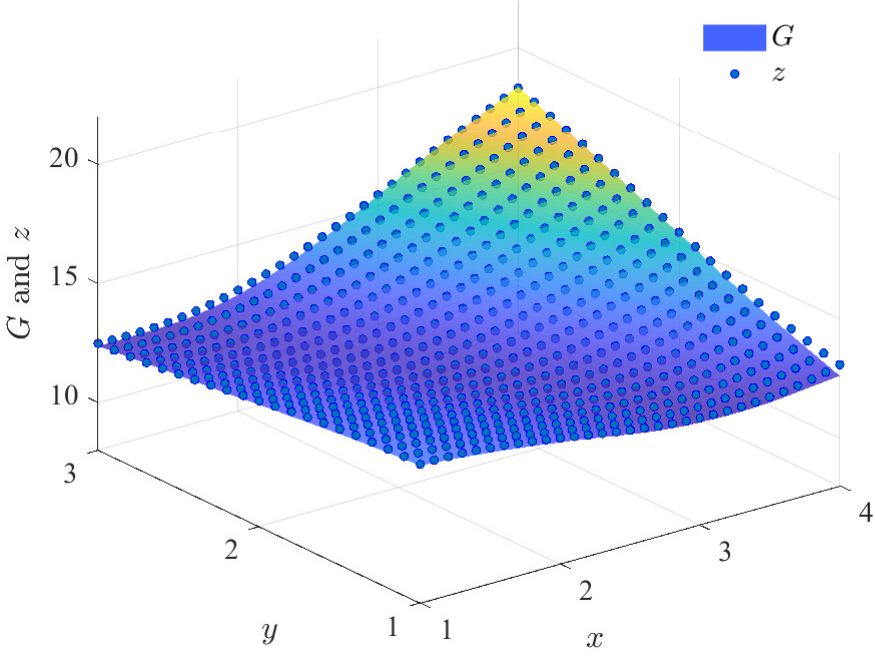}}\\
	\subfloat[Median filtering.]{\includegraphics[width=.47\columnwidth]{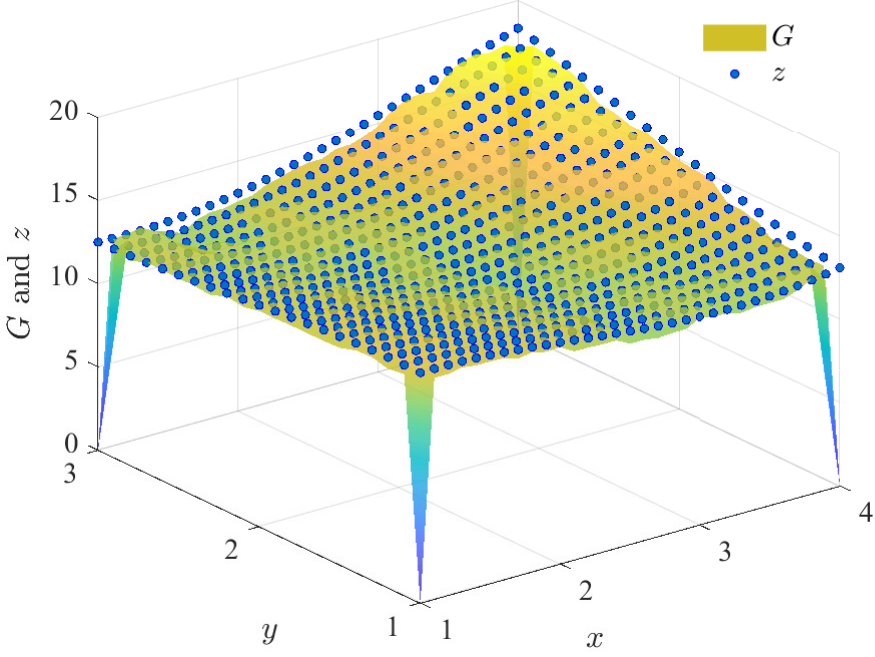}}\hspace{5pt}
	\subfloat[Mean filtering.]{\includegraphics[width=.47\columnwidth]{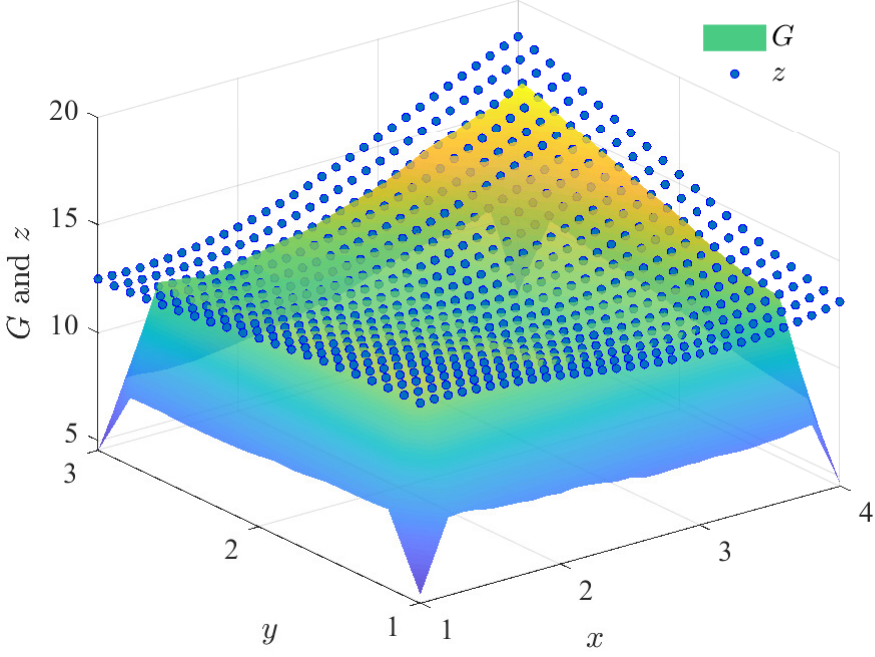}}\\
	\subfloat[Gaussian filtering.]{\includegraphics[width=.47\columnwidth]{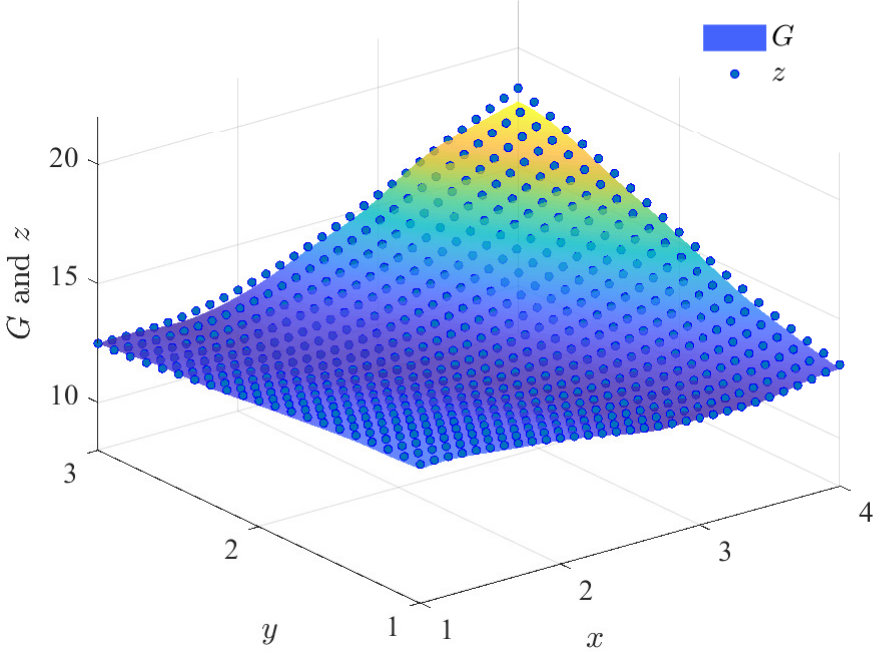}}\hspace{5pt}
	\subfloat[2D adaptive Wiener filtering.]{\includegraphics[width=.47\columnwidth]{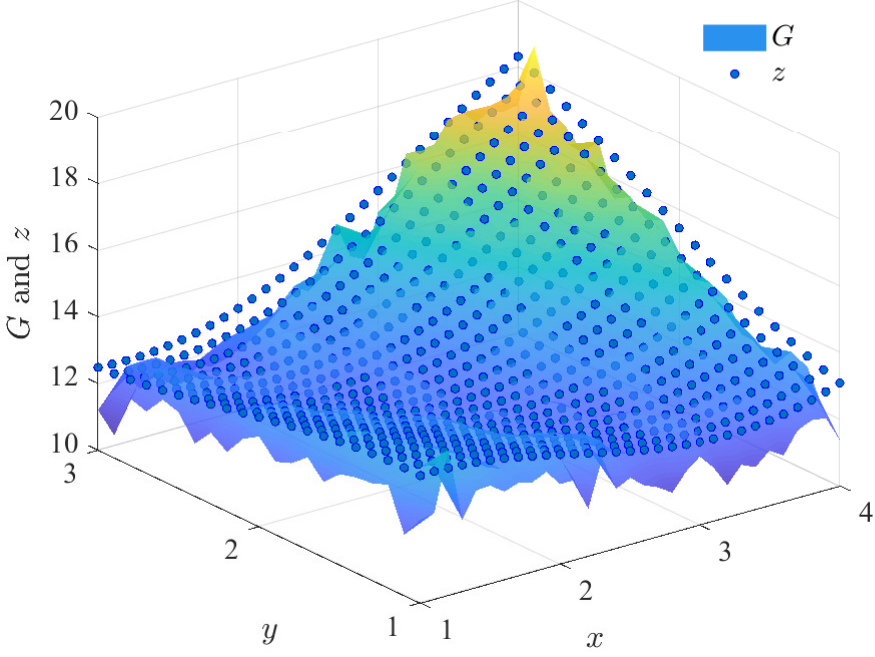}}\\
	\caption{Denoising results of filtering algorithms under MWGN.}\label{Fig9_3}
\end{figure}

We assume that the noise $ W(x,y) $ follows a Gaussian distribution with mean of $ 0 $ and standard deviation of $ 0.04 $. The other details remain unchanged, and $ \ Z(x,y)=z(x,y)+z(x,y)\times W(x,y) $. The effects of different algorithms are shown in Fig.\ref{Fig9_3}.

As shown in Fig.\ref{Fig9_3}, compared with other filtering algorithms, the TDS algorithm can effectively smooth the original sequence well under MWGN interference.

\subsubsection{Salt-and-pepper noise (SPN)}\label{sec 5.1.4}

\begin{figure}[htbp]
	\centering
	\subfloat[The measured experimental data.]{\includegraphics[width=.47\columnwidth]{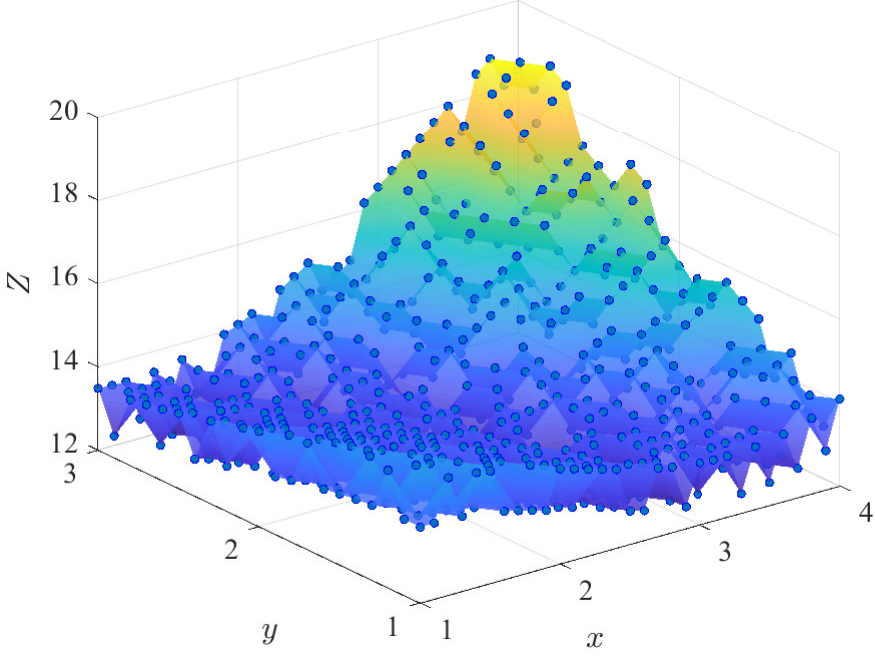}}\hspace{5pt}
	\subfloat[TDS.]{\includegraphics[width=.47\columnwidth]{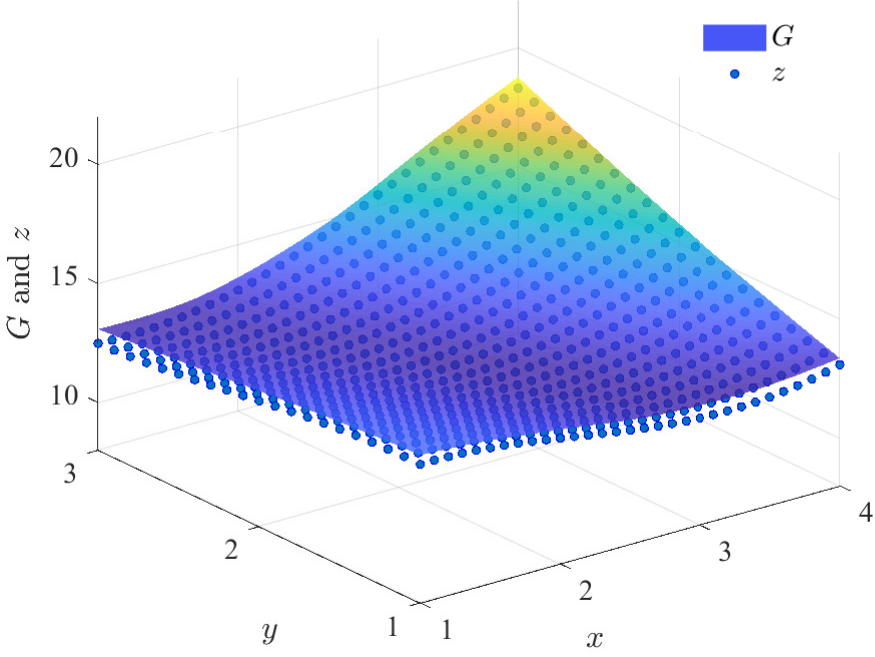}}\\
	\subfloat[Median filtering.]{\includegraphics[width=.47\columnwidth]{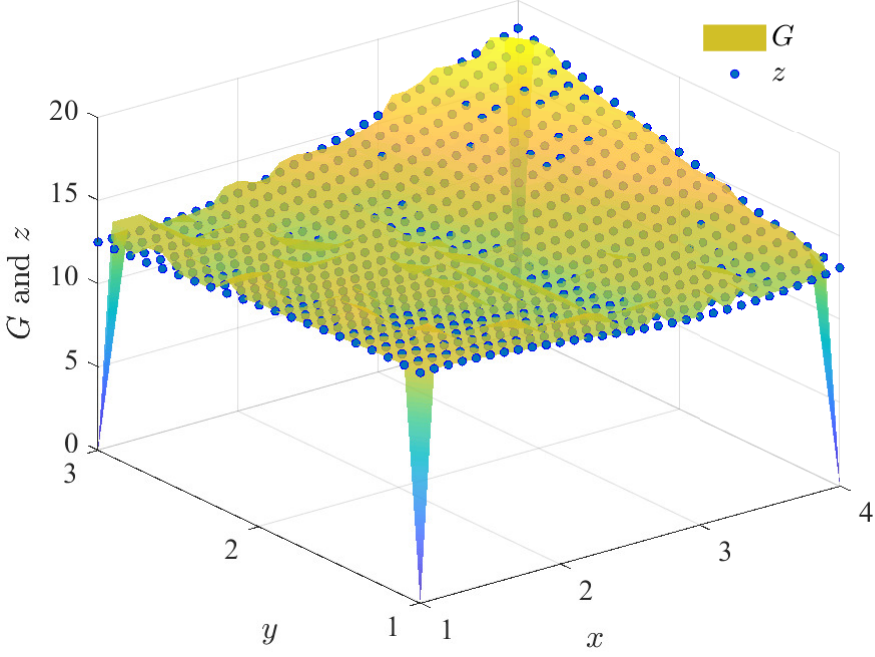}}\hspace{5pt}
	\subfloat[Mean filtering.]{\includegraphics[width=.47\columnwidth]{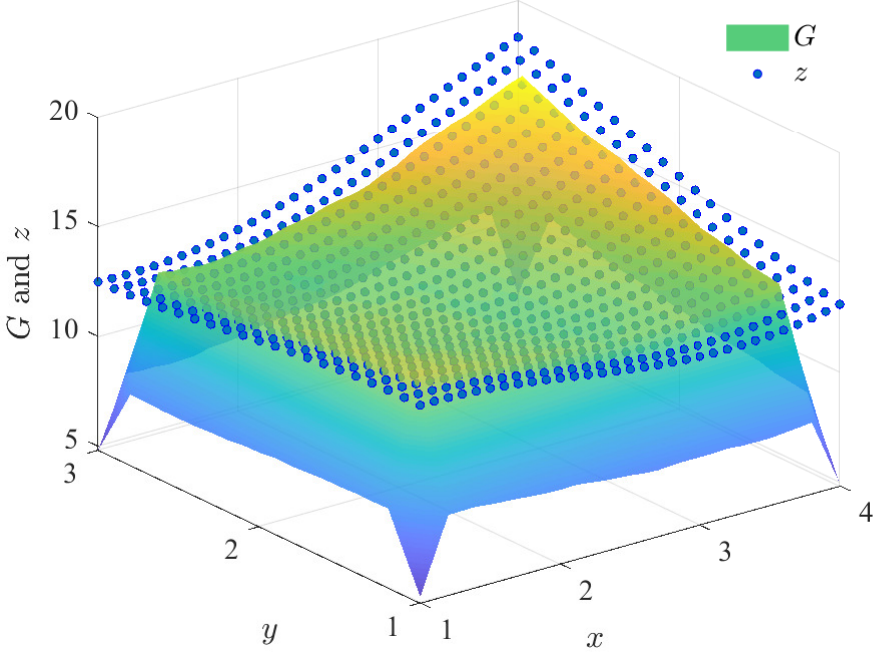}}\\
	\subfloat[Gaussian filtering.]{\includegraphics[width=.47\columnwidth]{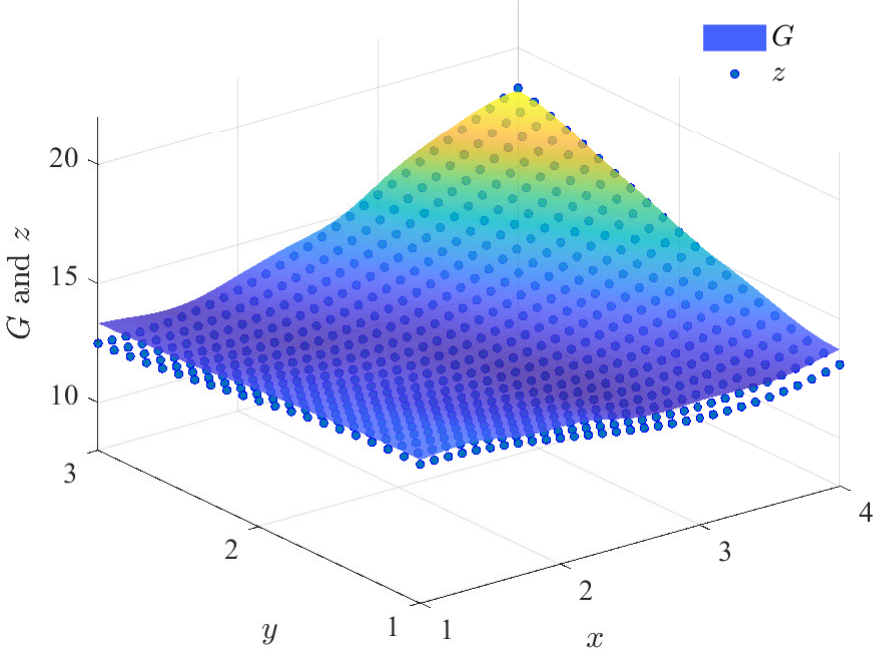}}\hspace{5pt}
	\subfloat[2D adaptive Wiener filtering.]{\includegraphics[width=.47\columnwidth]{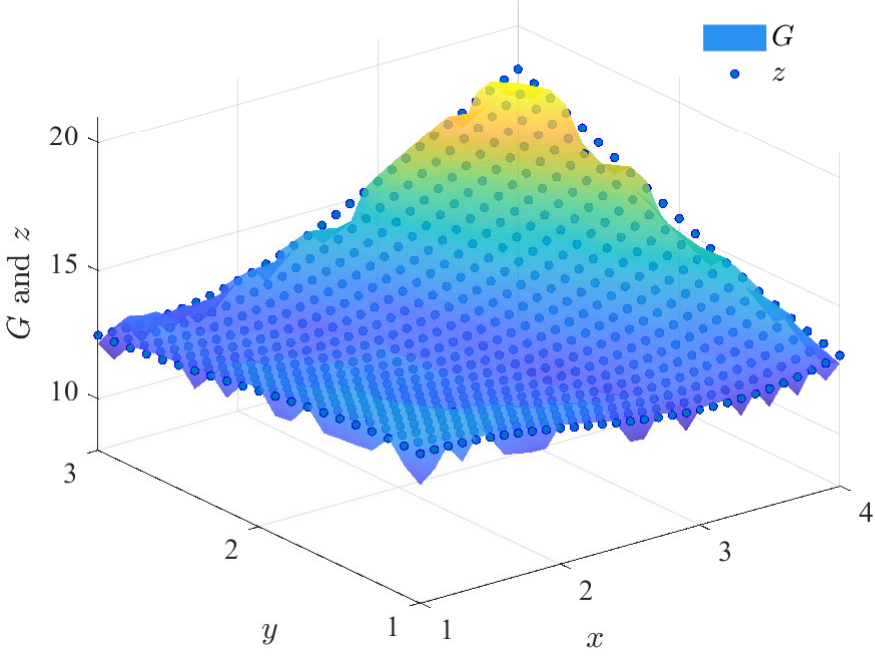}}\\
	\caption{Denoising results of filtering algorithms under SPN.}\label{Fig9_4}
\end{figure}

Assuming that the noise $ S(x,y) $ is SPN with an intensity of $ 0.9 $ and is added to the ideal experimental data $ z(x,y) $, and other details remain unchanged. The effects of different algorithms are shown in Fig.\ref{Fig9_4}.

As shown in Fig.\ref{Fig9_4}, compared with other filtering algorithms, the TDS algorithm can also still smooth the original sequence well under SPN interference.

\subsubsection{Comparative analysis based on quality assessment metrics} \label{sec 5.1.5}

We measure the execution time of the core code of each filtering algorithm to verify the execution efficiency of different filtering algorithms. Moreover, the quality of each filtering effect is measured using MATLAB R2018b in terms of a set of quality assessment metrics given as follows:

\paragraph{Mean Square Error (MSE)} It is used to evaluate the difference between the filtered image and the original image at the pixel level. The smaller the MSE value, the better the filtering effect \cite{SHAH2022505}.
\paragraph{Peak Signal to Noise Ratio (PSNR)} The similarity between the filtered image and the original image is evaluated. The larger the PSNR value, the better the filtering effect \cite{SHAH2022505}.
\paragraph{Structural Similarity Index (SSIM)} The SSIM considers the differences in brightness, contrast, structure, etc., and pays more attention to the human eye's perception of visual quality. The closer the SSIM value is to $ 1 $, the better the filtering effect \cite{venkat2014assessment}.

\textit{Remark 1}: Computer configuration: CPU i5-9300H, GPU GTX1660ti, Memory 16G.

\textit{Remark 2}: Based on corollary 1, the trend sequence \textit{\textbf{G}} of the sequence \textit{\textbf{Z}} in the TDS algorithm is the solution of a certain kind Sylvester equation. Since $ \textit{\textbf{T}} $ and $ \textit{\textbf{H}} $ are real symmetric matrices, matrices $ \textit{\textbf{A}} $ and $ \textit{\textbf{B}} $ must be diagonalization decomposition. Furthermore, we can obtain the analytical solution of formula \eqref{19}. In this paper, we use this method to solve the Sylvester equation.

\textit{Remark 3}: As mentioned above, $ \textit{\textbf{T}} $ and $ \textit{\textbf{H}} $ are also sparse matrices, which provides an important information, that is, it is also a good method to solve Sylvester equation by numerical methods. Currently, there are many effective methods for solving large sparse matrix equations such as Cholesky decomposition, sparse LU factorization method \cite{lee2018dynamic,hager2021applied}.

According to the above metrics, the difference between the real sequence $ z $ and the filtered sequence $ G $ is calculated. It can be clearly seen from Table.\ref{tab1} that the TDS algorithm is optimal in terms of MSE, PSNR and SSIM metrics at various noise types. Moreover, the TDS algorithm is also significantly superior to other filtering algorithms in terms of execution efficiency. In addition, at AWGN interference, MSE, PSNR and SSIM are equal to $ 0.06 $, $ 37.17 $, and $ 0.76 $, respectively when $ \lambda=735 $. Although $ \lambda = 735 $ is not optimal in these metrics, it is optimal when the noise distribution after filtering is used as a metric.
\begin{table*}[htbp]
	\setlength\tabcolsep{3pt}
	\centering
	\caption{MSE, PSNR and SSIM of each filtering algorithm at various noise types ($ \lambda=100 $).}
	\label{tab1}
    \resizebox{\textwidth}{!}{
    \begin{tabular}{cccccccccccccc}
    \toprule
	\multirow{2}[4]{*}{} & \multicolumn{3}{c}{AWGN} & \multicolumn{3}{c}{CN} & \multicolumn{3}{c}{MWGN} & \multicolumn{3}{c}{SPN} & \multirow{2}[4]{*}{Time $ \downarrow $} \\
	\cmidrule{2-13}          & MSE $ \downarrow $  & PSNR $ \uparrow $ & SSIM $ \uparrow $ & MSE $ \downarrow $  & PSNR $ \uparrow $ & SSIM $ \uparrow $ & MSE $ \downarrow $  & PSNR $ \uparrow $ & SSIM $ \uparrow $ & MSE $ \downarrow $  & PSNR $ \uparrow $ & SSIM $ \uparrow $ &  \\
	\midrule
		TDS   & \textbf{0.02} & \textbf{41.51} & \textbf{0.90} & \textbf{0.09} & \textbf{35.56} & \textbf{0.66} & \textbf{0.01(05)} & \textbf{45.05} & \textbf{0.94} & \textbf{0.28} & \textbf{30.74} & \textbf{0.96} & $ \textless\textbf{0.001} \mathrm{s}$ \\
		Median filtering & 1.55  & 23.36 & 0.32  & 1.88  & 22.52 & 0.20   & 1.39  & 23.84 & 0.51  & 1.72  & 22.91 & 0.34  & $ \textless0.068 \mathrm{s}$ \\
		Mean filtering & 7.09  & 16.75 & 0.18  & 7.53  & 16.49 & 0.09  & 7.00     & 16.81 & 0.22  & 6.28  & 17.28 & 0.20   & $ \textless0.025 \mathrm{s}$ \\
		Gaussian filtering & 0.03  & 40.57 & 0.80   & 0.13  & 33.99 & 0.51  & 0.01(19) & 44.50  & 0.91  & 0.30   & 30.47 & 0.91  & $ \textless0.038 \mathrm{s}$ \\
		2D adaptive Wiener filtering & 0.33  & 30.01 & 0.35  & 1.91  & 22.45 & 0.13  & 0.17  & 32.89 & 0.50   & 0.30   & 30.52 & 0.50   & $ \textless0.016 \mathrm{s}$ \\
		\bottomrule
	\end{tabular}%
    }
\end{table*}%

\subsubsection{Sensitivity analysis}\label{sec 5.1.6}
For the above simulation analysis, $ \lambda=100 $ is not a deliberately chosen number. Therefore, in order to deeply compare the performance of each filtering algorithms, we conduct sensitivity analysis on filtering algorithms at different noises.

As shown in Table.\ref{tab2}, we set the following parameters for different filtering algorithms.

\begin{itemize}
	\item \textbf{TDS}: The global smoothing parameters $ \lambda $ are taken as $ 10 $, $ 20 $, $ \cdots $, and $ 310 $ for the TDS algorithm, respectively.
	\item \textbf{Median filtering}: Templates of sizes $ 3\times3 $, $ \cdots $, and $ 5\times5 $ are selected for the median filtering, respectively.
	\item \textbf{Mean filtering}: Templates $ \mathrm{ones(3,3)}/9 $ and $ \mathrm{ones(5,5)}/25 $ are selected for the mean filtering algorithm, where $ \mathrm{ones}(m,n) $ represents a $ m\times n $ matrix whose elements are all $ 1 $.
	\item \textbf{Gaussian filtering}: The standard deviation values of $ 1.2 $, $ 1.4 $, $ \cdots $, and $ 5.0 $ are selected for the Gaussian filtering algorithm, respectively.
	\item \textbf{2D adaptive Wiener filtering}: Templates of sizes $ 3\times3 $, $ \cdots $, and $ 5\times5 $ are selected for the 2D adaptive Wiener filtering, respectively.
\end{itemize} 
	
The implementation and characteristic of different filtering algorithms are shown in Table.\ref{filtering_methods}.

\begin{table*}[htbp]
		\setlength\tabcolsep{12.5pt}
		\centering
		\caption{Characteristics and implementation methods of different filtering algorithms.}
        \resizebox{\textwidth}{!}{
		\begin{threeparttable}[b]
		\begin{tabular}{>{\centering\arraybackslash}m{3.5cm} >{\centering\arraybackslash}m{6cm} >{\centering\arraybackslash}m{6cm}}
			\toprule
			Filtering Method & Characteristics & Implementation Method \\
			\midrule
			TDS & 
			\begin{itemize}
				\item Image denoising, suitable for images with varied noise.
				\item Rich image processing application scenarios (image smoothing, image sharpening, etc).
				\item Trend extraction of general two-dimensional sequence.
				\item Simple and fast, low computational complexity, but more sensitive to the amount of data.
			\end{itemize}
			& 
			\begin{itemize}
				\item Considering the overall characteristics of the sequence, a loss function is defined.
				\item A general two-dimensional sequence can be decomposed into a trend sequence and a fluctuation sequence.
				\item The trend sequence is identified as the solution when this loss function takes a minimum value.
				\item The trend sequence contains the main features of the sequence and the fluctuation sequence contains the detailed features or noise interference of the sequence.
			\end{itemize}
			\\
			\midrule
			Median Filtering & 
			\begin{itemize}
				\item Non-linear filter.
				\item Excellent for removing sharp noise (e.g., salt-and-pepper).
				\item Preserves edges.
				\item Higher computational complexity, especially with larger windows.
			\end{itemize}
			& 
			\begin{itemize}
				\item For each pixel, select a window around it.
				\item Sort the pixel values within the window.
				\item Replace the central pixel with the median value.
				\item Typically uses a $ 3\times3 $ or larger window.
			\end{itemize}
			\\
			\midrule
			Mean Filtering & 
			\begin{itemize}
				\item Linear filter.
				\item Smooths the image, reducing random noise.
				\item Simple and fast, low computational complexity.
				\item Easy to implement, but may blur edges.
			\end{itemize}
			& 
			\begin{itemize}
				\item For each pixel, select a window around it.
				\item Calculate the average value of the pixel values within the window.
				\item Replace the central pixel with the average value.
				\item Typically uses a $ 3\times3 $ or larger window.
			\end{itemize}
			\\
			\midrule
			Gaussian Filtering & 
			\begin{itemize}
				\item Linear filter.
				\item Weighs neighboring pixels using a Gaussian function.
				\item Smooths the image, effectively removes Gaussian noise.
				\item Less effective at preserving edges, especially with high standard deviation $\sigma$.
			\end{itemize}
			& 
			\begin{itemize}
				\item Create a Gaussian kernel.
				\item For each pixel, apply the Gaussian kernel to the window of pixel values.
				\item Replace the central pixel with the weighted average value.
				\item Kernel size and standard deviation $\sigma$ are user-defined.
			\end{itemize}
			\\
			\midrule
			2D Adaptive Wiener Filtering & 
			\begin{itemize}
				\item Non-linear filter.
				\item Adaptive adjustment of filter parameters.
				\item Balances noise reduction and edge detail preservation.
				\item Higher computational complexity, suitable for images with varied noise.
			\end{itemize}
			& 
			\begin{itemize}
				\item For each pixel, select a window around it.
				\item Calculate the local mean and variance of the pixels within the window.
				\item Adaptively adjust filter parameters to balance noise removal and detail preservation.
				\item Calculate the central pixel with the adjusted mean value.
			\end{itemize}
			\\
			\bottomrule
		\end{tabular}
		\begin{tablenotes}
			\footnotesize
			\item[1] For the filtering task, the noise sequence can be regarded as the original sequence, then the trend sequence is the filtered sequence. This fact will be verified in Sections \ref{sec 5.1} and \ref{sec 5.2}.
			\item[2] More generally, the TDS algorithm can decompose the original sequence into a trend sequence and a fluctuation sequence. According to this property, we can achieve rich image processing tasks. This fact will be verified in Section \ref{sec 5.3}.
		\end{tablenotes}
	\end{threeparttable}
            }
		\label{filtering_methods}
\end{table*}

\begin{table*}[htbp]
	\setlength\tabcolsep{4.5pt}
	\centering
	\caption{Sensitivity analysis of filtering algorithms.}
	\label{tab2}
        \resizebox{\textwidth}{!}{
	\begin{tabular}{ccccccccccccccc}
		\toprule
		\multirow{2}[4]{*}{} &   & \multicolumn{3}{c}{AWGN} & \multicolumn{3}{c}{CN} & \multicolumn{3}{c}{MWGN} & \multicolumn{3}{c}{SPN} & \multirow{2}[4]{*}{Time $ \downarrow $} \\
		\cmidrule{3-14}      &   & MSE $ \downarrow $ & PSNR $ \uparrow $  & SSIM $ \uparrow $ & MSE  $ \downarrow $ & PSNR $ \uparrow $ & SSIM $ \uparrow $ & MSE $ \downarrow $ & PSNR $ \uparrow $ & SSIM $ \uparrow $ & MSE $ \downarrow $ & PSNR $ \uparrow $ & SSIM $ \uparrow $ &  \\
		\midrule
		\multirow{25}[2]{*}{TDS} & 10 & 0.0307  & 40.3856  & 0.8092  & 0.1565  & 33.3116  & 0.5003  & 0.0093  & 45.5928  & 0.9243  & 0.2827  & 30.7421  & 0.9246  & \multirow{25}[2]{*}{$ \textless\textbf{0.001} \mathrm{s}$} \\
		& 20 & 0.0256  & 41.1735  & 0.8581  & 0.1255  & 34.2699  & 0.5513  & 0.0080  & 46.2111  & 0.9431  & 0.2810  & 30.7685  & 0.9517  &  \\
		& 30 & 0.0239  & 41.4807  & 0.8768  & 0.1133  & 34.7141  & 0.5797  & \textbf{0.0078} & \textbf{46.3260} & 0.9480  & 0.2806  & 30.7755  & 0.9615  &  \\
		& 40 & 0.0231  & 41.6247  & 0.8864  & 0.1065  & 34.9835  & 0.5995  & 0.0079  & 46.2705  & \textbf{0.9492} & \textbf{0.2805} & \textbf{30.7761} & 0.9654  &  \\
		& 50 & 0.0227  & 41.6873  & 0.8920  & 0.1021  & 35.1665  & 0.6147  & 0.0082  & 46.1327  & 0.9489  & 0.2807  & 30.7736  & \textbf{0.9664} &  \\
		& 60 & \textbf{0.0227} & \textbf{41.7008} & 0.8954  & 0.0990  & 35.2979  & 0.6269  & 0.0085  & 45.9491  & 0.9477  & 0.2810  & 30.7692  & 0.9659  &  \\
		& 70 & 0.0228  & 41.6817  & 0.8974  & 0.0968  & 35.3950  & 0.6370  & 0.0089  & 45.7391  & 0.9460  & 0.2813  & 30.7635  & 0.9643  &  \\
		& 80 & 0.0230  & 41.6396  & 0.8984  & 0.0952  & 35.4677  & 0.6456  & 0.0094  & 45.5143  & 0.9438  & 0.2818  & 30.7569  & 0.9621  &  \\
		& 90 & 0.0233  & 41.5811  & \textbf{0.8986} & 0.0940  & 35.5221  & 0.6530  & 0.0099  & 45.2821  & 0.9414  & 0.2822  & 30.7495  & 0.9594  &  \\
		& 100 & 0.0237  & 41.5103  & 0.8983  & 0.0932  & 35.5624  & 0.6594  & 0.0105  & 45.0476  & 0.9387  & 0.2828  & 30.7415  & 0.9564  &  \\
		& 110 & 0.0241  & 41.4307  & 0.8976  & 0.0926  & 35.5915  & 0.6650  & 0.0111  & 44.8140  & 0.9359  & 0.2833  & 30.7330  & 0.9532  &  \\
		& 120 & 0.0246  & 41.3445  & 0.8964  & 0.0921  & 35.6115  & 0.6699  & 0.0117  & 44.5837  & 0.9329  & 0.2839  & 30.7241  & 0.9498  &  \\
		& 130 & 0.0251  & 41.2536  & 0.8950  & 0.0919  & 35.6240  & 0.6743  & 0.0123  & 44.3581  & 0.9299  & 0.2845  & 30.7149  & 0.9463  &  \\
		& 140 & 0.0257  & 41.1595  & 0.8933  & 0.0917  & 35.6304  & 0.6782  & 0.0129  & 44.1382  & 0.9268  & 0.2851  & 30.7053  & 0.9427  &  \\
		& 150 & 0.0263  & 41.0631  & 0.8914  & \textbf{0.0917} & \textbf{35.6317} & 0.6816  & 0.0136  & 43.9244  & 0.9237  & 0.2858  & 30.6956  & 0.9391  &  \\
		& 160 & 0.0269  & 40.9655  & 0.8894  & 0.0918  & 35.6287  & 0.6847  & 0.0143  & 43.7172  & 0.9205  & 0.2864  & 30.6856  & 0.9355  &  \\
		& 170 & 0.0275  & 40.8673  & 0.8872  & 0.0919  & 35.6220  & 0.6874  & 0.0149  & 43.5167  & 0.9174  & 0.2871  & 30.6754  & 0.9318  &  \\
		& 180 & 0.0281  & 40.7690  & 0.8849  & 0.0921  & 35.6123  & 0.6898  & 0.0156  & 43.3228  & 0.9142  & 0.2878  & 30.6652  & 0.9281  &  \\
		& 190 & 0.0287  & 40.6710  & 0.8825  & 0.0924  & 35.5999  & 0.6919  & 0.0163  & 43.1354  & 0.9110  & 0.2885  & 30.6548  & 0.9245  &  \\
		& 200 & 0.0294  & 40.5736  & 0.8800  & 0.0927  & 35.5854  & 0.6938  & 0.0170  & 42.9544  & 0.9079  & 0.2892  & 30.6443  & 0.9209  &  \\
		& 270 & 0.0341  & 39.9259  & 0.8616  & 0.0958  & 35.4434  & 0.7007  & 0.0219  & 41.8475  & 0.8865  & 0.2942  & 30.5698  & 0.8965  &  \\
		& 280 & 0.0348  & 39.8393  & 0.8590  & 0.0963  & 35.4196  & 0.7010  & 0.0226  & 41.7094  & 0.8836  & 0.2949  & 30.5591  & 0.8932  &  \\
		& 290 & 0.0355  & 39.7542  & 0.8563  & 0.0968  & 35.3954  & \textbf{0.7011} & 0.0233  & 41.5756  & 0.8808  & 0.2956  & 30.5484  & 0.8900  &  \\
		& 300 & 0.0362  & 39.6707  & 0.8537  & 0.0974  & 35.3707  & 0.7011  & 0.0240  & 41.4460  & 0.8779  & 0.2963  & 30.5378  & 0.8868  &  \\
		& 310 & 0.0369  & 39.5888  & 0.8511  & 0.0979  & 35.3458  & 0.7010  & 0.0247  & 41.3202  & 0.8751  & 0.2971  & 30.5272  & 0.8836  &  \\
		\midrule
		\multicolumn{1}{c}{\multirow{7}[2]{*}{\makecell{Median\\filtering}}} & $ 3\times3 $ & \textbf{1.5464} & \textbf{23.3624} & \textbf{0.3170} & \textbf{1.8776} & \textbf{22.5197} & 0.2048  & \textbf{1.3865} & \textbf{23.8364} & \textbf{0.5083} & \textbf{1.7151} & \textbf{22.9127} & 0.3397  & \multirow{7}[2]{*}{$ \textless0.045 \mathrm{s} $} \\
		& $ 3\times4 $ & 3.4380  & 19.8926  & 0.2744  & 4.1218  & 19.1047  & 0.1850  & 3.1930  & 20.2137  & 0.4602  & 3.3637  & 19.9875  & 0.3077  &  \\
		& $ 4\times3 $ & 4.1196  & 19.1071  & 0.2475  & 4.9382  & 18.3199  & 0.1692  & 3.8192  & 19.4360  & 0.4161  & 3.9466  & 19.2934  & 0.2799  &  \\
		& $ 4\times4 $ & 6.8062  & 16.9266  & 0.2083  & 8.1185  & 16.1609  & 0.1543  & 6.4471  & 17.1620  & 0.3695  & 6.5186  & 17.1141  & 0.2294  &  \\
		& $ 4\times5 $ & 5.6699  & 17.7199  & 0.2649  & 6.9971  & 16.8065  & 0.1885  & 5.3785  & 17.9491  & 0.4260  & 5.5186  & 17.8374  & 0.2817  &  \\
		& $ 5\times4 $ & 5.0090  & 18.2582  & 0.2657  & 5.8729  & 17.5671  & 0.1869  & 4.7648  & 18.4752  & 0.4411  & 4.9257  & 18.3310  & 0.3068  &  \\
		& $ 5\times5 $ & 4.1851  & 19.0386  & 0.3105  & 4.6717  & 18.5609  & \textbf{0.2270} & 4.0186  & 19.2150  & 0.4791  & 4.2526  & 18.9691  & \textbf{0.3483} &  \\
		\midrule
		\multicolumn{1}{c}{\multirow{2}[2]{*}{\makecell{Mean\\filtering}}} & $ 3\times3 $ & \textbf{3.9313} & \textbf{19.3103} & 0.1758  & \textbf{4.5397} & \textbf{18.6854} & 0.0688  & \textbf{3.8105} & \textbf{19.4459} & \textbf{0.2557} & \textbf{3.4953} & \textbf{19.8208} & \textbf{0.2428} & \multirow{2}[2]{*}{$ \textless0.025 \mathrm{s}$} \\
		& $ 5\times5 $ & 7.0872  & 16.7509  & \textbf{0.1825} & 7.5314  & 16.4869  & \textbf{0.0939} & 6.9952  & 16.8077  & 0.2238  & 6.2755  & 17.2792  & 0.2043  &  \\
		\midrule
		\multicolumn{1}{c}{\multirow{16}[2]{*}{\makecell{Gaussian\\filtering}}} & 1.2 & 0.0573  & 37.6773  & 0.6360  & 0.3081  & 30.3682  & 0.3459  & 0.0176  & 42.7968  & 0.8154  & 0.2948  & 30.5599  & 0.8247  & \multirow{16}[2]{*}{$ \textless0.038 \mathrm{s}$} \\
		& 1.4 & 0.0449  & 38.7308  & 0.6935  & 0.2339  & 31.5662  & 0.4012  & 0.0144  & 43.6749  & 0.8524  & \textbf{0.2942} & \textbf{30.5692} & 0.8556  &  \\
		& 1.6 & 0.0362  & 39.6679  & 0.7444  & 0.1800  & 32.7036  & 0.4533  & 0.0125  & 44.2714  & 0.8822  & 0.2961  & 30.5406  & 0.8805  &  \\
		& 1.8 & 0.0319  & 40.2216  & 0.7782  & 0.1515  & 33.4526  & 0.4873  & 0.0119 & 44.4845  & 0.9000  & 0.2983  & 30.5090  & 0.8952  &  \\
		& 2.0 & 0.0294  & 40.5688  & 0.8033  & 0.1339  & 33.9884  & 0.5120  & \textbf{0.0119} & \textbf{44.4968} & 0.9117  & 0.3008  & 30.4733  & 0.9054  &  \\
		& 2.2 & 0.0277  & 40.8347  & 0.8286  & 0.1167  & 34.5850  & 0.5463  & 0.0129  & 44.1653  & 0.9206  & 0.3074  & 30.3793  & 0.9173  &  \\
		& 2.4 & \textbf{0.0272} & \textbf{40.9080} & 0.8424  & 0.1087  & 34.8930  & 0.5655  & 0.0137  & 43.8800  & 0.9248  & 0.3114  & 30.3218  & 0.9231  &  \\
		& 2.6 & 0.0279  & 40.8073  & 0.8527  & 0.1013  & 35.2011  & 0.5912  & 0.0160  & 43.2209  & \textbf{0.9254} & 0.3205  & 30.1969  & 0.9267  &  \\
		& 2.8 & 0.0286  & 40.6926  & 0.8582  & 0.0978  & 35.3542  & 0.6082  & 0.0176  & 42.8007  & 0.9251  & 0.3261  & 30.1219  & \textbf{0.9280} &  \\
		& 3.0 & 0.0296  & 40.5469  & \textbf{0.8613} & 0.0956  & 35.4516  & 0.6222  & 0.0193  & 42.4069  & 0.9239  & 0.3315  & 30.0510  & 0.9279  &  \\
		& 3.2 & 0.0323  & 40.1607  & 0.8613  & 0.0934  & 35.5524  & 0.6421  & 0.0233  & 41.5787  & 0.9180  & 0.3445  & 29.8834  & 0.9217  &  \\
		& 3.4 & 0.0341  & 39.9322  & 0.8608  & \textbf{0.0929} & \textbf{35.5748} & 0.6522  & 0.0257  & 41.1598  & 0.9145  & 0.3513  & 29.7992  & 0.9180  &  \\
		& 3.6 & 0.0381  & 39.4442  & 0.8562  & 0.0933  & 35.5570  & 0.6639  & 0.0310  & 40.3474  & 0.9056  & 0.3664  & 29.6160  & 0.9077  &  \\
		& 3.8 & 0.0406  & 39.1694  & 0.8532  & 0.0940  & 35.5231  & 0.6701  & 0.0341  & 39.9324  & 0.9003  & 0.3745  & 29.5214  & 0.9014  &  \\
		& 4.7 & 0.0624  & 37.3073  & 0.8211  & 0.1067  & 34.9723  & \textbf{0.6823} & 0.0599  & 37.4816  & 0.8569  & 0.4360  & 28.8608  & 0.8476  &  \\
		& 5.0 & 0.0685  & 36.9002  & 0.8116  & 0.1111  & 34.7975  & 0.6815  & 0.0669  & 37.0007  & 0.8453  & 0.4509  & 28.7145  & 0.8333  &  \\
		\midrule
		\multicolumn{1}{c}{\multirow{7}[2]{*}{\makecell{2D\\adaptive\\Wiener\\filtering}}} & $ 3\times3 $ & \textbf{0.3349} & \textbf{30.0060} & \textbf{0.3497} & 1.9094  & 22.4466  & 0.1268  & \textbf{0.1726} & \textbf{32.8853} & \textbf{0.4955} & \textbf{0.2978} & \textbf{30.5167} & \textbf{0.5023} & \multirow{7}[2]{*}{$ \textless0.016 \mathrm{s}$} \\
		& $ 3\times4 $ & 0.4107  & 29.1204  & 0.3476  & \textbf{1.7899} & \textbf{22.7273} & 0.1251  & 0.2509  & 31.2604  & 0.4732  & 0.3542  & 29.7635  & 0.4784  &  \\
		& $ 4\times3 $ & 0.4501  & 28.7223  & 0.3375  & 1.8555  & 22.5711  & 0.1409  & 0.2857  & 30.6962  & 0.4575  & 0.3596  & 29.6978  & 0.4546  &  \\
		& $ 4\times4 $ & 0.5701  & 27.6965  & 0.3215  & 1.8061  & 22.6881  & 0.1366  & 0.4057  & 29.1735  & 0.4175  & 0.4440  & 28.7821  & 0.4164  &  \\
		& $ 4\times5 $ & 0.6587  & 27.0689  & 0.3328  & 1.8510  & 22.5815  & 0.1526  & 0.4876  & 28.3749  & 0.4166  & 0.4448  & 28.7744  & 0.4037  &  \\
		& $ 5\times4 $ & 0.7259  & 26.6468  & 0.3247  & 1.9446  & 22.3672  & 0.1584  & 0.5400  & 27.9319  & 0.4039  & 0.4631  & 28.5988  & 0.3927  &  \\
		& $ 5\times5 $ & 0.8386  & 26.0203  & 0.3301  & 2.0325  & 22.1753  & \textbf{0.1775} & 0.6462  & 27.1518  & 0.3989  & 0.4863  & 28.3865  & 0.3798  &  \\
		\midrule
		\makecell{TDS\\ranking} &   & 1/5 & 1/5 & 1/5 & 1/5 & 1/5 & 1/5 & 1/5 & 1/5 & 1/5 & 1/5 & 1/5 & 1/5 &  1/5\\
		\bottomrule
	\end{tabular}%
    }
\end{table*}%

For various noise types, Table.\ref{tab2} clearly shows that the TDS algorithm excels in all metrics, exhibiting the fastest execution efficiency. Compared to the Gaussian filtering algorithm, a notable advantage of the TDS algorithm lies in its ability to achieve superior filtering results by selecting the global smoothing parameter $\lambda$ from a larger interval. Moreover, the TDS algorithm overcomes the limitations of a fixed filtering window, allowing for the flexible selection of the global smoothing parameter $ \lambda $.

While the analysis presented is in the context of numerical simulations, the essence of the problem in various fields is consistent. In the given numerical case, it is evident that the TDS algorithm can accurately recover the real data by selecting the appropriate global smoothing parameter. Moreover, when the global smoothing parameter is large, its variation has little impact on the smoothing result, indicating that the global smoothing parameter significantly influences the result only within a limited range (As shown in Table.\ref{tab2}, on the interval $ (0,20] $, it can be seen from the quality evaluation metric that the change of the global smoothing parameter $ \lambda $ has a significant impact on the filtering results. In contrast, on the interval $ (20,+\infty) $, this effect is not significant.). The significance of this result is profound, which shows that when the global smoothing parameter increases to a certain extent, the influence of its change on the smoothing result is no longer significant (As shown in Table.\ref{tab2}, e.g., for MWGN, when $ \lambda=30 $, $ \lambda $ needs to change by $ 10 $ for the MSE to change by $ 0.0001 $.), i.e., the TDS algorithm has better robustness at this time. Moreover, the implementation of the TDS algorithm is simple, and only changing a single parameter can flexibly achieve different smoothing effects without noise prior information. Additionally, the TDS algorithm effectively filters out noise even under complex interference with the generalization ability.

\textit{Remark 4}: Robustness refers to the ability of a system to maintain certain performance under certain (structural, dimensional) parameter disturbances \cite{sastry2011adaptive}. By analogy, robustness can be understood as the ability of the quality assessment metric to resist changes when the parameter changes. Therefore, near the optimal quality evaluation metric, the stronger the robustness, the easier it is to adjust the parameters.

\subsubsection{Application of modified TDS algorithm}\label{sec 5.1.7}

The above analysis of a large number of numerical simulation cases shows that the TDS algorithm can achieve satisfactory results in various challenging scenarios. As previously mentioned, the modified TDS algorithm such as TDS-\uppercase\expandafter{\romannumeral1}, TDS-\uppercase\expandafter{\romannumeral2}, and TDS-\uppercase\expandafter{\romannumeral3}, can fine-tune the results of the TDS algorithm. Without loss of generality, we only take the application of TDS-\uppercase\expandafter{\romannumeral1} algorithm in the presence of AWGN as an example.

According to previous analysis, when $ \gamma=\delta $, the TDS-\uppercase\expandafter{\romannumeral1} degenerates into the TDS algorithm. Then, the quality assessment metrics for the TDS-\uppercase\expandafter{\romannumeral1} algorithm are shown in Table.\ref{tab3}.

\begin{table}[htbp]
    \setlength\tabcolsep{14.6pt}
	\centering
	\caption{Quality assessment metrics of TDS-\uppercase\expandafter{\romannumeral1} algorithm.}
	\label{tab3}
	\begin{tabular}{cccccc}
		\toprule
		&   & MSE $ \downarrow $ & PSNR $ \uparrow $ & SSIM $ \uparrow $ & Time $ \downarrow $ \\
		\midrule
		\multicolumn{1}{c}{\multirow{2}[2]{*}{TDS}} & $ \gamma=\delta=6 $0 & 0.0227  & 41.7008  & 0.8954  & \multirow{6}[8]{*}{$ \textless0.001 \mathrm{s}$ } \\
		& $ \gamma=\delta=90 $ & 0.0233  & 41.5811  & 0.8986  &  \\
		\cmidrule{1-5}    \multicolumn{1}{c}{\multirow{4}[6]{*}{TDS-\uppercase\expandafter{\romannumeral1}}} & $ \gamma=100 $, $ \delta=60 $ & 0.0224  & 41.7454  & \textbf{0.9035}  &  \\
		& $ \gamma=60 $, $ \delta=50 $ & 0.0225  & 41.7382  & 0.8958  &  \\
		\cmidrule{2-5}      & $ \gamma=90 $, $ \delta=40 $ & \textbf{0.0220}  & \textbf{41.8250}  & 0.9026  &  \\
        & $ \gamma=120 $, $ \delta=90 $ & 0.0235  & 41.5362  & 0.9013  &  \\
		\bottomrule
	\end{tabular}%
\end{table}%

As shown in Table.\ref{tab2}, when $ \lambda $ is equal to $ 60 $ and $ 90 $ respectively, the TDS algorithm is optimal in MSE (PSNR) and SSIM respectively. Then, we use TDS-\uppercase\expandafter{\romannumeral1} to adjust the results of TDS. As can be seen from Table.\ref{tab3}, the results of TDS-\uppercase\expandafter{\romannumeral1} are always better than those of TDS. This is because TDS is just a special case of TDS-\uppercase\expandafter{\romannumeral1} in theory. Consequently, the modified TDS algorithm serves as a supplementary approach to enhance the results achieved by the TDS algorithm.

\subsection{Image Filtering Application}\label{sec 5.2}

In section \ref{sec 5.1}, we extensively compared various filtering algorithms through numerical simulation analysis. The results show that for various complex noises, both the TDS algorithm and the modified TDS algorithm effectively filter noise and exhibit excellent performance in terms of MSE, PSNR, and SSIM. In this section, we further explore the application of the TDS algorithm in image filtering.

\begin{figure}[!ht]
	\centering
	\subfloat[Original image.]{\label{Fig14-1:a}\includegraphics[width=.47\columnwidth]{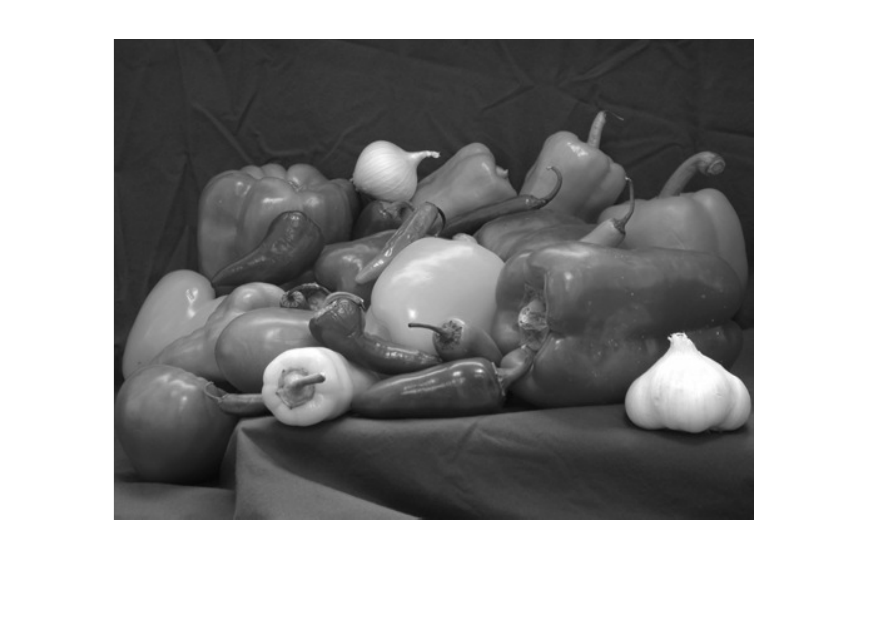}}\hspace{5pt}
	\subfloat[Noise image with AWGN.]{\label{Fig14-1:b}\includegraphics[width=.47\columnwidth]{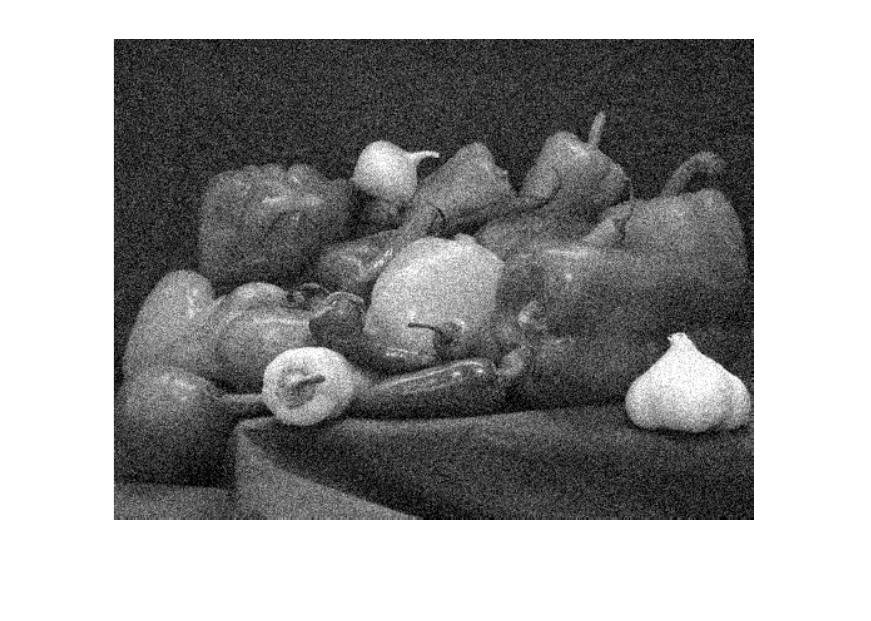}}\\
	\subfloat[Noise image with SPN.]{\label{Fig14-1:c}\includegraphics[width=.47\columnwidth]{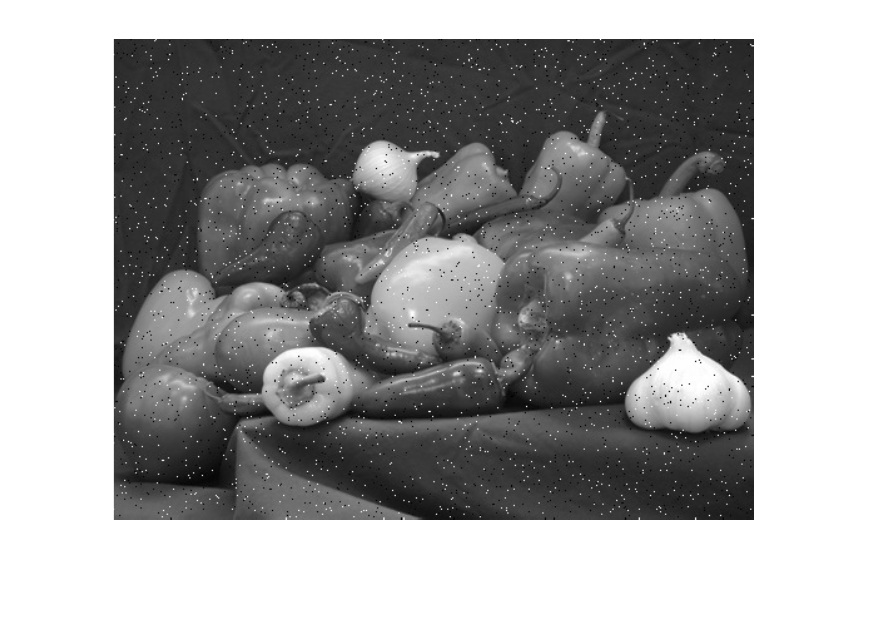}}\hspace{5pt}
	\subfloat[Noise image with Poisson noise.]{\label{Fig14-1:d}\includegraphics[width=.47\columnwidth]{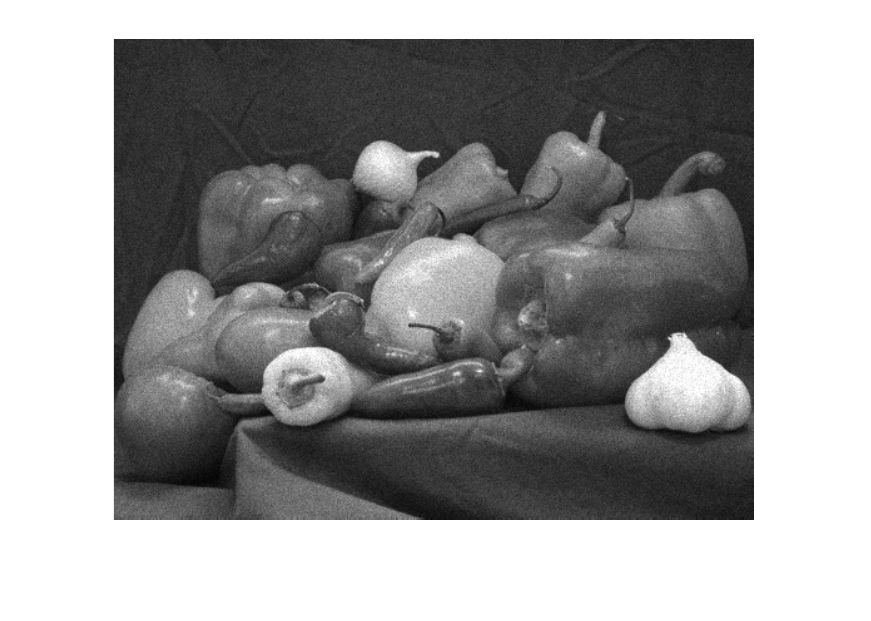}}\\
	\caption{Original image and noise image.}
	\label{14-1}
\end{figure}

The original image is shown in Fig.\ref{14-1}\subref{Fig14-1:a}. Then, AWGN with mean $ 0 $ and variance of $ 0.01 $, Poisson noise, and SPN with intensity of $ 0.02 $ are added to the original image respectively, as shown in Figs.\ref{14-1}\subref{Fig14-1:b}, \subref{Fig14-1:c}, \subref{Fig14-1:d}, respectively.

\begin{figure}[htbp]
	\centering
	\subfloat[TDS ($ \lambda=4 $, MSE, PSNR).]{\includegraphics[width=.47\columnwidth]{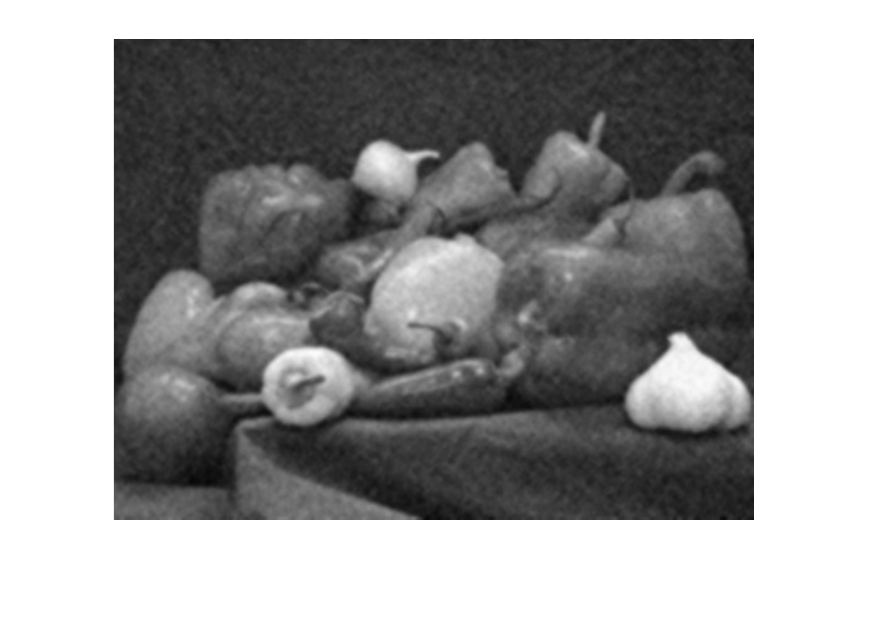}}\hspace{5pt}
	\subfloat[TDS ($ \lambda=18 $, SSIM).]{\includegraphics[width=.47\columnwidth]{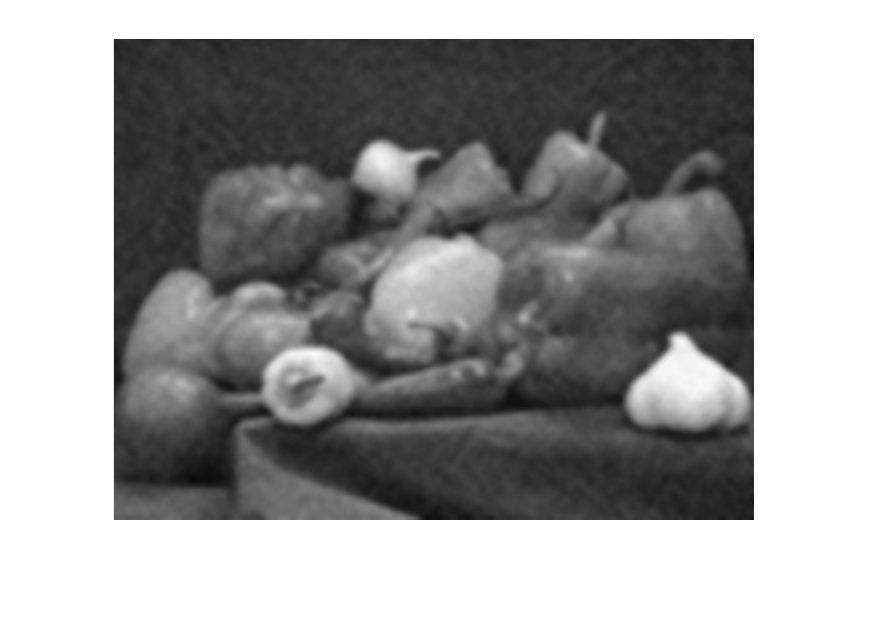}}\\
	\subfloat[TDS-\uppercase\expandafter{\romannumeral1} ($ \gamma=10 $, $ \delta=2 $, MSE, PSNR).]{\includegraphics[width=.47\columnwidth]{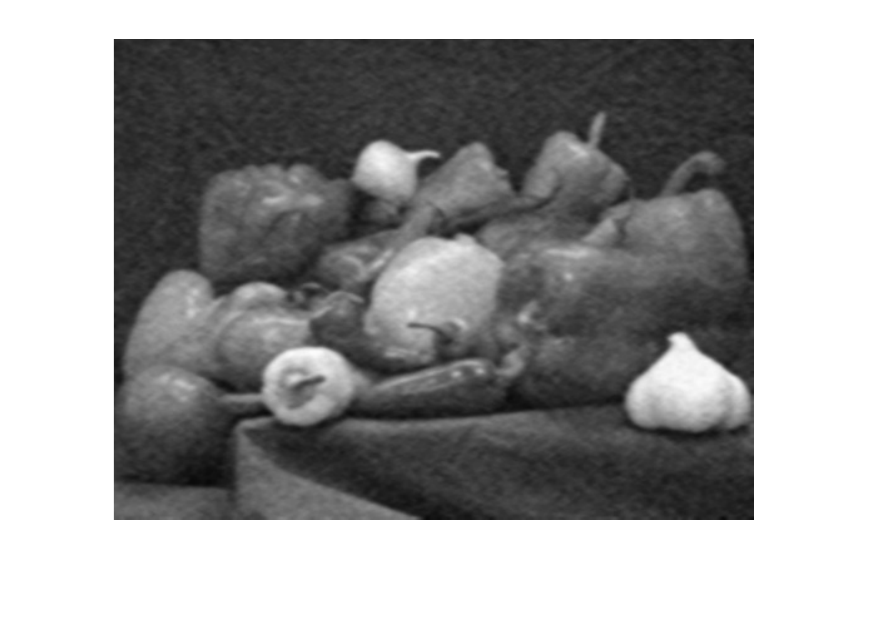}}\hspace{5pt}
	\subfloat[TDS-\uppercase\expandafter{\romannumeral1} ($ \gamma=33 $, $ \delta=10 $, SSIM).]{\includegraphics[width=.47\columnwidth]{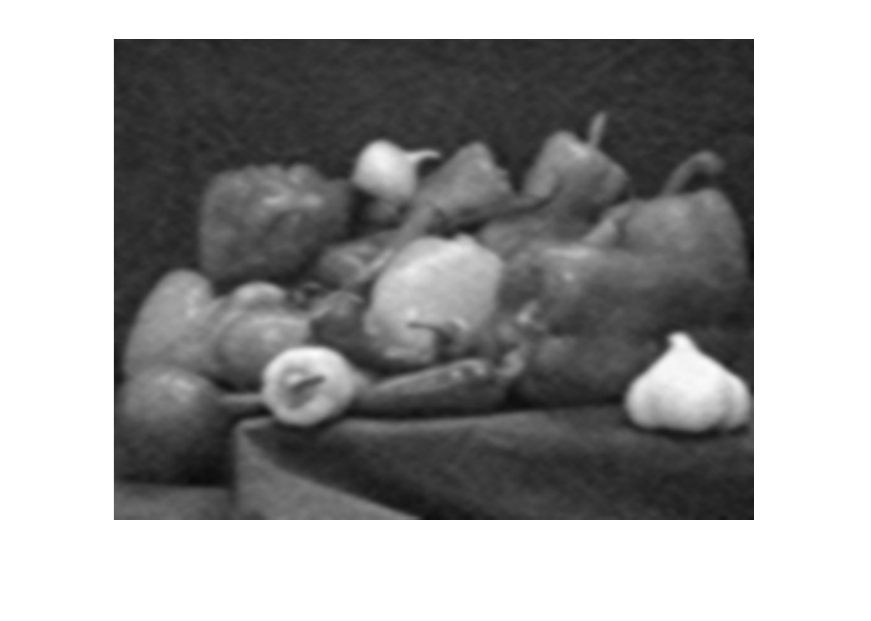}}\\
	\caption{Denoised image under AWGN.}
	\label{Fig15-1}
\end{figure}

\begin{figure}[htbp]
	\centering
	\subfloat[TDS ($ \lambda=3 $, MSE, PSNR).]{\includegraphics[width=.47\columnwidth]{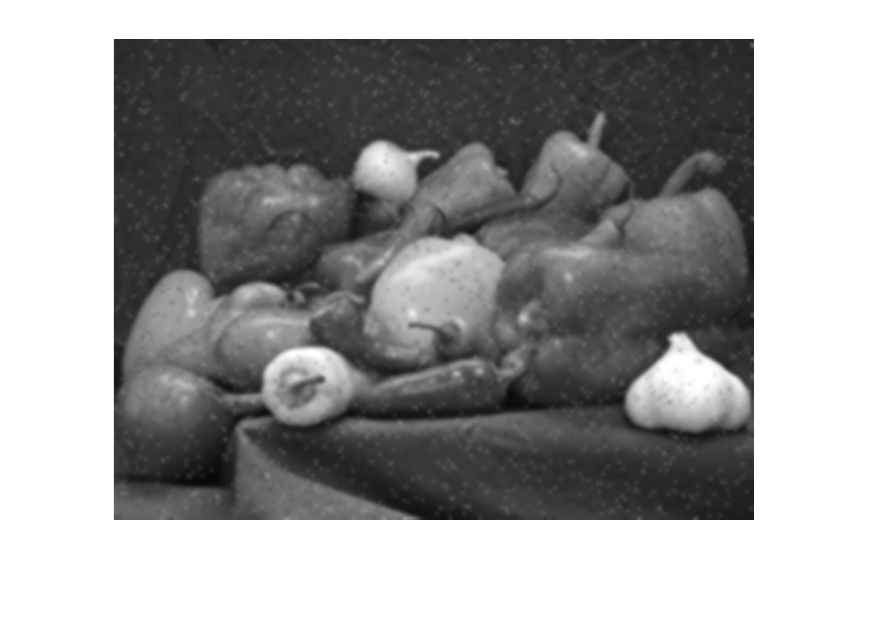}}\hspace{5pt}
	\subfloat[TDS ($ \lambda=10 $, SSIM).]{\includegraphics[width=.47\columnwidth]{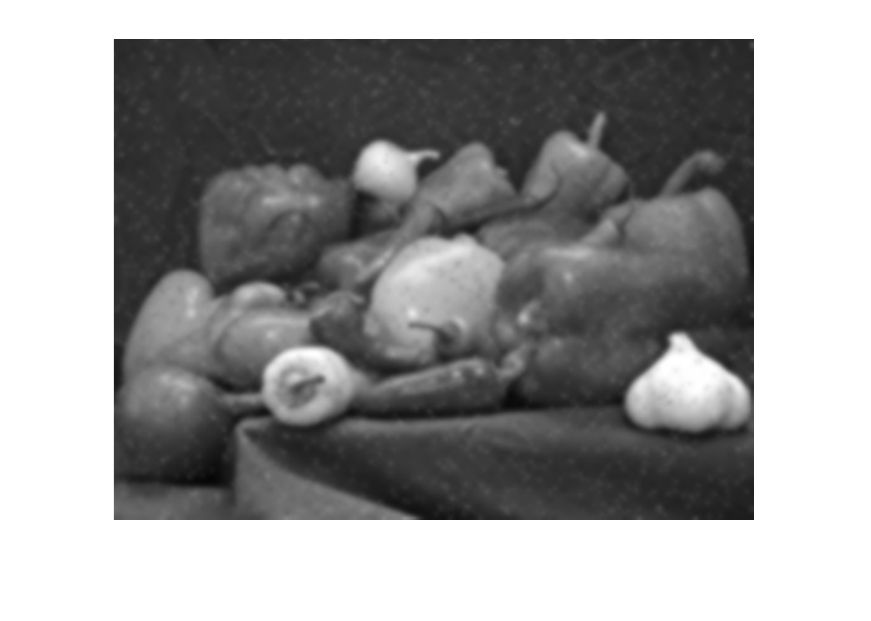}}\\
	\subfloat[TDS-\uppercase\expandafter{\romannumeral1} ($ \gamma=5 $, $ \delta=2 $, MSE, PSNR).]{\includegraphics[width=.47\columnwidth]{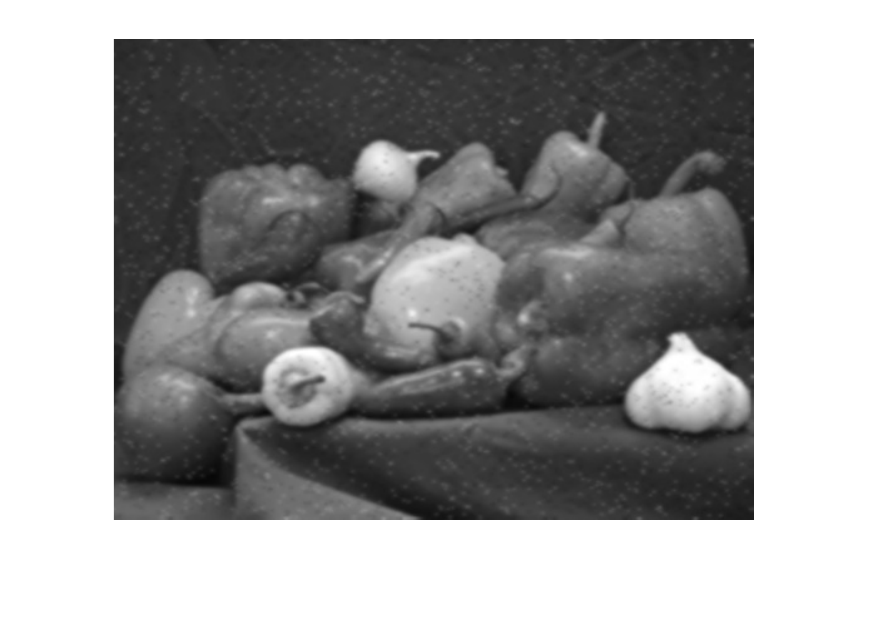}}\hspace{5pt}
	\subfloat[TDS-\uppercase\expandafter{\romannumeral1} ($ \gamma=19 $, $ \delta=6 $, SSIM).]{\includegraphics[width=.47\columnwidth]{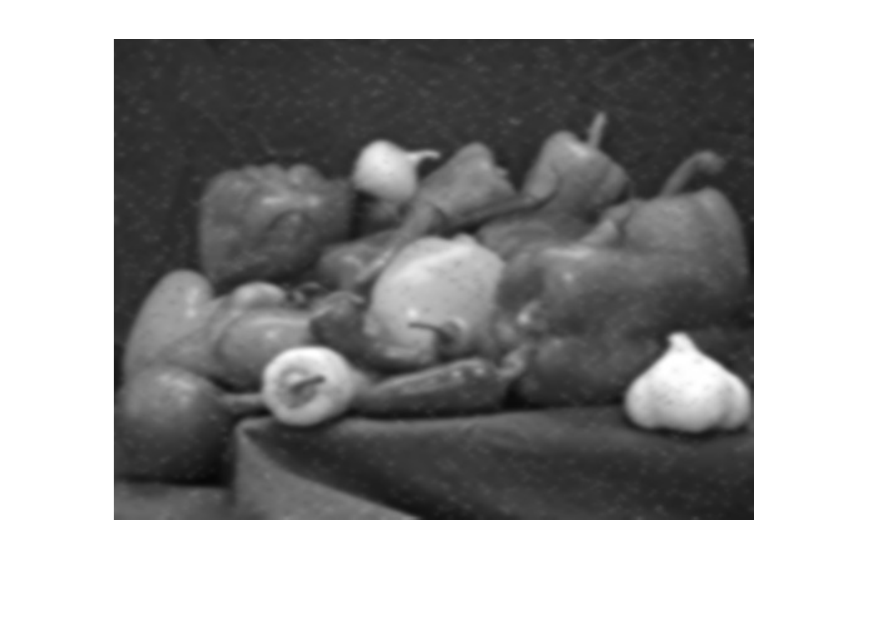}}\\
	\caption{Denoised image under SPN.}
	\label{Fig16-1}
\end{figure}

\begin{figure}[htbp]
	\centering
	\subfloat[TDS ($ \lambda=0.5 $, MSE, PSNR).]{\includegraphics[width=.47\columnwidth]{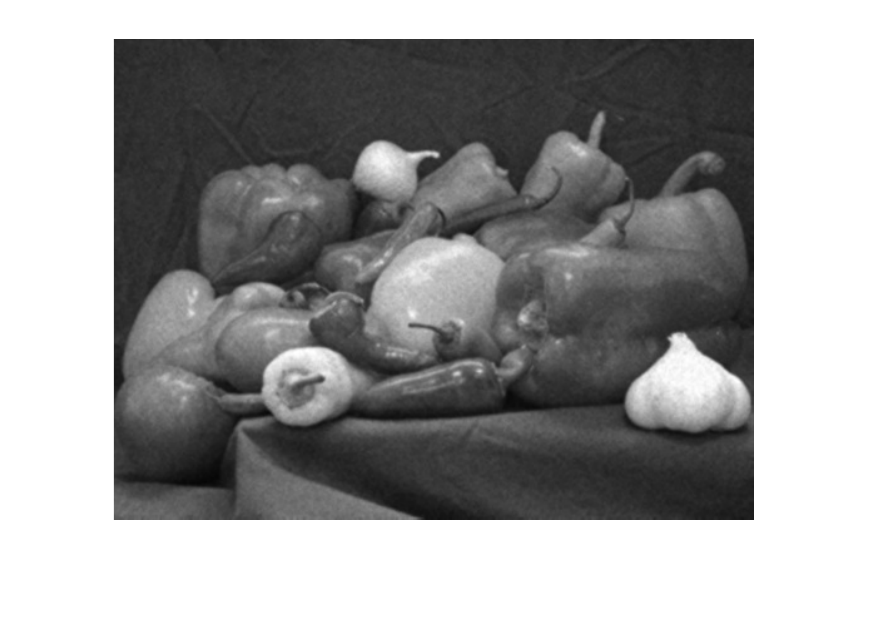}}\hspace{5pt}
	\subfloat[TDS ($ \lambda=1.8 $, SSIM).]{\includegraphics[width=.47\columnwidth]{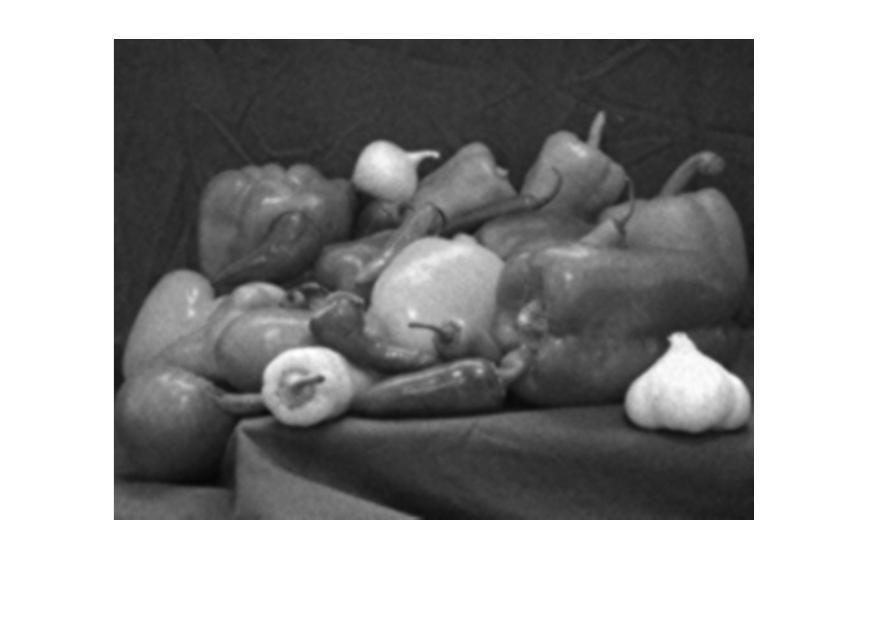}}\\
	\subfloat[TDS-\uppercase\expandafter{\romannumeral1} ($ \gamma=0.8 $, $ \delta=0.3 $, MSE, PSNR).]{\includegraphics[width=.47\columnwidth]{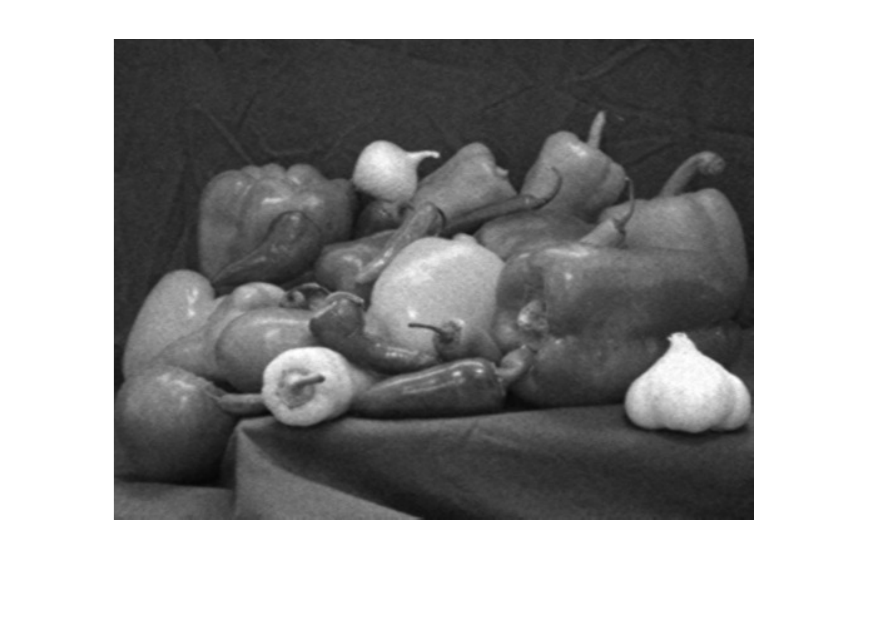}}\hspace{5pt}
	\subfloat[TDS-\uppercase\expandafter{\romannumeral1} ($ \gamma=3 $, $ \delta=1.2 $, SSIM).]{\includegraphics[width=.47\columnwidth]{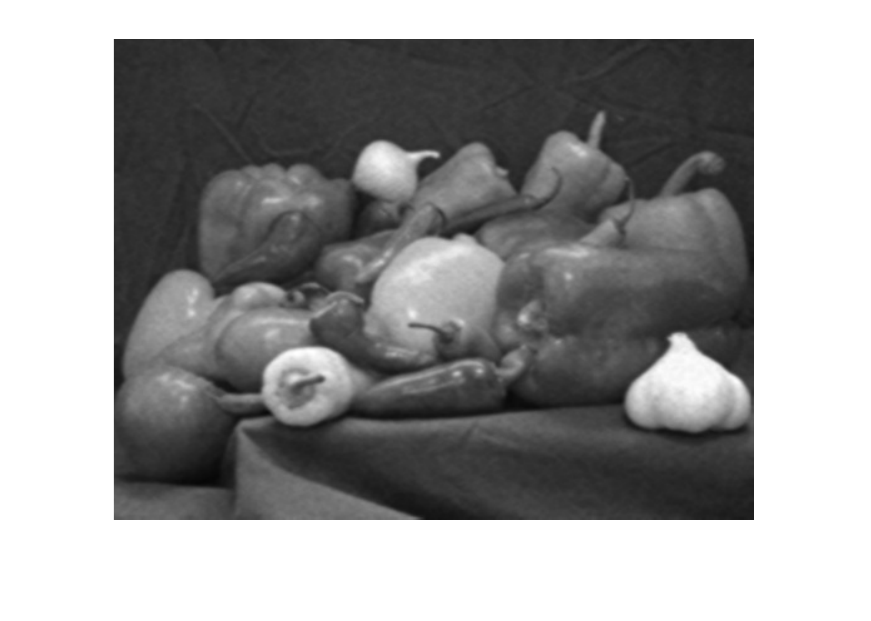}}\\
	\caption{Denoised image under Poisson noise.}
	\label{Fig17-1}
\end{figure}

In the image filtering, the value of the parameters in the filtering algorithm and corresponding metrics are shown in Table.\ref{tab4}. For different types of noise, when TDS-\uppercase\expandafter{\romannumeral1} and TDS algorithms reach optimal results under each quality assessment metric, the filtered images are shown in Figs.\ref{Fig15-1}, \ref{Fig16-1}, and \ref{Fig17-1}, respectively. 

As shown in Table.\ref{tab4}, we have the following conclusions for this example.
\begin{itemize}
	\item For AWGN, the TDS-\uppercase\expandafter{\romannumeral1} algorithm ranks first in all metrics, followed by the TDS algorithm. The TDS algorithm is second only to the Gaussian filtering algorithm in one metric.
	\item For SPN, the median filtering algorithm ranks first in all metrics, followed by the TDS-\uppercase\expandafter{\romannumeral1} algorithm. The TDS algorithm is inferior to the TDS-\uppercase\expandafter{\romannumeral1} algorithm, and only one metric is inferior to the Gaussian filtering algorithm. Here, it is necessary to point out that the SPN is approximately equal in amplitude but randomly distributed in different locations. There are both noise-contaminated points and clean points in the image. However, according to the principle of the TDS algorithm, all data points are involved in the calculation of the TDS algorithm, so the TDS algorithm is not suitable for filtering out SPN. In section \ref{sec 5.1.5}, the reason why the TDS algorithm has excellent performance is that SPN does not change the trend of the original data to a large extent. In fact, the combination of the TDS algorithm and median filtering algorithm can further improve the filtering effect of SPN. That is, for the image contaminated by SPN, the decomposition of the contaminated image is first performed by the TDS algorithm, and then processed by the median filtering algorithm, we no longer provide examples of this.
	\item For Poisson noise, the TDS-\uppercase\expandafter{\romannumeral1} algorithm ranks first in one metric, and the remaining metrics are ranked second. Additionally, all the metrics of the TDS algorithm are second to TDS-\uppercase\expandafter{\romannumeral1} and rank third.
\end{itemize}

In sections \ref{sec 5.1} and \ref{sec 5.2}, we present numerical simulation cases and image filtering applications of the TDS algorithm and the modified TDS algorithm, and its effectiveness and advantages are sufficient demonstrated. In fact, filtering is only one of their applications. In section \ref{sec 5.3}, we give the application of TDS algorithm in image smoothing and sharpening.

\begin{table*}[htbp]
	\setlength\tabcolsep{3.3pt}
	\centering
	\caption{Comparison of image filtering quality assessment metrics of different algorithms.}
	\label{tab4}
    \resizebox{\textwidth}{!}{
	\begin{tabular}{cccccccccccccc}
		\toprule
		\multirow{2}[4]{*}{} & \multirow{2}[4]{*}{} & \multicolumn{3}{c}{AWGN} &   & \multicolumn{3}{c}{SPN} &   & \multicolumn{3}{c}{Poisson noise} & \multirow{2}[4]{*}{Time $ \downarrow $} \\
		\cmidrule{3-13}      &   & MSE $ \downarrow $ & PSNR $ \uparrow $ & SSIM $ \uparrow $ &   & MSE $ \downarrow $ & PSNR $ \uparrow $ & SSIM $ \uparrow $ &   & MSE $ \downarrow $ & PSNR $ \uparrow $ & SSIM $ \uparrow $ &  \\
		\midrule
		\multirow{10}[2]{*}{TDS} & 2 & 8.7153e-04 & 30.5972  & 0.7842  & 2  & 7.0844e-04 & 31.4970  & 0.8466  & 0.2  & 3.3798e-04 & 34.7111  & 0.8700  & \multirow{10}[2]{*}{$ \textless\textbf{0.01} \mathrm{s} $} \\
		& 4 & \textbf{8.2148e-04} & \textbf{30.8540} & 0.8304  & 3  & \textbf{7.0427e-04} & \textbf{31.5226} & 0.8597  & 0.4  & 2.9470e-04 & 35.3063  & 0.9047  &  \\
		& 6 & 8.3220e-04 & 30.7977  & 0.8474  & 4  & 7.1621e-04 & 31.4496  & 0.8667  & 0.5  & \textbf{2.9252e-04} & \textbf{35.3385} & 0.9125  &  \\
		& 8 & 8.5542e-04 & 30.6782  & 0.8556  & 5  & 7.3331e-04 & 31.3472  & 0.8709  & 0.6  & 2.9446e-04 & 35.3097  & 0.9176  &  \\
		& 10 & 8.8186e-04 & 30.5460  & 0.8599  & 6  & 7.5210e-04 & 31.2373  & 0.8734  & 1.0  & 3.1595e-04 & 35.0039  & 0.9271  &  \\
		& 12 & 9.0876e-04 & 30.4155  & 0.8622  & 7  & 7.7126e-04 & 31.1280  & 0.8749  & 1.2  & 3.2885e-04 & 34.8300  & 0.9288  &  \\
		& 14 & 9.3515e-04 & 30.2912  & 0.8634  & 8  & 7.9024e-04 & 31.0224  & 0.8758  & 1.4  & 3.4177e-04 & 34.6626  & 0.9298  &  \\
		& 16 & 9.6070e-04 & 30.1741  & 0.8639  & 9  & 8.0879e-04 & 30.9217  & 0.8762  & 1.6  & 3.5441e-04 & 34.5050  & 0.9302  &  \\
		& 18 & 9.8528e-04 & 30.0644  & \textbf{0.8640} & 10  & 8.2680e-04 & 30.8260  & \textbf{0.8764} & 1.8  & 3.6664e-04 & 34.3576  & \textbf{0.9303} &  \\
		& 20 & 1.0089e-03 & 29.9615  & 0.8638  & 11  & 8.4424e-04 & 30.7353  & 0.8763  & 2.0  & 3.7844e-04 & 34.2200  & 0.9301  &  \\
		\midrule
		\multirow{4}[4]{*}{TDS-\uppercase\expandafter{\romannumeral1}} & $ \gamma=10 $ & \multirow{2}[2]{*}{\textbf{7.9869e-04}} & \multirow{2}[2]{*}{\textbf{30.9762}} & \multirow{2}[2]{*}{0.8345} & $ \gamma=5 $ & \multirow{2}[2]{*}{\textbf{6.8562e-04}} & \multirow{2}[2]{*}{\textbf{31.6392}} & \multirow{2}[2]{*}{0.8620} & $ \gamma=0.8 $ & \multirow{2}[2]{*}{\textbf{2.8570e-04}} & \multirow{2}[2]{*}{\textbf{35.4409}} & \multirow{2}[2]{*}{0.9119} & \multirow{4}[4]{*}{$ \textless\textbf{0.01}  \mathrm{s}$} \\
		& $ \delta=2 $ &   &   &   & $ \delta=2 $ &   &   &   & $ \delta=0.3 $ &   &   &   &  \\
		\cmidrule{2-13}      & $ \gamma=33 $ & \multirow{2}[2]{*}{9.4596e-04} & \multirow{2}[2]{*}{30.2413} & \multirow{2}[2]{*}{\textbf{0.8663}} & $ \gamma=19 $ & \multirow{2}[2]{*}{8.0355e-04} & \multirow{2}[2]{*}{30.9499} & \multirow{2}[2]{*}{\textbf{0.8785}} & $ \gamma=3.0 $ & \multirow{2}[2]{*}{3.5637e-04} & \multirow{2}[2]{*}{34.4810} & \multirow{2}[2]{*}{\textbf{0.9314}} &  \\
		&  $ \delta=10 $ &   &   &   & $ \delta=6 $ &   &   &   & $ \delta=1.2 $ &   &   &   &  \\
		\midrule
		\multicolumn{1}{c}{\multirow{10}[2]{*}{\makecell{Median\\filtering}}} & $ 3\times3 $ & 1.8913e-03 & 27.2323  & 0.5431  & $ 3\times3 $ & \textbf{1.0348e-04} & \textbf{39.8515} & \textbf{0.9784} & $ 3\times3 $ & \textbf{3.4522e-04} & \textbf{34.6191} & 0.8824  & \multirow{10}[2]{*}{$ \textless0.25  \mathrm{s}$} \\
		& $ 4\times4 $ & 1.6026e-03 & 27.9517  & 0.6621  & $ 4\times4 $ & 5.6737e-04 & 32.4613  & 0.9348  & $ 4\times4 $ & 6.7743e-04 & 31.6914  & 0.8877  &  \\
		& $ 5\times5 $ & \textbf{1.1315e-03} & \textbf{29.4634} & 0.7317  & $ 5\times5 $ & 2.8676e-04 & 35.4248  & 0.9508  & $ 5\times5 $ & 4.0809e-04 & 33.8924  & \textbf{0.9107} &  \\
		& $ 6\times6 $ & 1.3803e-03 & 28.6002  & 0.7688  & $ 6\times6 $ & 7.4672e-04 & 31.2684  & 0.9134  & $ 6\times6 $ & 7.8883e-04 & 31.0302  & 0.8901  &  \\
		& $ 7\times7 $ & 1.1492e-03 & 29.3960  & 0.7930  & $ 7\times7 $ & 5.0575e-04 & 32.9606  & 0.9253  & $ 7\times7 $ & 5.9609e-04 & 32.2469  & 0.8997  &  \\
		& $ 8\times8 $ & 1.4896e-03 & 28.2693  & 0.7995  & $ 8\times8 $ & 9.6720e-04 & 30.1449  & 0.8924  & $ 8\times8 $ & 9.7505e-04 & 30.1097  & 0.8771  &  \\
		& $ 9\times9 $ & 1.3573e-03 & 28.6732  & \textbf{0.8057} & $ 9\times9 $ & 7.6685e-04 & 31.1529  & 0.9021  & $ 9\times9 $ & 8.3277e-04 & 30.7947  & 0.8825  &  \\
		& $ 10\times10 $ & 1.7160e-03 & 27.6550  & 0.8025  & $ 10\times10 $ & 1.2187e-03 & 29.1410  & 0.8732  & $ 10\times10 $ & 1.1990e-03 & 29.2116  & 0.8615  &  \\
		& $ 11\times11 $ & 1.6422e-03 & 27.8457  & 0.8020  & $ 11\times11 $ & 1.0315e-03 & 29.8652  & 0.8824  & $ 11\times11 $ & 1.0970e-03 & 29.5981  & 0.8657  &  \\
		& $ 12\times12 $ & 2.0025e-03 & 26.9842  & 0.7957  & $ 12\times12 $ & 1.4907e-03 & 28.2661  & 0.8564  & $ 12\times12 $ & 1.4510e-03 & 28.3834  & 0.8467  &  \\
		\midrule
		\multicolumn{1}{c}{\multirow{2}[2]{*}{\makecell{Mean\\filtering}}} & $ 3\times3 $ & 1.3217e-03 & 28.7887  & 0.6364  & $ 3\times3 $ & 9.7864e-04 & 30.0938  & 0.7711  & $ 3\times3 $ & \textbf{3.7880e-04} & \textbf{34.2159} & \textbf{0.9000} & \multirow{2}[2]{*}{$ \textless0.03 \mathrm{s} $} \\
		& $ 5\times5 $ & \textbf{1.0239e-03} & \textbf{29.8975} & \textbf{0.7929} & $ 5\times5 $ & \textbf{9.0928e-04} & \textbf{30.4130} & \textbf{0.8320} & $ 5\times5 $ & 6.6944e-04 & 31.7429  & 0.8965  &  \\
		\midrule
		\multicolumn{1}{c}{\multirow{12}[2]{*}{\makecell{Gaussian\\filtering}}} & 1.00  & 1.0281e-03 & 29.8795  & 0.7102  & 1.0  & 7.8367e-04 & 31.0587  & 0.8169  & 0.6  & 3.6566e-04 & 34.3693  & 0.8555  &  \\
		& 1.20  & 9.4596e-04 & 30.6075  & 0.7796  & 1.1  & 7.3252e-04 & 31.3518  & 0.8342  & 0.7  & 3.0639e-04 & 35.1373  & 0.8920  & \multirow{11}[1]{*}{$ \textless0.04 \mathrm{s} $} \\
		& 1.40  & 8.2631e-04 & 30.8286  & 0.8172  & 1.2  & 7.1318e-04 & 31.4680  & 0.8459  & 0.8  & \textbf{2.9281e-04} & \textbf{35.3342} & 0.9112  &  \\
		& 1.47  & 8.2346e-04 & 30.8436  & 0.8259  & 1.3  & \textbf{7.0893e-04} & \textbf{31.4940} & 0.8550  & 0.9  & 2.9948e-04 & 35.2363  & 0.9216  &  \\
		& 1.48  & \textbf{8.2340e-04} & \textbf{30.8439} & 0.8270  & 1.4  & 7.1461e-04 & 31.4593  & 0.8618  & 1.0  & 3.1589e-04 & 35.0046  & 0.9272  &  \\
		& 1.49  & 8.2341e-04 & 30.8438  & 0.8281  & 1.5  & 7.2654e-04 & 31.3874  & 0.8668  & 1.1  & 3.4816e-04 & 34.5822  & 0.9306  &  \\
		& 1.60  & 8.3360e-04 & 30.7904  & 0.8440  & 1.6  & 7.5788e-04 & 31.2040  & 0.8731  & 1.2  & 3.8015e-04 & 34.2005  & \textbf{0.9312} &  \\
		& 1.80  & 8.6841e-04 & 30.6127  & 0.8564  & 1.7  & 7.8419e-04 & 31.0558  & 0.8757  & 1.3  & 4.1442e-04 & 33.8256  & 0.9304  &  \\
		& 2.00  & 9.1423e-04 & 30.3894  & 0.8621  & 1.8  & 8.1270e-04 & 30.9007  & 0.8772  & 1.4  & 4.4922e-04 & 33.4754  & 0.9287  &  \\
		& 2.10  & 9.6723e-04 & 30.1447  & 0.8651  & 1.9  & 8.4223e-04 & 30.7457  & 0.8779  & 1.5  & 4.8329e-04 & 33.1579  & 0.9265  &  \\
		& 2.20  & 1.0004e-03 & 29.9982  & \textbf{0.8656} & 2.0  & 8.7187e-04 & 30.5955  & \textbf{0.8779} & 1.6  & 5.4448e-04 & 32.6402  & 0.9223  &  \\
		& 2.30  & 1.0339e-03 & 29.8553  & 0.8655  & 2.1  & 9.3615e-04 & 30.2865  & 0.8774  & 1.7  & 5.8650e-04 & 32.3173  & 0.9190  &  \\
		\midrule
		\multicolumn{1}{c}{\multirow{10}[2]{*}{\makecell{2D\\adaptive\\ Wiener\\filtering}}} & $ 3\times3 $ & 1.8007e-03 & 27.4455  & 0.5626  & $ 3\times3 $ & 4.9728e-03 & 23.0340  & 0.6129  & $ 3\times3 $ & 3.0375e-04 & 35.1749  & 0.8996  & \multirow{10}[2]{*}{$ \textless0.03 \mathrm{s} $} \\
		& $ 4\times4 $ & 1.2627e-03 & 28.9869  & 0.6848  & $ 4\times4 $ & 4.1956e-03 & 23.7721  & 0.6299  & $ 4\times4 $ & 2.9067e-04 & 35.3660  & 0.9192  &  \\
		& $ 5\times5 $ & 9.7788e-04 & 30.0972  & 0.7670  & $ 5\times5 $ & 3.3367e-03 & 24.7668  & 0.6591  & $ 5\times5 $ & \textbf{2.7248e-04} & \textbf{35.6467} & \textbf{0.9272} &  \\
		& $ 6\times6 $ & 9.5080e-04 & 30.2191  & 0.8064  & $ 6\times6 $ & 2.6639e-03 & 25.7448  & 0.6832  & $ 6\times6 $ & 3.2296e-04 & 34.9085  & 0.9214  &  \\
		& $ 7\times7 $ & \textbf{9.1864e-04} & \textbf{30.3686} & 0.8292  & $ 7\times7 $ & 2.0838e-03 & 26.8114  & 0.7210  & $ 7\times7 $ & 3.4811e-04 & 34.5828  & 0.9158  &  \\
		& $ 8\times8 $ & 9.9793e-04 & 30.0090  & 0.8336  & $ 8\times8 $ & 1.7795e-03 & 27.4971  & 0.7616  & $ 8\times8 $ & 4.1632e-04 & 33.8057  & 0.9064  &  \\
		& $ 9\times9 $ & 1.0323e-03 & 29.8619  & \textbf{0.8351} & $ 9\times9 $ & 1.6067e-03 & 27.9406  & 0.7769  & $ 9\times9 $ & 4.5959e-04 & 33.3763  & 0.8987  &  \\
		& $ 10\times10 $ & 1.1412e-03 & 29.4263  & 0.8298  & $ 10\times10 $ & 1.5122e-03 & 28.2039  & 0.7827  & $ 10\times10 $ & 5.3920e-04 & 32.6825  & 0.8896  &  \\
		& $ 11\times11 $ & 1.2004e-03 & 29.2068  & 0.8256  & $ 11\times11 $ & \textbf{1.4161e-03} & \textbf{28.4892} & 0.7975  & $ 11\times11 $ & 5.9483e-04 & 32.2561  & 0.8822  &  \\
		& $ 12\times12 $ & 1.3204e-03 & 28.7928  & 0.8187  & $ 12\times12 $ & 1.4258e-03 & 28.4593  & \textbf{0.8007} & $ 12\times12 $ & 6.8332e-04 & 31.6537  & 0.8741  &  \\
		\midrule
		\multicolumn{1}{c}{\multirow{0.8}[1]{*}{\makecell{TDS ranking}}} & $   $ & 2/6 & 2/6  & 3/6  & $   $ & 3/6 & 3/6  & 4/6  &   & 3/6 & 3/6  & 3/6  & 1/6  \\
		\midrule
		\multicolumn{1}{c}{\multirow{0.8}[1]{*}{\makecell{TDS-\uppercase\expandafter{\romannumeral1} ranking}}} & $   $ & 1/6 & 1/6  & 1/6  & $   $ & 2/6 & 2/6  & 2/6  &   & 2/6 & 2/6  & 1/6  & 1/6  \\
		\bottomrule
	\end{tabular}%
    }
\end{table*}%
\vspace{-0.4cm}

\subsection{Image Processing}\label{sec 5.3}
\subsubsection{Image Smoothing}

The grayscale image's pixel data can be represented as a matrix, whose rows and columns correspond to the height and width of the image respectively, and the value of the matrix element is the gray value of the pixel. We regard the matrix corresponding to the grayscale image as the original data \textit{\textbf{Z}}, then we can calculate the trend matrix \textit{\textbf{G}} of the original data \textit{\textbf{Z}} based on the TDS algorithm. In this case, \textit{\textbf{G}} is the matrix of the grayscale image after smoothing. The following we present a specific case.
\begin{figure}[!ht]
	\centering
	\subfloat[Original image.]{\includegraphics[width=.47\columnwidth]{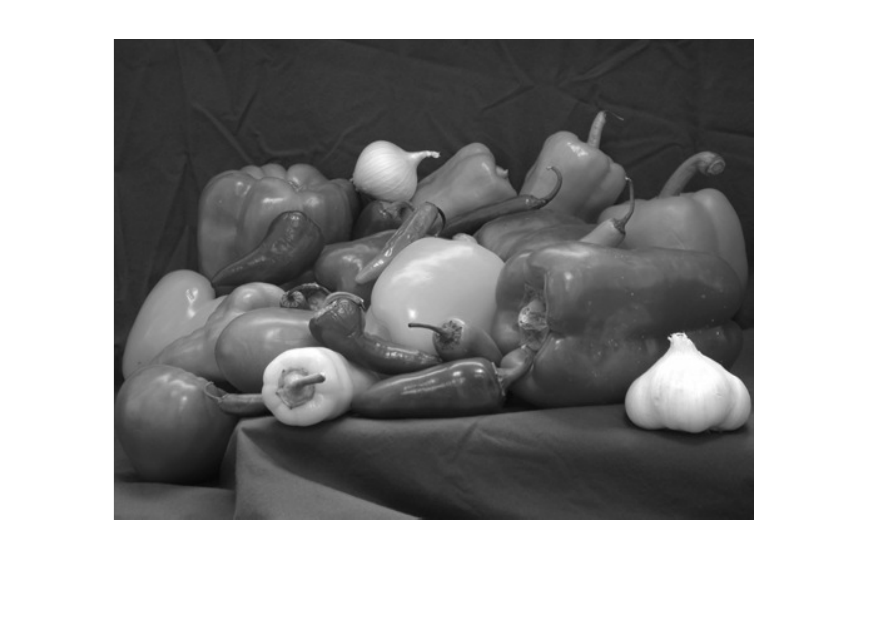}}\hspace{5pt}
	\subfloat[$\lambda=0.2$.]{\includegraphics[width=.47\columnwidth]{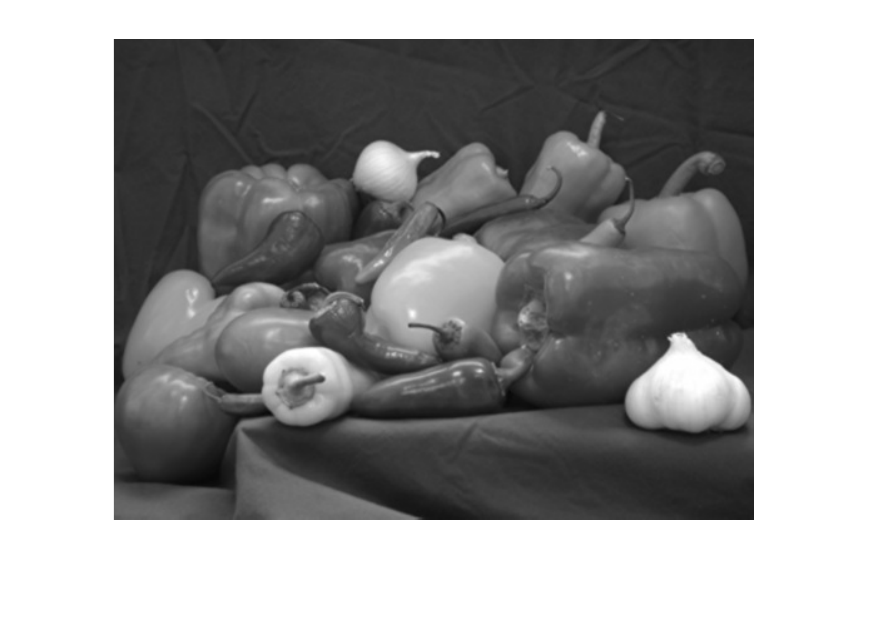}}\\
	\subfloat[$\lambda=0.6$.]{\includegraphics[width=.47\columnwidth]{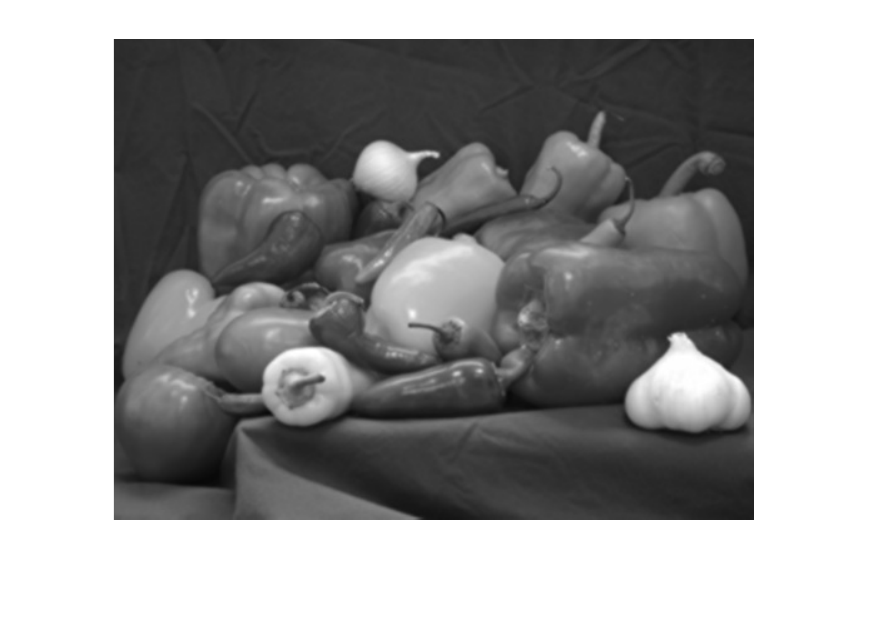}}\hspace{5pt}
	\subfloat[$\lambda=1.0$.]{\includegraphics[width=.47\columnwidth]{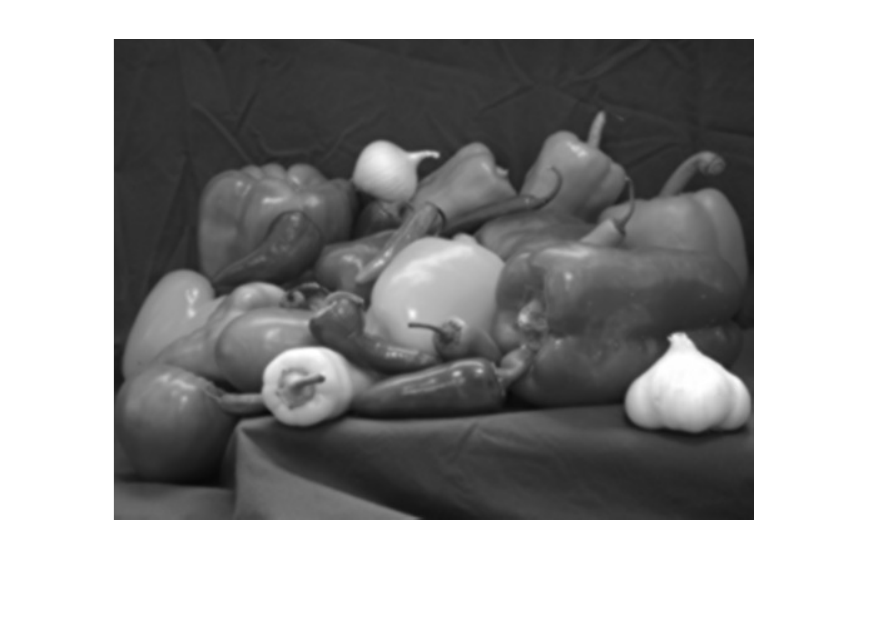}}\\
	\subfloat[$\lambda=1.4$.]{\includegraphics[width=.47\columnwidth]{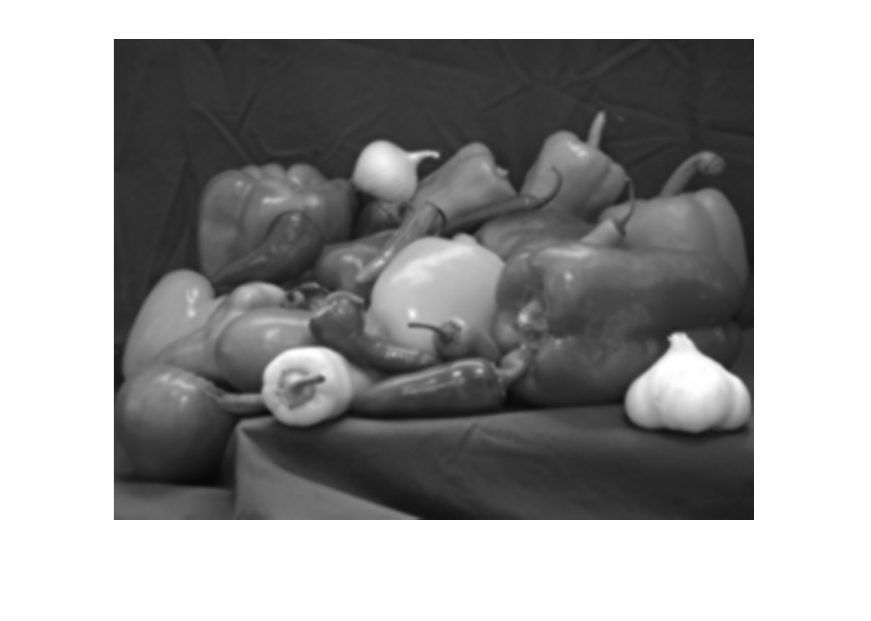}}\hspace{5pt}
	\subfloat[$\lambda=1.8$.]{\includegraphics[width=.47\columnwidth]{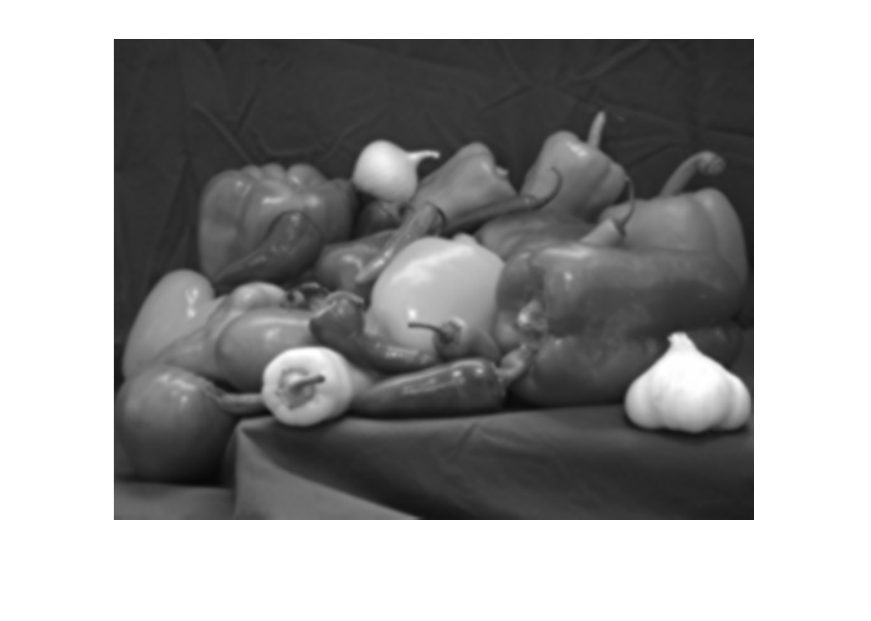}}
	\caption{Image smoothing.}\label{Fig12}
\end{figure}

As shown in Fig.\ref{Fig12}, the smoothing effect is gradually strengthened with the increase of the smoothing parameters $ \lambda $, and the image smoothing is better realized by the TDS algorithm.

\subsubsection{Image Sharpening}

Considering another case, we hope to compensate for the contour of the image by sharpening the image when the image clarity is poor, enhancing the edge of the image and the part of the gray jump, so that the image becomes clear. We regard the matrix corresponding to the grayscale image as the matrix \textit{\textbf{G}}, and we solve the original matrix \textit{\textbf{Z}} of the matrix \textit{\textbf{G}} based on the TDS algorithm. In this case, \textit{\textbf{Z}} is the matrix of the grayscale image after sharpening. Likewise, we present a specific case.
\begin{figure}[!ht]
	\centering
	\subfloat[Original image.]{\includegraphics[width=.47\columnwidth]{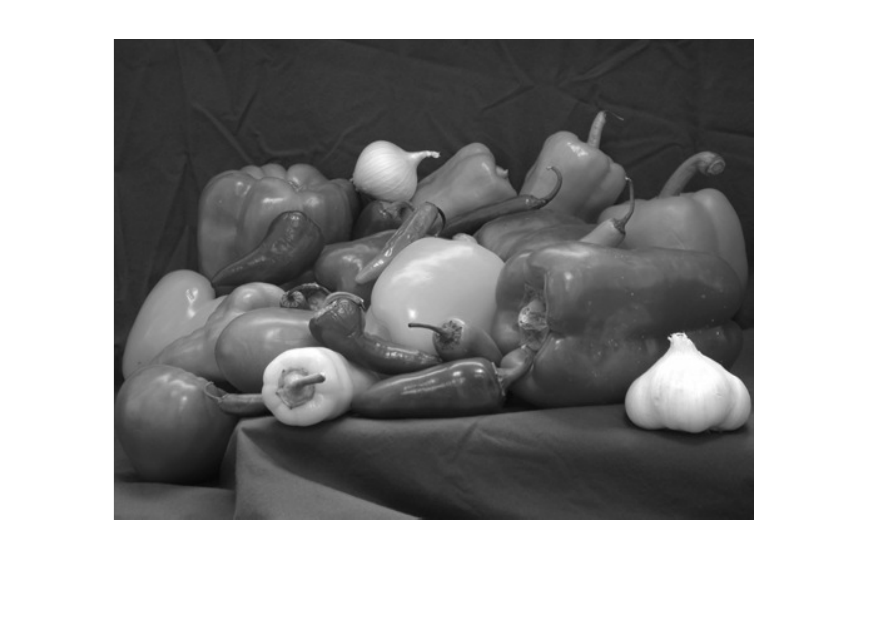}}\hspace{5pt}
	\subfloat[$\lambda=0.2$.]{\includegraphics[width=.47\columnwidth]{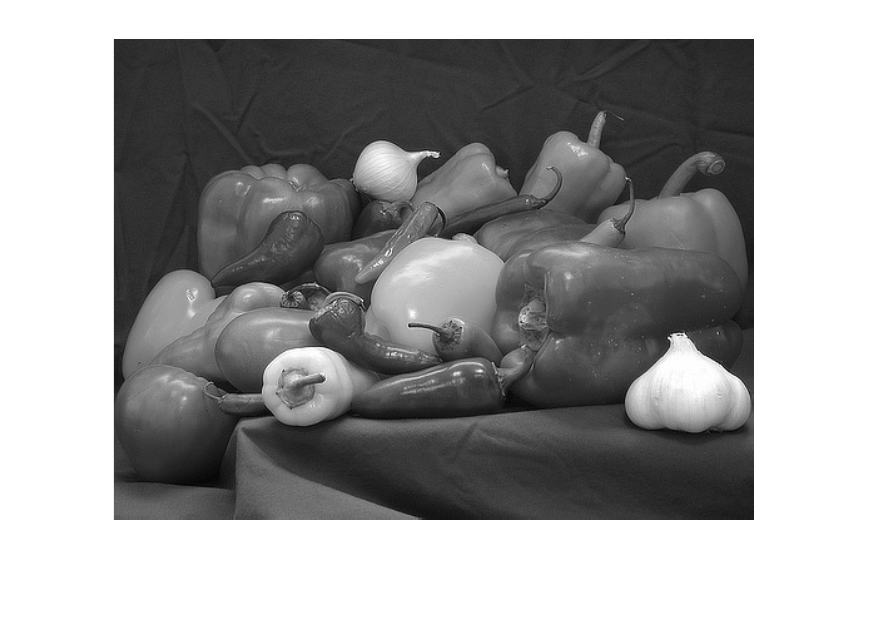}}\\
	\subfloat[$\lambda=0.4$.]{\includegraphics[width=.47\columnwidth]{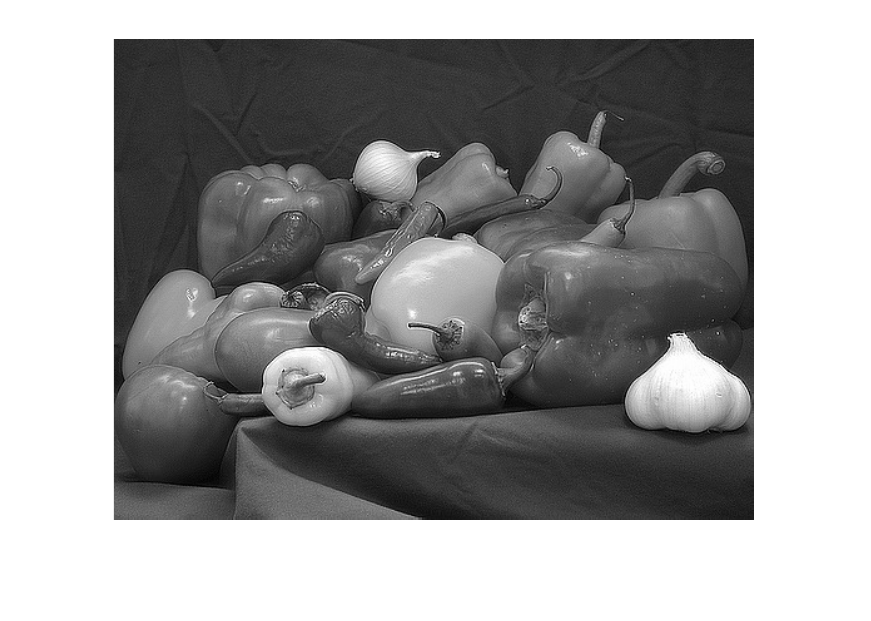}}\hspace{5pt}
	\subfloat[$\lambda=0.6$.]{\includegraphics[width=.47\columnwidth]{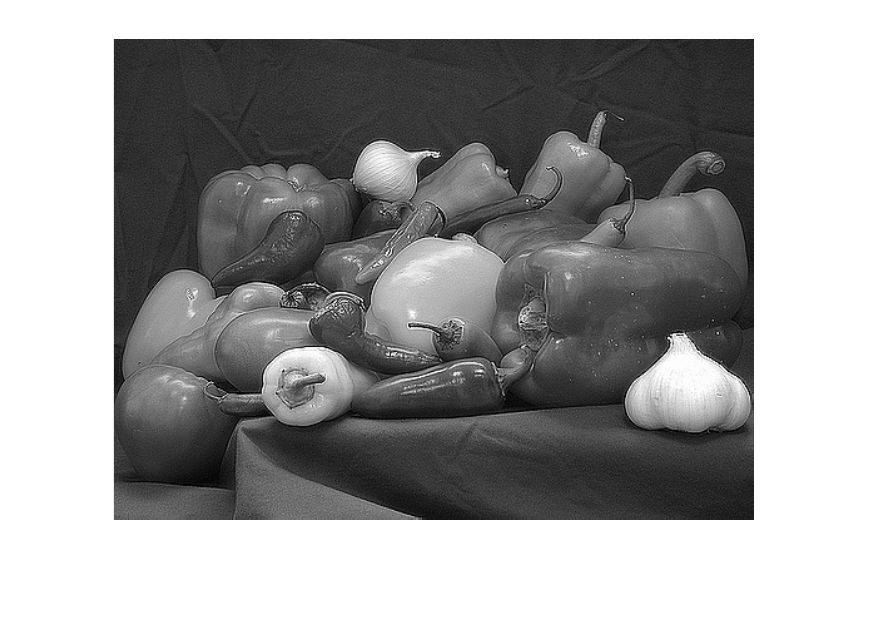}}\\
	\subfloat[$\lambda=0.8$.]{\includegraphics[width=.47\columnwidth]{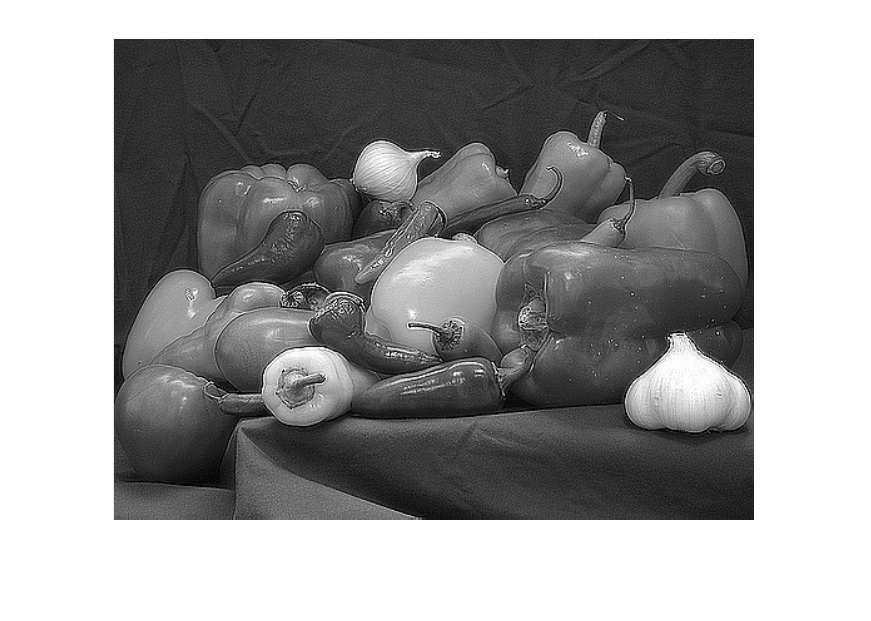}}\hspace{5pt}
	\subfloat[$\lambda=1.0$.]{\includegraphics[width=.47\columnwidth]{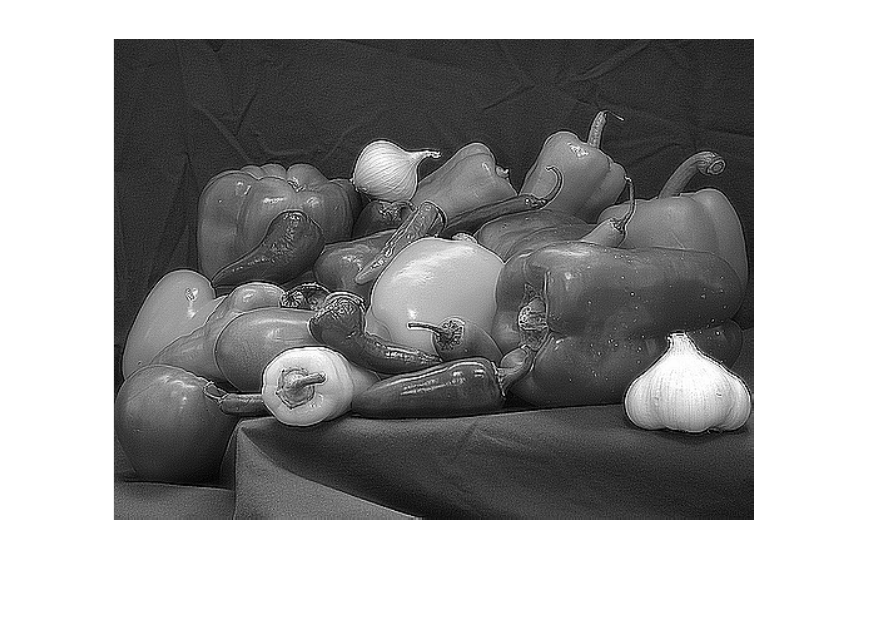}}
	\caption{Image sharpening.}\label{Fig13}
\end{figure}

As shown in Fig.\ref{Fig13}, the sharpening effect is gradually strengthened with the increase of the smoothing parameters $ \lambda $, and the image sharpening is better realized by the TDS algorithm.

\section{Conclusion} \label{sec 6}

In this paper, we propose a novel TDS algorithm for the smoothing and filtering of two-dimensional sequences. Distinguished from conventional filtering methods, the TDS algorithm takes into account the overall characteristics of the two-dimensional sequence, eliminating the need to assume a specific type of noise distribution. It is suitable for various types of noise reduction, including both additive and multiplicative noise, demonstrating excellent performance compared to traditional filtering algorithms across various challenging scenarios. Additionally, we extend the TDS algorithm by proposing three modified versions: TDS-\uppercase\expandafter{\romannumeral1}, TDS-\uppercase\expandafter{\romannumeral2}, and TDS-\uppercase\expandafter{\romannumeral3} algorithms. The TDS algorithm also exhibits better robustness. Numerical simulation results highlight its superior smoothing effect in terms of MSE, PSNR, and SSIM metrics under various noise types compared to traditional filtering algorithms. Furthermore, the proposed TDS algorithm shows promising applications in future research areas such as signal processing, pattern recognition, wireless communication, computer vision, and image processing. It is worth noting that the correlation matrices in the matrix equation for solving the trend sequence are sparse matrices, significantly reducing the complexity of the TDS algorithm. Future work will concentrate on estimating the parameter in the TDS algorithm and numerically solving the correlation matrix equation, a crucial step for its practical application.

\section*{Appendix}

\subsection*{Appendix A. Proof of Equation \eqref{14}}\label{Appendix A}
\vspace{0.5em}

We take partial derivatives of the loss function $ M $ with respect to $ g_{i,j} $, then

\begin{align}
	\frac{\partial M}{\partial g_{i,j}}=\frac{\partial R}{\partial g_{i,j}}+\lambda\frac{\partial P}{\partial g_{i,j}}+\lambda\frac{\partial Q}{\partial g_{i,j}}
\end{align}

\noindent where $ R $ takes partial derivatives of $ g_{i,j} $, then

\begin{align}
	\frac{\partial R}{\partial g_{i,j}}=\frac{\partial r\left(i,j\right)}{\partial g_{i,j}}=-2\left(z_{i,j}-g_{i,j}\right)
\end{align}

\noindent where $ P $ and $ Q $ takes the partial derivatives of $ g_{i,j} $ respectively, then

\begin{enumerate}
	\item 	When $ m=3 $, $ n=3 $
	
	\begin{align}
		\begin{split}
			\frac{\partial P}{\partial g_{i,1}}&=\frac{\partial p\left(i,3\right)}{\partial g_{i,1}}=2\left(g_{i,1}-2g_{i,2}+g_{i,3}\right)
			\\\frac{\partial P}{\partial g_{i,2}}&=\frac{\partial p\left(i,3\right)}{\partial g_{i,2}}=2\left(-2g_{i,1}+4g_{i,2}-2g_{i,3}\right)
			\\\frac{\partial P}{\partial g_{i,3}}&=\frac{\partial p\left(i,3\right)}{\partial g_{i,3}}=2\left(g_{i,1}-2g_{i,2}+g_{i,3}\right)
		\end{split}
	\end{align}
	\begin{align}
		\begin{split}
			\frac{\partial Q}{\partial g_{1,j}}=&\frac{\partial q\left(3,j\right)}{\partial g_{1,j}}=2\left(g_{1,j}-2g_{2,j}+g_{3,j}\right)
			\\\frac{\partial Q}{\partial g_{2,j}}=&\frac{\partial q\left(3,j\right)}{\partial g_{2,j}}=2\left(-2g_{1,j}+4g_{2,j}-2g_{3,j}\right)
			\\\frac{\partial Q}{\partial g_{3,j}}=&\frac{\partial q\left(3,j\right)}{\partial g_{3,j}}=2\left(g_{1,j}-2g_{2,j}+g_{3,j}\right)
		\end{split}
	\end{align}

	\item 	When $ m\geq4 $, $ n\geq4 $
	
	\begin{align}
		\begin{split}
			\frac{\partial P}{\partial g_{i,1}}=&\frac{\partial p\left(i,3\right)}{\partial g_{i,1}}=2\left(g_{i,1}-2g_{i,2}+g_{i,3}\right)
			\\\frac{\partial P}{\partial g_{i,2}}=&\frac{\partial p\left(i,3\right)}{\partial g_{i,2}}+\frac{\partial p\left(i,4\right)}{\partial g_{i,2}}\\
			=&2\left(-2g_{i,1}+5g_{i,2}-4g_{i,3}+g_{i,4}\right)		
			\\\frac{\partial P}{\partial g_{i,j}}=&\frac{\partial p\left(i,j\right)}{\partial g_{i,j}}+\frac{\partial p\left(i,j+1\right)}{\partial g_{i,j}}+\frac{\partial p\left(i,j+2\right)}{\partial g_{i,j}}
			\\=&2\Big(g_{i,j-2}-4g_{i,j-1}+6g_{i,j}\\-&4g_{i,j+1}+g_{i,j+2}\Big)
			\\\frac{\partial P}{\partial g_{i,n-1}}=&\frac{\partial p\left(i,n-1\right)}{\partial g_{i,n-1}}+\frac{\partial p\left(i,n\right)}{\partial g_{i,n-1}}
			\\=&2\left(g_{i,n-3}-4g_{i,n-2}+5g_{i,n-1}-2g_{i,n}\right)
			\\\frac{\partial P}{\partial g_{i,n}}=&\frac{\partial p\left(i,n\right)}{\partial g_{i,n}}=2\left(g_{i,n-2}-2g_{i,n-1}+g_{i,n}\right)
		\end{split}
	\end{align}
	\begin{align}
		\begin{split}
			\frac{\partial Q}{\partial g_{1,j}}=&\frac{\partial q\left(3,j\right)}{\partial g_{1,j}}=2\left(g_{1,j}-2g_{2,j}+g_{3,j}\right)
			\\\frac{\partial Q}{\partial g_{2,j}}=&\frac{\partial q\left(3,j\right)}{\partial g_{2,j}}+\frac{\partial q\left(4,j\right)}{\partial g_{2,j}}
			\\=&2\left(-2g_{1,j}+5g_{2,j}-4g_{3,j}+g_{4,j}\right)
			\\\frac{\partial Q}{\partial g_{i,j}}=&\frac{\partial q\left(i,j\right)}{\partial g_{i,j}}+\frac{\partial q\left(i+1,j\right)}{\partial g_{i,j}}+\frac{\partial q\left(i+2,j\right)}{\partial g_{i,j}}
			\\=&2\Big(g_{i-2,j}-4g_{i-1,j}+6g_{i,j}\\-&4g_{i+1,j}+g_{i+2,j}\Big)	
			\\\frac{\partial Q}{\partial g_{m-1,j}}=&\frac{\partial q\left(m-1,j\right)}{\partial g_{m-1,j}}+\frac{\partial q\left(m,j\right)}{\partial g_{m-1,j}}
			\\=&2\left(g_{m-3,j}-4g_{m-2,j}+5g_{m-1,j}-2g_{m,j}\right)	
			\\\frac{\partial Q}{\partial g_{m,j}}=&\frac{\partial q\left(m,j\right)}{\partial g_{m,j}}=2\left(g_{m-2,j}-2g_{m-1,j}+g_{m,j}\right)
		\end{split}
	\end{align}
\end{enumerate}

Normally (i.e., $  m\geq5 $ and $n\geq5 $.), the solution of $ \partial M/\partial g_{i,j} $ can be divided into the following large five categories (\uppercase\expandafter{\romannumeral1}, \uppercase\expandafter{\romannumeral2}, \uppercase\expandafter{\romannumeral3}, \uppercase\expandafter{\romannumeral4}, \uppercase\expandafter{\romannumeral5}. Reference Fig.3(b), Fig.4(a) and Fig.4(b).), including $ 25 $ subcategories. Under different subcategories, the number of partial derivatives is shown in Table.\ref{tab5}.

\begin{table*}[htbp]
	\centering
	\caption{The number of partial derivatives.}
	\label{tab5}
	\begin{tabular}{cccccc}
		\toprule
		&\uppercase\expandafter{\romannumeral1} ($ i=1 $)&	\uppercase\expandafter{\romannumeral2} ($ i=2 $)&	\uppercase\expandafter{\romannumeral3} ($ 3\le i\le m-2 $)&	\uppercase\expandafter{\romannumeral4} ($ i=m-1 $)&	\uppercase\expandafter{\romannumeral5} ($ i=m $)\\
		\midrule
		$ j=1 $	&	1	&	1	&	$ m-4 $	&	1	&	1\\
		$ j=2 $	&	1	&	1	&	$ m-4 $	&	1	&	1\\
		$ 3\le j\le n-2 $	&	$ n-4 $	&	$ n-4 $	&	$ \left(n-4\right)\left(m-4\right) $	&	$ n-4 $	&	$ n-4 $\\
		$ j=n-1 $	&	1	&	1	&	$ m-4 $	&	1	&	1\\
		$ j=n $	&	1	&	1	&	$ m-4 $	&	1	&	1\\
		\bottomrule
	\end{tabular}
\end{table*}

At the same time, in order to obtain the extreme value of $ M $ and set $ \partial M/\partial g_{i,j}=0 $, then

\begin{enumerate}[I]
	
	\item When $ i=1 $
	
	\begin{itemize}
		\item 	When $ i=1 $, $ j=1 $, $ \partial M/\partial g_{1,1}$  is
		\begin{align*}
			\frac{\partial M}{\partial g_{1,1}}=\frac{\partial r\left(1,1\right)}{\partial g_{1,1}}+\lambda\frac{\partial p\left(1,3\right)}{\partial g_{1,1}}+\lambda\frac{\partial q\left(3,1\right)}{\partial g_{1,1}}
		\end{align*}
		
		Let $ \partial M/\partial g_{1,1}=0 $, then 
		\begin{subequations}
			\begin{align}
				\begin{split}\label{73a}
					g_{1,1}+&\lambda\left(g_{1,1}-2g_{1,2}+g_{1,3}\right)
					\\+&\lambda\left(g_{1,1}-2g_{2,1}+g_{3,1}\right)=z_{1,1}
				\end{split}
			\end{align}
			
			\item 	When $ i=1 $, $ j=2 $, $ \partial M/\partial g_{1,2} $ is
			\begin{align*}
				\frac{\partial M}{\partial g_{1,2}}=&\frac{\partial r\left(1,2\right)}{\partial g_{1,2}}+\lambda\left(\frac{\partial p\left(1,3\right)}{\partial g_{1,2}}+\frac{\partial p\left(1,4\right)}{\partial g_{1,2}}\right)\\+
				&\lambda\frac{\partial q\left(3,2\right)}{\partial g_{1,2}}
			\end{align*}
			
			Let $ \partial M/\partial g_{1,2}=0 $, then
			\begin{align}
				\begin{split}
					g_{1,2}+&\lambda\left(-2g_{1,1}+5g_{1,2}-4g_{1,3}+g_{1,4}\right)\\
					+&\lambda\left(g_{1,2}-2g_{2,2}+g_{3,2}\right)=z_{1,2}
				\end{split}
			\end{align}
			
			\item 	When $ i=1 $, $ 3\le j\le n-2 $, $ \partial M/\partial g_{1,j} $ is
			\begin{align*}
				\frac{\partial M}{\partial g_{1,j}}=&\frac{\partial r\left(1,j\right)}{\partial g_{1,j}}\\&+\lambda\Big(\frac{\partial p\left(1,j\right)}{\partial g_{1,j}}+\frac{\partial p\left(1,j+1\right)}{\partial g_{1,j}}\\&+\frac{\partial p\left(1,j+2\right)}{\partial g_{1,j}}\Big)+\lambda\frac{\partial q\left(3,j\right)}{\partial g_{1,j}}
			\end{align*}
			
			Let $ \partial M/\partial g_{1,j}=0 $, then
			\begin{align}
				\begin{split}
					g_{1,j}+&\lambda\left(g_{1,j-2}-4g_{1,j-1}+6g_{1,j}-4g_{1,j+1}+g_{1,j+2}\right)\\+&\lambda\left(g_{1,j}-2g_{2,j}+g_{3,j}\right)=z_{1,j}
				\end{split}
			\end{align}
			
			\item 	When $ i=1 $, $ j=n-1 $, $ \partial M/\partial g_{1,n-1} $ is
			\begin{align*}
				\begin{split}
					\frac{\partial M}{\partial g_{1,n-1}}=&\frac{\partial r\left(1,n-1\right)}{\partial g_{1,n-1}}\\+&\lambda\left(\frac{\partial p\left(1,n-1\right)}{\partial g_{1,n-1}}+\frac{\partial p\left(1,n\right)}{\partial g_{1,n-1}}\right)\\+&\lambda\frac{\partial q\left(3,n-1\right)}{\partial g_{1,n-1}}
				\end{split}
			\end{align*}
			
			Let $ \partial M/\partial g_{1,n-1}=0 $, then
			\begin{align}
				\begin{split}
					g_{1,n-1}+&\lambda\left(g_{1,n-3}-4g_{1,n-2}+5g_{1,n-1}-2g_{1,n}\right)\\+&\lambda\left(g_{1,n-1}-2g_{2,n-1}+g_{3,n-1}\right)=z_{1,n-1}
				\end{split}
			\end{align}
			
			\item 	When $ i=1 $, $ j=n $, $ \partial M/\partial g_{1,n} $ is
			\begin{align*}
				\frac{\partial M}{\partial g_{1,n}}=\frac{\partial r\left(1,n\right)}{\partial g_{1,n}}+\lambda\frac{\partial p\left(1,n\right)}{\partial g_{1,n}}+\lambda\frac{\partial q\left(3,n\right)}{\partial g_{1,n}}
			\end{align*}
			
			Let $ \partial M/\ \partial g_{1,n}=0 $, then
			\begin{align}
				\begin{split}
					g_{1,n}+&\lambda\left(g_{1,n-2}-2g_{1,n-1}+g_{1,n}\right)\\+&\lambda\left(g_{1,n}-2g_{2,n}+g_{3,n}\right)=z_{1,n}
				\end{split}
			\end{align}
		\end{subequations}
	\end{itemize}
	\item When $ i=2 $
	\begin{itemize}
		\item 	When $ i=2 $, $ j=1 $, $ \partial M/\partial g_{2,1} $ is
		\begin{align*}
			\begin{split}
				\frac{\partial M}{\partial g_{2,1}}=&\frac{\partial r\left(2,1\right)}{\partial g_{2,1}}+\lambda\frac{\partial p\left(2,3\right)}{\partial g_{2,1}}\\+&\lambda\left(\frac{\partial q\left(3,1\right)}{\partial g_{2,1}}+\frac{\partial q\left(4,1\right)}{\partial g_{2,1}}\right)
			\end{split}
		\end{align*}
		\begin{subequations}
			Let $ \partial M/\partial g_{2,1}=0 $, then
			\begin{align}
				\begin{split}
					g_{2,1}+&\lambda\left(g_{2,1}-2g_{2,2}+g_{2,3}\right)\\+&\lambda\left(-2g_{1,1}+5g_{2,1}-4g_{3,1}+g_{4,1}\right)=z_{2,1}
				\end{split}
			\end{align}
			
			\item 	When $ i=2 $, $ j=2 $, $ \partial M/\partial g_{2,2} $ is
			\begin{align*}
				\frac{\partial M}{\partial g_{2,2}}=&\frac{\partial r\left(2,2\right)}{\partial g_{2,2}}\\+&\lambda\left(\frac{\partial p\left(2,3\right)}{\partial g_{2,2}}+\frac{\partial p\left(2,4\right)}{\partial g_{2,2}}\right)\\+&\lambda\left(\frac{\partial q\left(3,2\right)}{\partial g_{2,2}}+\frac{\partial q\left(4,2\right)}{\partial g_{2,2}}\right)
			\end{align*}
			
			Let $ \partial M/\partial g_{2,2}=0 $, then
			\begin{align}
				\begin{split}
					g_{2,2}+&\lambda\left(-2g_{2,1}+5g_{2,2}-4g_{2,3}+g_{2,4}\right)\\+&\lambda\left(-2g_{1,2}+5g_{2,2}-4g_{3,2}+g_{4,2}\right)=z_{2,2}
				\end{split}
			\end{align}
			
			\item	When $ i=2 $, $ 3\le j\le n-2 $, $ \partial M/\partial g_{2,j} $ is
			\begin{align*}
				\begin{split}
					\frac{\partial M}{\partial g_{2,j}}=&\frac{\partial r\left(2,j\right)}{\partial g_{2,j}}\\+&\lambda\Big(\frac{\partial p\left(2,j\right)}{\partial g_{2,j}}+\frac{\partial p\left(2,j+1\right)}{\partial g_{2,j}}+\frac{\partial p\left(2,j+2\right)}{\partial g_{2,j}}\Big)\\+&\lambda\left(\frac{\partial q\left(3,j\right)}{\partial g_{2,j}}+\frac{\partial q\left(4,j\right)}{\partial g_{2,j}}\right)
				\end{split}
			\end{align*}
			
			Let $ \partial M/\partial g_{2,j}=0 $, then
			\begin{align}
				\begin{split}
					g_{2,j}+&\lambda\left(g_{2,j-2}-4g_{2,j-1}+6g_{2,j}-4g_{2,j+1}+g_{2,j+2}\right)\\+&\lambda\left(-2g_{1,j}+5g_{2,j}-4g_{3,j}+g_{4,j}\right)=z_{2,j}
				\end{split}
			\end{align}
			
			\item 	When $ i=2 $, $ j=n-1 $, $ \partial M/\partial g_{2,n-1} $ is
			\begin{align*}
				\begin{split}
					\frac{\partial M}{\partial g_{2,n-1}}=&\frac{\partial r\left(2,n-1\right)}{\partial g_{2,n-1}}\\+&\lambda\left(\frac{\partial p\left(2,n-1\right)}{\partial g_{2,n-1}}+\frac{\partial p\left(2,n\right)}{\partial g_{2,n-1}}\right)\\+&\lambda\left(\frac{\partial q\left(3,n-1\right)}{\partial g_{2,n-1}}+\frac{\partial q\left(4,n-1\right)}{\partial g_{2,n-1}}\right)
				\end{split}
			\end{align*}
			
			Let $ \partial M/\partial g_{2,n-1}=0 $, then
			\begin{align}
				\begin{split}
					g_{2,n-1}+&\lambda\left(g_{2,n-3}-4g_{2,n-2}+5g_{2,n-1}-2g_{2,n}\right)\\+&\lambda\left(-2g_{1,n-1}+5g_{2,n-1}-4g_{3,n-1}+g_{4,n-1}\right)\\=&z_{2,n-1}
				\end{split}
			\end{align}
			
			\item 	When $ i=2 $, $ j=n $, $ \partial M/\partial g_{2,n} $ is 
			\begin{align*}
				\begin{split}
					\frac{\partial M}{\partial g_{2,n}}=&\frac{\partial r\left(2,n\right)}{\partial g_{2,n}}+\lambda\frac{\partial p\left(2,n\right)}{\partial g_{2,n}}\\+&\lambda\left(\frac{\partial q\left(3,n\right)}{\partial g_{2,n}}+\frac{\partial q\left(4,n\right)}{\partial g_{2,n}}\right)
				\end{split}
			\end{align*}
			
			Let $ \partial M/\partial g_{2,n}=0 $, then
			\begin{align}
				\begin{split}
					g_{2,n}+&\lambda\left(g_{2,n-2}-2g_{2,n-1}+g_{2,n}\right)\\+&\lambda\left(-2g_{1,n}+5g_{2,n}-4g_{3,n}+g_{4,n}\right)=z_{2,n}
				\end{split}
			\end{align}
		\end{subequations}
	\end{itemize}
	\item When $ 3\le i\le m-2 $
	\begin{itemize}
		\item 	When $ 3\le i\le m-2 $, $ j=1 $, $ \partial M/\partial g_{i,1 } $ is
		\begin{align*}
			\begin{split}
				\frac{\partial M}{\partial g_{i,1}}=&\frac{\partial r\left(i,1\right)}{\partial g_{i,1}}+\lambda\frac{\partial p\left(i,3\right)}{\partial g_{i,1}}\\+&\lambda\left(\frac{\partial q\left(i,1\right)}{\partial g_{i,1}}+\frac{\partial q\left(i+1,1\right)}{\partial g_{i,1}}+\frac{\partial q\left(i+2,1\right)}{\partial g_{i,1}}\right)
			\end{split}
		\end{align*}
		
		Let $ \partial M/\partial g_{i,1}=0 $, then
		\begin{subequations}
			\begin{align}
				\begin{split}
					g_{i,1}+&\lambda\left(g_{i,1}-2g_{i,2}+g_{i,3}\right)\\+&\lambda\left(g_{i-2,j}-4g_{i-1,j}+6g_{i,j}-4g_{i+1,j}+g_{i+2,j}\right)\\=&z_{i,1}
				\end{split}
			\end{align}
			
			\item 	When $ 3\le i\le m-2 $, $ j=2 $, $ \partial M/\partial g_{i,2} $ is
			\begin{align*}
				\begin{split}
					\frac{\partial M}{\partial g_{i,2}}=&\frac{\partial r\left(i,2\right)}{\partial g_{i,2}}+\lambda\left(\frac{\partial p\left(i,3\right)}{\partial g_{i,2}}+\frac{\partial p\left(i,4\right)}{\partial g_{i,2}}\right)\\+&\lambda\left(\frac{\partial q\left(i,2\right)}{\partial g_{i,2}}+\frac{\partial q\left(i+1,2\right)}{\partial g_{i,2}}+\frac{\partial q\left(i+2,2\right)}{\partial g_{i,2}}\right)
				\end{split}
			\end{align*}
			
			Let $ \partial M/\partial g_{i,2}=0 $, then
			\begin{align}
				\begin{split}
					g_{i,2}+&\lambda\left(-2g_{i,1}+5g_{i,2}-4g_{i,3}+g_{i,4}\right)\\+&\lambda\left(g_{i-2,2}-4g_{i-1,2}+6g_{i,2}-4g_{i+1,2}+g_{i+2,2}\right)\\=&z_{i,2}
				\end{split}
			\end{align}
			\item	When $ 3\le i\le m-2 $, $ 3\le j\le n-2 $, $ \partial M/\partial g_{i,j} $ is
			\begin{align*}
				\begin{split}
					\frac{\partial M}{\partial g_{i,j}}=&\frac{\partial r\left(i,j\right)}{\partial g_{i,j}}\\+&\lambda\left(\frac{\partial p\left(i,j\right)}{\partial g_{i,j}}+\frac{\partial p\left(i,j+1\right)}{\partial g_{i,j}}+\frac{\partial p\left(i,j+2\right)}{\partial g_{i,j}}\right)\\+&\lambda\left(\frac{\partial q\left(i,j\right)}{\partial g_{i,j}}+\frac{\partial q\left(i+1,j\right)}{\partial g_{i,j}}+\frac{\partial q\left(i+2,j\right)}{\partial g_{i,j}}\right)
				\end{split}
			\end{align*}
			
			Let $ \partial M/\partial g_{i,j}=0 $, then
			\begin{align}
				\begin{split}
					g_{i,j}+&\lambda\left(g_{i,j-2}-4g_{i,j-1}+6g_{i,j}-4g_{i,j+1}+g_{i,j+2}\right)\\+&\lambda\left(g_{i-2,j}-4g_{i-1,j}+6g_{i,j}-4g_{i+1,j}+g_{i+2,j}\right)\\=&z_{i,j}
				\end{split}
			\end{align}
			
			\item 	When $ 3\le i\le m-2, j=n-1 $, $ \partial M/\partial g_{i,n-1} $ is
			\begin{align*}
				\begin{split}
					\frac{\partial M}{\partial g_{i,n-1}}=&\frac{\partial r\left(i,n-1\right)}{\partial g_{i,n-1}}\\+&\lambda\left(\frac{\partial p\left(i,n-1\right)}{\partial g_{i,n-1}}+\frac{\partial p\left(i,n\right)}{\partial g_{i,n-1}}\right)\\+&\lambda\Big(\frac{\partial q\left(i,n-1\right)}{\partial g_{i,n-1}}+\frac{\partial q\left(i+1,n-1\right)}{\partial g_{i,n-1}}\\+&\frac{\partial q\left(i+2,n-1\right)}{\partial g_{i,n-1}}\Big)
				\end{split}
			\end{align*}
			
			Let $ \partial M/\partial g_{i,n-1}=0 $, then
			\begin{align}
				\begin{split}
					g_{i,n-1}+&\lambda\left(g_{i,n-3}-4g_{i,n-2}+5g_{i,n-1}-2g_{i,n}\right)\\+&\lambda\Big(g_{i-2,n-1}-4g_{i-1,n-1}+6g_{i,n-1}\\-&4g_{i+1,n-1}+g_{i+2,n-1}\Big)=z_{i,n-1}
				\end{split}
			\end{align}
			\item 	When $ 3\le i\le m-2, j=n $, $ \partial M/\partial g_{i,n} $ is
			\begin{align*}
				\begin{split}
					\frac{\partial M}{\partial g_{i,n}}=&\frac{\partial r\left(i,n\right)}{\partial g_{i,n}}+\lambda\frac{\partial p\left(i,n\right)}{\partial g_{i,n}}\\+&\lambda\Big(\frac{\partial q\left(i,n\right)}{\partial g_{i,n}}+\frac{\partial q\left(i+1,n\right)}{\partial g_{i,n}}+\frac{\partial q\left(i+2,n\right)}{\partial g_{i,n}}\Big)
				\end{split}
			\end{align*}
			
			Let $ \partial M/\partial g_{i,n}=0 $, then
			\begin{align}
				\begin{split}
					g_{i,n}+&\lambda\left(g_{i,n-2}-2g_{i,n-1}+g_{i,n}\right)\\+&\lambda\left(g_{i-2,n}-4g_{i-1,n}+6g_{i,n}-4g_{i+1,n}+g_{i+2,n}\right)\\=&z_{i,n}
				\end{split}
			\end{align}
		\end{subequations}
	\end{itemize}
	\item When $ i=m-1 $
	\begin{itemize}
		\item 	When $ i=m-1 $, $ j=1 $, $ \partial M/\partial g_{m-1,1} $ is
		\begin{align*}
			\begin{split}
				\frac{\partial M}{\partial g_{m-1,1}}=&\frac{\partial r\left(m-1,1\right)}{\partial g_{m-1,1}}+\lambda\frac{\partial p\left(m-1,3\right)}{\partial g_{m-1,1}}\\+&\lambda\left(\frac{\partial q\left(m-1,1\right)}{\partial g_{m-1,1}}+\frac{\partial q\left(m,1\right)}{\partial g_{m-1,1}}\right)
			\end{split}
		\end{align*}
		\begin{subequations}
			Let $ \partial M/\partial g_{m-1,1}=0 $, then
			
			\begin{align}
				\begin{split}
					g_{m-1,1}+&\lambda\left(g_{m-1,1}-2g_{m-1,2}+g_{m-1,3}\right)\\+&\lambda\left(g_{m-3,1}-4g_{m-2,1}+5g_{m-1,1}-2g_{m,1}\right)\\=&z_{m-1,1}
				\end{split}
			\end{align}
			
			\item		When $ i=m-1 $, $ j=2 $,  $\partial M/\partial g_{m-1,2} $ is
			\begin{align*}
				\begin{split}
					\frac{\partial M}{\partial g_{m-1,2}}=&\frac{\partial r\left(m-1,2\right)}{\partial g_{m-1,2}}\\+&\lambda\left(\frac{\partial p\left(m-1,3\right)}{\partial g_{m-1,2}}+\frac{\partial p\left(m-1,4\right)}{\partial g_{m-1,2}}\right)\\+&\lambda\left(\frac{\partial q\left(m-1,2\right)}{\partial g_{m-1,2}}+\frac{\partial q\left(m,2\right)}{\partial g_{m-1,2}}\right)
				\end{split}
			\end{align*}
			
			Let $ \partial M/\partial g_{m-1,2}=0 $, then
			\begin{align}
				\begin{split}
					g_{m-1,2}+&\lambda\Big(-2g_{m-1,1}+5g_{m-1,2}\\-&4g_{m-1,3}+g_{m-1,4}\Big)\\+&\lambda\left(g_{m-3,2}-4g_{m-2,2}+5g_{m-1,2}-2g_{m,2}\right)\\=&z_{m-1,2}
				\end{split}
			\end{align}
			\item 	When $ i=m-1, 3\le j\le n-2 $, $ \partial M/\partial g_{m-1,j} $ is
			\begin{align*}
				\frac{\partial M}{\partial g_{m-1,j}}=&\frac{\partial r\left(m-1,j\right)}{\partial g_{m-1,j}}\\+&\lambda\Big(\frac{\partial p\left(m-1,j\right)}{\partial g_{m-1,j}}+\frac{\partial p\left(m-1,j+1\right)}{\partial g_{m-1,j}}\\+&\frac{\partial p\left(m-1,j+2\right)}{\partial g_{m-1,j}}\Big)\\+&\lambda\left(\frac{\partial q\left(m-1,j\right)}{\partial g_{m-1,j}}+\frac{\partial q\left(m,j\right)}{\partial g_{m-1,j}}\right)
			\end{align*}
			
			Let $ \partial M/\partial g_{m-1,j}=0 $, then
			\begin{align}
				\begin{split}
					g_{m-1,j}+&\lambda\Big(g_{m-1,j-2}-4g_{m-1,j-1}\\+&6g_{m-1,j}-4g_{m-1,j+1}+g_{m-1,j+2}\Big)\\+&\lambda\left(g_{m-3,j}-4g_{m-2,j}+5g_{m-1,j}-2g_{m,j}\right)\\=&z_{m-1,j}
				\end{split}
			\end{align}
			
			\item 	When $ i=m-1 $, $ j=n-1 $, $ \partial M/\partial g_{m-1,n-1} $ is
			\begin{align*}
				\begin{split}
					\frac{\partial M}{\partial g_{m-1,n-1}}=&\frac{\partial r\left(m-1,n-1\right)}{\partial g_{m-1,n-1}}\\+&\lambda\Big(\frac{\partial p\left(m-1,n-1\right)}{\partial g_{m-1,n-1}}+\frac{\partial p\left(m-1,n\right)}{\partial g_{m-1,n-1}}\Big)\\+&\lambda\Big(\frac{\partial q\left(m-1,n-1\right)}{\partial g_{m-1,n-1}}+\frac{\partial q\left(m,n-1\right)}{\partial g_{m-1,n-1}}\Big)
				\end{split}
			\end{align*}
			
			Let $ \partial M/\partial g_{m-1,n-1}=0 $, then
			
			\begin{align}
				\begin{split}
					g_{m-1,n-1}+&\lambda\Big(g_{m-1,n-3}-4g_{m-1,n-2}\\+&5g_{m-1,n-1}-2g_{m-1,n}\Big)\\+&\lambda\Big(g_{m-3,n-1}-4g_{m-2,n-1}\\+&5g_{m-1,n-1}-2g_{m,n-1}\Big)\\=&z_{m-1,n-1}
				\end{split}
			\end{align}
			
			\item 	When $ i=m-1 $, $ j=n $, $ \partial M/\partial g_{m-1,n} $ is 
			\begin{align*}
				\begin{split}
					\frac{\partial M}{\partial g_{m-1,n}}=&\frac{\partial r\left(m-1,n\right)}{\partial g_{m-1,n}}+\lambda\frac{\partial p\left(m-1,n\right)}{\partial g_{m-1,n}}\\+&\lambda\left(\frac{\partial q\left(m-1,n\right)}{\partial g_{m-1,n}}+\frac{\partial q\left(m,n\right)}{\partial g_{m-1,n}}\right)
				\end{split}
			\end{align*}
			
			Let $ \partial M/\partial g_{m-1,n}=0 $, then
			\begin{align}
				\begin{split}
					g_{m-1,n}+&\lambda\left(g_{m-1,n-2}-2g_{m-1,n-1}+g_{m-1,n}\right)\\+&\lambda\left(g_{m-3,n}-4g_{m-2,n}+5g_{m-1,n}-2g_{m,n}\right)\\=&z_{m-1,n}
				\end{split}
			\end{align}
		\end{subequations}
	\end{itemize}
	\item When $ i=m $
	\begin{itemize}
		\item  When $ i=m $, $ j=1 $, $ \partial M/\partial g_{m,1} $ is
		\begin{align*}
			\frac{\partial M}{\partial g_{m,1}}=\frac{\partial r\left(m,1\right)}{\partial g_{m,1}}+\lambda\frac{\partial p\left(m,3\right)}{\partial g_{m,1}}+\lambda\frac{\partial q\left(m,1\right)}{\partial g_{m,1}}
		\end{align*}
		
		Let $ \partial M/\partial g_{m,1}=0 $, then
		\begin{subequations}
			\begin{align}
				\begin{split}
					g_{m,1}+&\lambda\left(g_{m,1}-2g_{m,2}+g_{m,3}\right)\\+&\lambda\left(g_{m-2,1}-2g_{m-1,1}+g_{m,1}\right)=z_{m,1}
				\end{split}
			\end{align}
			\item 	When $ i=m $, $ j=2 $, $ \partial M/\partial g_{m,2} $ is		
			\begin{align*}
				\begin{split}
					\frac{\partial M}{\partial g_{m,2}}=&\frac{\partial r\left(m,2\right)}{\partial g_{m,2}}\\+&\lambda\left(\frac{\partial p\left(m,3\right)}{\partial g_{m,2}}+\frac{\partial p\left(m,4\right)}{\partial g_{m,2}}\right)+\lambda\frac{\partial q\left(m,2\right)}{\partial g_{m,2}}
				\end{split}
			\end{align*}
			
			Let $ \partial M/\partial g_{m,2}=0 $, then
			\begin{align}
				\begin{split}
					g_{m,2}+&\lambda\left(-2g_{m,1}+5g_{m,2}-4g_{m,3}+g_{m,4}\right)\\+&\lambda\left(g_{m-2,2}-2g_{m-1,2}+g_{m,2}\right)=z_{m,2}
				\end{split}
			\end{align}
			\item 	When $ i=m $, $ 3\le j\le n-2 $, $ \partial M/\partial g_{m,j} $ is
			\begin{align*}
				\begin{split}
					\frac{\partial M}{\partial g_{m,j}}=&\frac{\partial r\left(m,j\right)}{\partial g_{m,j}}\\+&\lambda\Big(\frac{\partial p\left(m,j\right)}{\partial g_{m,j}}+\frac{\partial p\left(m,j+1\right)}{\partial g_{m,j}}\\+&\frac{\partial p\left(m,j+2\right)}{\partial g_{m,j}}\Big)+\lambda\frac{\partial q\left(m,j\right)}{\partial g_{m,j}}
				\end{split}
			\end{align*}
			
			Let $ \partial M/\partial g_{m,j}=0 $, then
			\begin{align}
				\begin{split}
					g_{m,j}+&\lambda\Big(g_{m,j-2}-4g_{m,j-1}+6g_{m,j}\\-&4g_{m,j+1}+g_{m,j+2}\Big)\\+&\lambda\left(g_{m-2,j}-2g_{m-1,j}+g_{m,j}\right)=z_{m,j}
				\end{split}
			\end{align}
			\item  When $ i=m $, $ j=n-1 $, $ \partial M/\partial g_{m,n-1} $ is
			\begin{align*}
				\begin{split}
					\frac{\partial M}{\partial g_{m,n-1}}=&\frac{\partial r\left(m,n-1\right)}{\partial g_{m,n-1}}\\+&\lambda\left(\frac{\partial p\left(m,n-1\right)}{\partial g_{m,n-1}}+\frac{\partial p\left(m,n\right)}{\partial g_{m,n-1}}\right)\\+&\lambda\frac{\partial q\left(m,n-1\right)}{\partial g_{m,n-1}}
				\end{split}
			\end{align*}
			
			Let $ \partial M/\partial g_{m,n-1}=0 $, then
			\begin{align}
				\begin{split}
					g_{m,n-1}+&\lambda\left(g_{m,n-3}-4g_{m,n-2}+5g_{m,n-1}-2g_{m,n}\right)\\+&\lambda\left(g_{m-2,n-1}-2g_{m-1,n-1}+g_{m,n-1}\right)\\=&z_{m,n-1}
				\end{split}
			\end{align}
			\item 	When $ i=m $, $ j=n $, $ \partial M/\partial g_{m,n} $ is
			\begin{align*}
				\frac{\partial M}{\partial g_{m,n}}=\frac{\partial r\left(m,n\right)}{\partial g_{m,n}}+\lambda\frac{\partial p\left(m,n\right)}{\partial g_{m,n}}+\lambda\frac{\partial q\left(m,n\right)}{\partial g_{m,n}}
			\end{align*}
			
			Let $ \partial M/\partial g_{m,n}=0 $, then
			\begin{align}
				\begin{split}\label{77e}
					g_{m,n}+&\lambda\left(g_{m,n-2}-2g_{m,n-1}+g_{m,n}\right)\\+&\lambda\left(g_{m-2,n}-2g_{m-1,n}+g_{m,n}\right)=z_{m,n}
				\end{split}
			\end{align}
		\end{subequations}
	\end{itemize}
\end{enumerate}

Therefore, combining \textit{Eqn}.\eqref{73a} to \textit{Eqn}.\eqref{77e}, the final result can be equivalently expressed as

\begin{align*}
	\textit{\textbf{G}}+\lambda\left(\textit{\textbf{GT}}+\textit{\textbf{HG}}\right)=\textit{\textbf{Z}}
\end{align*}

\subsection*{Appendix B. Positive Semidefinite Proof of Matrix \textit{\textbf{N}}}\label{Appendix B}

Assuming any $ \textbf{\textit{x}}=\left[x_1,x_2,\cdots,x_k\right]^T\neq \mathbf{0}\in \mathbb{R}^k $, and $ \textbf{\textit{N}}\in \mathbb{R}^{k\times k} $ is a real symmetry matrix, then let $ f\left(\textbf{\textit{x}}\right) $ be

\begin{align}
	f\left(\textit{\textbf{x}}\right)=\textit{\textbf{x}}^T\textit{\textbf{Nx}}
\end{align}

Then

\begin{align}
	f\left(\textit{\textbf{x}}\right)=\sum_{i=3}^{k}\left(x_i-2x_{i-1}+x_{i-2}\right)^2\geq0
\end{align}

Therefore, $ \textit{\textbf{N}} $ is a positive semidefinite and real symmetric matrix. Finally, we estimate the eigenvalue of matrix\textit{\textbf{ N}}. The $ k $ Gerschgorin circles of the matrix \textit{\textbf{N}} are, respectively

\begin{align}
	\begin{split}
		S_{1,k}=&\left\{z\in \mathbb{R}|\left|z-1\right|\le3\right\}
		\\	S_{2,k-1}=&\left\{z\in \mathbb{R}|\left|z-5\right|\le7\right\}
		\\	S_{3,\cdots,k-2}=&\left\{z\in \mathbb{R}|\left|z-6\right|\le10\right\}		
	\end{split}
\end{align}

Meanwhile, since \textit{\textbf{N}} is a positive semidefinite matrix, and therefore the eigenvalues $ \zeta $ of matrix \textit{\textbf{N}} satisfy $ 0\le\zeta\le16 $.

\bibliographystyle{Bibliography/IEEEtranTIE}
\bibliography{Bibliography/BIB_xx-TIE-xxxx}

\end{document}